\documentclass[twocolumn]{aastex63}
\usepackage{ulem}

\shorttitle{the radial IMF of M89}
\shortauthors{Lonoce et al.}

\begin{document}

\title{The Stellar Initial Mass Function and Population Properties of M89 from Optical and NIR Spectroscopy: Addressing Biases in Spectral Index Analysis\footnote{This paper includes data gathered with the 6.5 meter Magellan Telescopes located at Las Campanas Observatory, Chile.}}

\correspondingauthor{Ilaria Lonoce}
\email{ilonoce@uchicago.edu}

\author[0000-0001-8421-1005]{I. Lonoce}
\affiliation{Department of Astronomy $\&$ Astrophysics, The University of Chicago, 5640 South Ellis Avenue, Chicago, IL 60637, USA}

\author[0000-0002-0160-7221]{A. Feldmeier-Krause}
\affiliation{Department of Astronomy $\&$ Astrophysics, The University of Chicago, 5640 South Ellis Avenue, Chicago, IL 60637, USA}
\affiliation{Max-Planck-Institut f\"ur Astronomie, K\"onigstuhl 17, 69117, Heidelberg, Germany}

\author[0000-0003-3431-9135]{Wendy L. Freedman}
\affiliation{Department of Astronomy $\&$ Astrophysics \& the Kavli Institute for Cosmological Physics, The University of Chicago, 5640 South Ellis Avenue, Chicago, IL 60637, USA}



\begin{abstract}

The complexity of constraining the stellar initial mass function (IMF) in early-type galaxies cannot be overstated, given the necessity of both very high signal-to-noise (S/N) data and the difficulty of breaking the strong degeneracies that occur among several stellar population parameters including age, metallicity and elemental abundances. With this paper, the second in a series, we present a detailed analysis of the biases that can occur when retrieving the IMF shape by exploiting both optical and NIR IMF sensitive spectral indices. As a test case, here we analyze data for the nearby galaxy M89, for which we have high S/N spectroscopic data that cover the $3500-9000$\AA\space spectral region and allow us to study the radial variation of the stellar population properties out to $1$R$_e$. Carrying out parallel simulations that mimic the retrieval of all the explored stellar parameters from a known input model, we quantify the amount of bias at each step of our analysis. From more general simulations we conclude that to accurately retrieve the IMF, it is necessary not only to retrieve accurate estimates of the age and metallicity, but also of all the elemental abundances that the spectral index fits are sensitive to. With our analysis technique applied to M89, we find consistency with a bottom-heavy IMF with a negative gradient from the center to half R$_e$ when using the \citet{conroy18} as well as \citet{vazdekis16} EMILES stellar population models. We find agreement both with a parallel full spectral fitting of the same data and with literature results.

\end{abstract}



\section{Introduction} 
\label{sec:intro}

The determination of the stellar initial mass function (IMF) is one of the greatest challenges to our understanding of the formation and evolution of galaxies. 
Perhaps one of the more interesting aspects in studying the IMF, is to understand if the initial physical processes that give rise to the initial mass distribution of stars are universal for all stellar systems or if they, for example, depend on their environment, on overall mass or chemical composition. Over the past decade, a growing number of observational studies have concluded that there is evidence for a non universality of the IMF amongst galaxies, and even within the same galaxy (\citealt{larson98, larson05}; \citealt{cappellari12}; \citealt{hopkins13}; \citealt{ferreras13}; \citealt{labarbera13}; \citealt{labarbera19}; \citealt{chabrier14}; \citealt{conroy14}; \citealt{spiniello14}; \citealt{martin-navarro15}; \citealt{vandokkum11, vandokkum17}; \citealt{sarzi18}; \citealt{barbosa20,barbosa21}). This means that theoretical models must be able to explain a difference in the formation processes not only from galaxy to galaxy, but also within the different regions of the same galaxy. 

There are several approaches to deriving the IMF in unresolved stellar populations (SP): by means of strong gravitational lenses (\citealt{treu10}; \citealt{auger10}; \citealt{leier16}) or stellar kinematics (\citealt{cappellari12, cappellari13}; \citealt{thomasj11c}; \citealt{tortora13}); both probes are able to constrain the mass-to-light ratio and then the IMF with an assumption about the dark matter profile of the stellar system. An independent method is to analyze spectral features that are sensitive to the ratio between dwarf and giant stars (\citealt{cenarro03}; \citealt{vandokkum10, vandokkum12}; \citealt{smith12}; \citealt{ferreras13}; \citealt{labarbera13, labarbera17, labarbera19}; \citealt{spiniello14}; \citealt{martin-navarro15}; \citealt{sarzi18}; \citealt{parikh18}). This last method has been widely applied to analyze the IMF of local early-type galaxies, which have been found to be generally more \textit{bottom-heavy}, that is, with a higher contribution of low-mass stars with respect to the Milky Way mass distribution; in particular, in their centers (\citealt{martin-navarro15}; \citealt{labarbera16, labarbera17, labarbera19}; \citealt{sarzi18};  \citealt{barbosa20}). Negative IMF slope gradients in early-type galaxies are not, however,  always confirmed, as shown in \citet{alton17} and \citet{vaughan18}. \\

In the few last years, many authors have taken advantages of the efficient full spectral fitting technique to derive the SP properties, thus extracting information not only from the regions of individual sensitive features, but making use of the entire spectrum (e.g. \citealt{vandokkum17,conroy17,vaughan18}). This method allows the derivation of many parameters (for example, elemental abundances, multiple stellar population components, age, IMF, velocity distributions, etc.) because of the large number of degrees of freedom in the fit. In parallel, other authors (e.g. \citealt{spiniello14,martin-navarro15,labarbera16,parikh18}) still choose to focus on the high-density information indicators, i.e. spectral indices, for their analysis of SP properties.
The advantage of deriving the IMF shape exploiting gravity-sensitive spectral features is that, in principle, with only availability of a small region of the spectrum, one is able to quantify the prevalence of low or high mass stars in the underlying SP just by analyzing a few important features like TiO1, TiO2 or the calcium triplet CaT (e.g. \citealt{martin-navarro15}; \citealt{labarbera17}; \citealt{sarzi18}). However, the IMF-sensitive features are also sensitive to other important SP properties such as age, stellar metallicity, and many elemental abundances, as already pointed out by many authors \citep[e.g.,][]{zieleniewski15, mcconnell16, vandokkum17, vaughan18}. The level of sensitivity to each SP property varies from index to index, and it also varies depending on the values of the other SP parameters. Moreover, the variation of indices due to variations of the IMF is generally mild and not sufficiently high to exclude the variation of other SP parameters. 
If not carefully taken into account, this additional complexity naturally gives rise to biases in the IMF determination due to the high level of degeneracy amongst stellar parameters.

For this reason, a common approach when constraining the IMF fitting spectral features, is to focus the analysis on a small number of indices in order to limit the effect of degeneracy among parameters (e.g. \citealt{martin-navarro15}; \citealt{parikh18}; \citealt{dominguez19}; \citealt{eftekhari19}). 
However, some authors have concluded that their results on IMF gradients could alternatively be explained by trends of some elemental abundances, insufficiently constrained due to the lack of enough indicators (\citealt{ ASL17,zieleniewski17,vaughan18}). Moreover, fitting small sets of indices results in a poorer constraint on the other important stellar parameters such as age and metallicity. In \citet{vandokkum17}, after a complete analysis of the radial chemical composition of their galaxy sample, the authors considered whether it is nevertheless possible that their IMF measures could be further affected by leftover degeneracies. A first approach to test the accuracy of the IMF retrieval has been performed by \citet{conroy17} by means of mock spectra; however, in that case, the main purpose was to verify the minimum required S/N to obtain unbiased results, and no exploration of biases due to elemental abundance retrieval is addressed. Moreover, the \citet{conroy17} analysis used the full spectral fitting technique and not a set of spectral indices.
It is thus highly valuable to obtain a clearer picture and to explore quantitatively and systematically the underlying degeneracies that affect the retrieval of the IMF when dealing with spectral indices.

In the first paper of this series, i.e. Feldmeier-Krause et al. (2020; hereafter, F20), we presented our approach to the IMF determination of a local early-type galaxy exploiting high-quality spectroscopic data. In particular, we explored the complexity of determining the IMF using different methodologies all based on fitting IMF sensitive spectral features. Our main result was that the IMF result depended on the adopted method, highlighting the challenging nature of this problem, and likely hinting that each method is affected in a different way by degeneracy. Having a deep knowledge of the biases underlying each applied method is thus essential to reliably constrain the IMF. This is also necessary to compare the results coming from different analyses that are intrinsically different due to, for example, data availability, and thus gaining the desired statistical significance of the overall IMF trend.\\

In paper F20 we began an investigation of the biases that affect the IMF retrieval by performing a full spectral fitting and spectral index fitting, building customized simulations that reproduced the same procedures as used to analyze real data for each applied method. \\
In this second paper, we present an expanded version of this tool, providing simulation results that can be useful for a wide range of spectral indices analysis, in order to help quantify the effects of degeneracies.

In particular, as a case test, in this paper we focus on the retrieval of the radial variation of the IMF in the local early-type M89 by means of spectral indices analysis. We make use of new, excellent-quality  spectroscopic data in the optical and NIR bands from the Inamori-Magellan Areal Camera \& Spectrograph (IMACS) on the Magellan-Baade telescope. This complete spectroscopic data set allows us to use all of the information enclosed in the spectral features of full high-S/N spectra from $3500$ to $7000$\AA\space and in the CaT region without gaps. Full spectral fitting of the M89 data set has been performed as well, but in this paper it will be used only for comparison purposes as discussed in Section \ref{sec:fsf}.
Parallel simulations allow us to quantify and take into account the real biases present at each step of the analysis, as hinted above. In turn, we are able to deal with a large number of optical and NIR (from $4000$\AA\space to $9000$\AA) spectral indices, which depend simultaneously on many stellar parameters, and thus derive a solid measure of the IMF shape as a function of the galactic radius, as well as other important properties such as age, metallicity, and many elemental abundances. The choice of our target, M89,  also allows us to compare our final results with those of two independent analyses in the literature (i.e. \citealt{martin-navarro15,vandokkum17}).

The paper is organized as follows: in Section \ref{sec:data} we present the optical and NIR spectroscopic data of our target galaxy M89, together with the details of the data reduction; in Section \ref{sec:analysis} we discuss our method analysis for retrieval of the IMF and the results for the kinematic profile of the object; details about  the simulations are described in Subsection \ref{sec:sim} and a full description of the steps of the analysis is given in Subsection \ref{sec:steps}; the final results of the analysis are discussed in Section \ref{sec:discussion}, and the summary and our conclusions are given in Section \ref{sec:conclusions}.


\section{M89 Data}
\label{sec:data}

M89 is a massive nearby S0 galaxy belonging to the Virgo cluster, with a V band magnitude of $9.75$.
The target has been observed with IMACS (Inamori-Magellan Areal Camera \& Spectrograph, \citealt{dressler06}) on the Magellan-Baade $6.5$m telescope at Las Campanas Observatory (Chile) in two observing runs, one in May 2018 and one in April-May 2019. We collected longslit spectroscopic data with 3 different grating angles (GA) to cover the wavelength range from about $3500$\AA\space up to $10000$\AA\space with a constant spectral resolution FWHM$\sim5.5$\AA\space (2\farcs5 slit). For the optical range we chose the grating 600-8.6 selecting both GA$=10$\fdg$46$ and GA$=9$\fdg$71$, which allowed us to cover the entire spectral range from $3500$ to $7100$\AA. The IMACS CCD is divided in four chips with wavelength gaps in between. Our observing strategy was to ensure that all spectral indices sensitive to IMF variations were covered.
For the NIR region we used the 600-13.0 grating with GA$=17$\fdg$11$ to cover the CaT region around $8500$\AA. We integrated for 1h for each GA configuration. Details of the observations are summarized in Table \ref{tab:infoM89}. With the above described setup we were able to reach exquisite signal-to-noise ratios (S/N) both in the center of the galaxy (up to $500$/\AA) and in the outer region around $1$ effective radius (R$_e$), where the S/N is not lower than $100$/\AA.

\begin{table*}
 \caption{Information on the Observations for each GA Configuration: Period of Observation, Grating, Slit Width, Spectral Range, and Total Exposure Time.}
 \label{tab:infoM89}
 \begin{tabular}{lccccccc}
 \hline
 \hline
 Grating Angle  & Period & Grating & Slit Width   & Spectral Range & Exposure Time \\
    ($\degr$)   &        &         &    (arcsec)  & (\AA)          & (sec)         \\
 \hline
 10.46    & May 2018 & 600-8.6 & 2.5 & 3900-7100 & 3600  \\
 9.71     & May 2018 & 600-8.6 & 2.5 & 3500-6700 & 3600  \\
 17.11    & April-May 2019 & 600-13.0 & 2.5 & 7500-10500 & 3600 \\
 \hline
\end{tabular}
\end{table*}

\subsection{Data Reduction}

The data reduction has been performed with both IRAF \citep{tody93}
tools and custom IDL scripts. Standard reduction steps have been performed on the raw data starting from bias, bad-pixel and cosmic-ray removal, and wavelength calibration (in air wavelengths). All of low level distortions of the 2D spectra have been rectified during the FITCOORD and TRANSFORM fits with IRAF. 
Because of the long length of the slit, i.e. $\sim15$ arcmin, it was possible to remove the sky background exploiting the scientific frames themselves, considering as background all of the flux beyond $2$R$_e$ from the center of the galaxy. Few pixels are still left with sky residuals in the fainter regions due to a different FWHM of the strong emission lines in the upper versus lower region of the CCD. Finally, the three $1200$s 2D frames were added together before being extracted. 

\begin{figure*}[ht!]
\begin{centering}
\includegraphics[width=18.5cm]{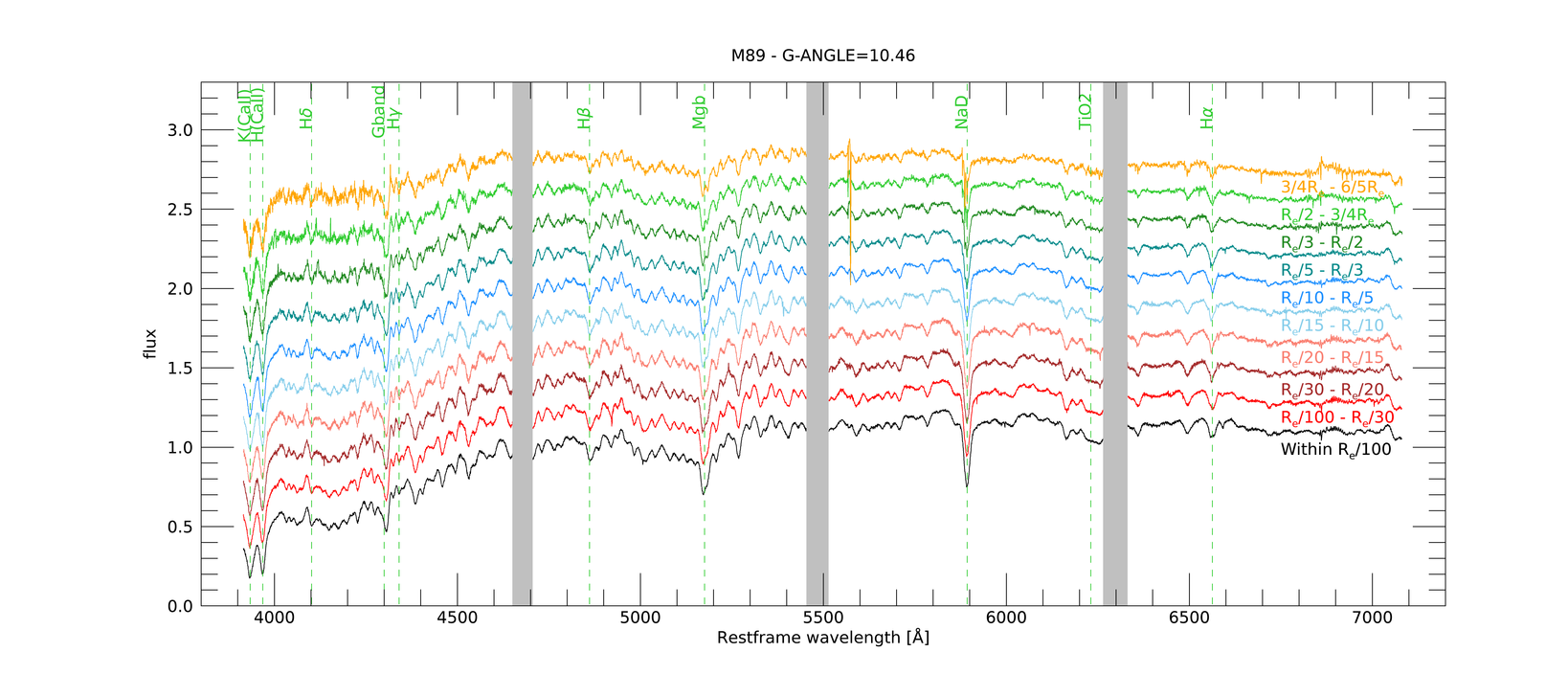}
\includegraphics[width=18.5cm]{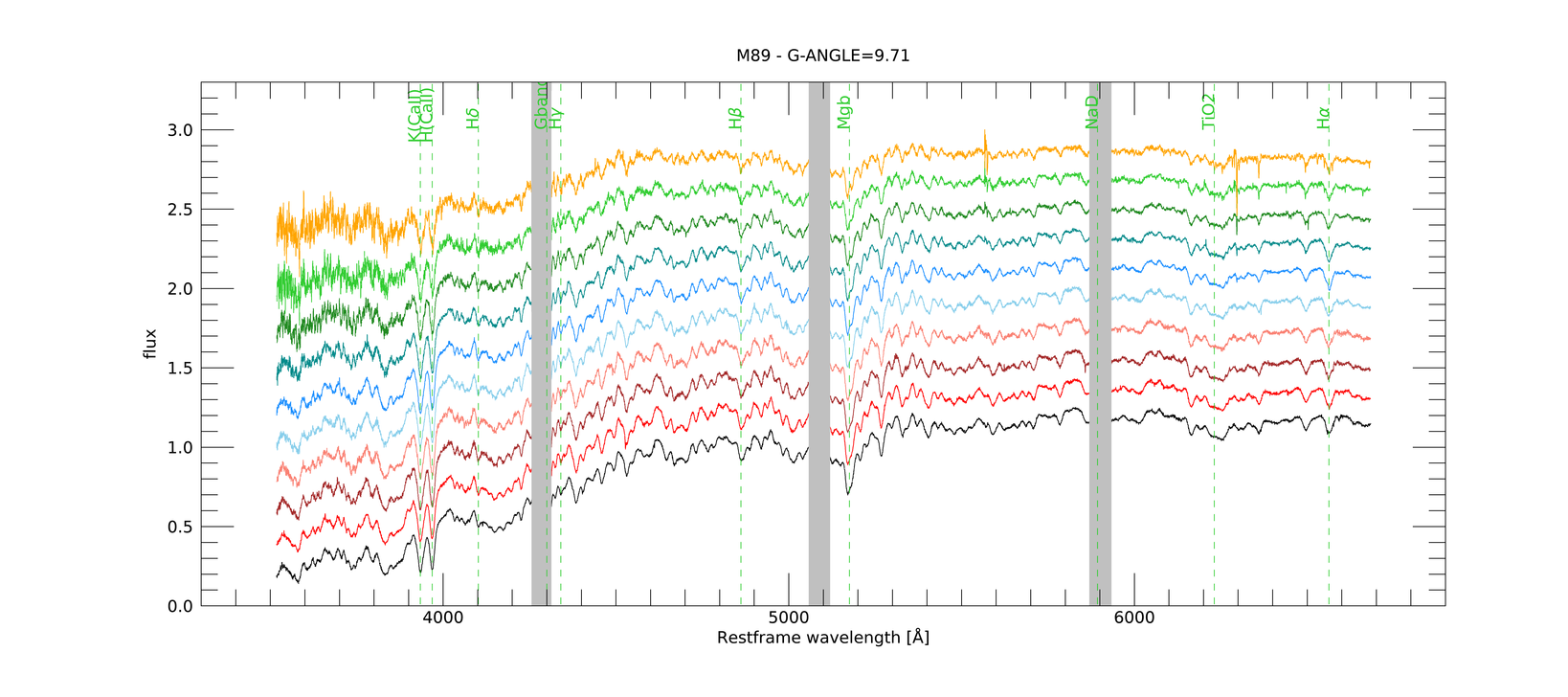}
\includegraphics[width=11cm]{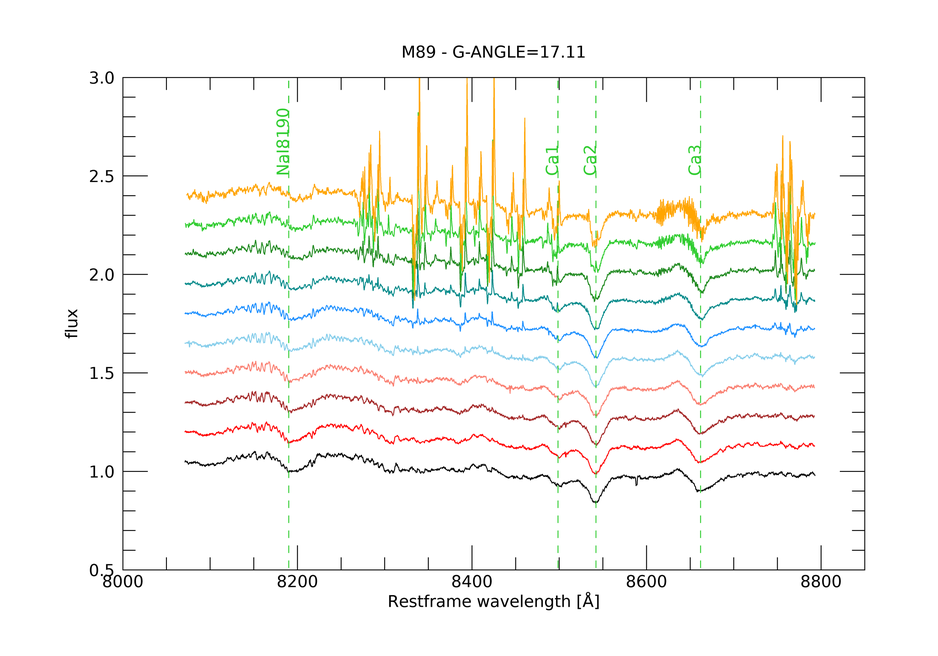}
\caption{\small{Reduced M89 spectra for the three GA configurations. Spectra from the central regions to the outer ones are shown in different colors, ordered from bottom to top. Vertical gray bands cover the chip gaps. Green vertical dashed lines indicate some reference spectral features.}}
\label{fig:spectra}
\end{centering}
\end{figure*}

Since the purpose of this work is to study the radial variation of the SP properties, the final longslit 2D spectrum was divided into symmetric galactocentric bands and each of them extracted and analyzed separately.
The 1D extraction in each band was performed using a custom IDL script that applied, in parallel, the respective spectroscopic flat field spectra extracted in the same way.
With the high quality of our data we were able to divide the light profile of the galaxy into both its northern and southern parts, into 10 regions with similar high S/N, up to $\sim1$R$_e$, as shown in Fig. \ref{fig:spectra} and in Table \ref{tab:results}. The half luminosity effective radius is R$_e=31.2$ arcsec in the I band \citep{vandokkum17}.

Relative flux calibration has been performed by means of standard stars observed for each GA on each night. Standard star spectra were reduced with the same procedure as the science spectra, with the exception of the 1D extraction which has been done using APSUM (IRAF). 
We performed the telluric removal on the spectra above $5500$\AA\space using the software MOLECFIT (\citealt{molecfit1,molecfit2}).

Due to a radial velocity gradient, before summing together the extracted spectra above and below the galaxy center, each radial spectrum had to be brought to the restframe. We have therefore studied the radial kinematics of M89 using the IDL version of the software PPXF \citep{cappellari17}. PPXF performs full spectral fitting on the 1D spectra using a set of SP models. Knowing the instrumental resolution and a proxy of the velocity dispersion it retrieves a best-fit solution providing the 4 Gauss-Hermite moments (radial velocity, velocity dispersion, h3 skewness and h4 kurtosis) with their respective errors. As input models we used the MILES models \citep{vazdekis10}, with a spectral resolution of $2.5$\AA\space and a fixed Chabrier IMF \citep{chabrier03}. The instrumental resolution of IMACS has been measured from the same arc lines frames used for the wavelength calibration, for each GA. Its values have a mean of $5.5$\AA\space and it is constant over the wavelength range. During the PPXF fit, residual bad pixels have been masked. We performed the fit on each CCD chip separately and despite small rigid shifts due probably to small biases of the wavelength calibration, all chips for all GA configurations show the same radial velocity trend: an inner core within $0.1-0.2$R$_e$ with a visible rotation which flattens in the outer regions. This is consistent with the 2D kinematics map of M89 of the ATLAS$^{3D}$ project \citep{krajnovic11}. The velocity dispersion profile will be discussed in the following sections where the PPXF fit has been run on the final spectra (i.e. where the two sides of the galaxy have been added together). The results are in agreement and consistent with the published trends, showing a very symmetric profile which assured us that a velocity dispersion correction to the  upper and lower spectra was negligible. Higher moments h3 and h4 have flat trends with values below $0.1$. All radial spectra have been then brought to rest-frame and the corresponding upper and lower regions of the galaxy summed together. The final spectra for each GA are shown in Figure \ref{fig:spectra}. 

\begin{figure*}[ht!]
\begin{centering}
\includegraphics[width=8.cm]{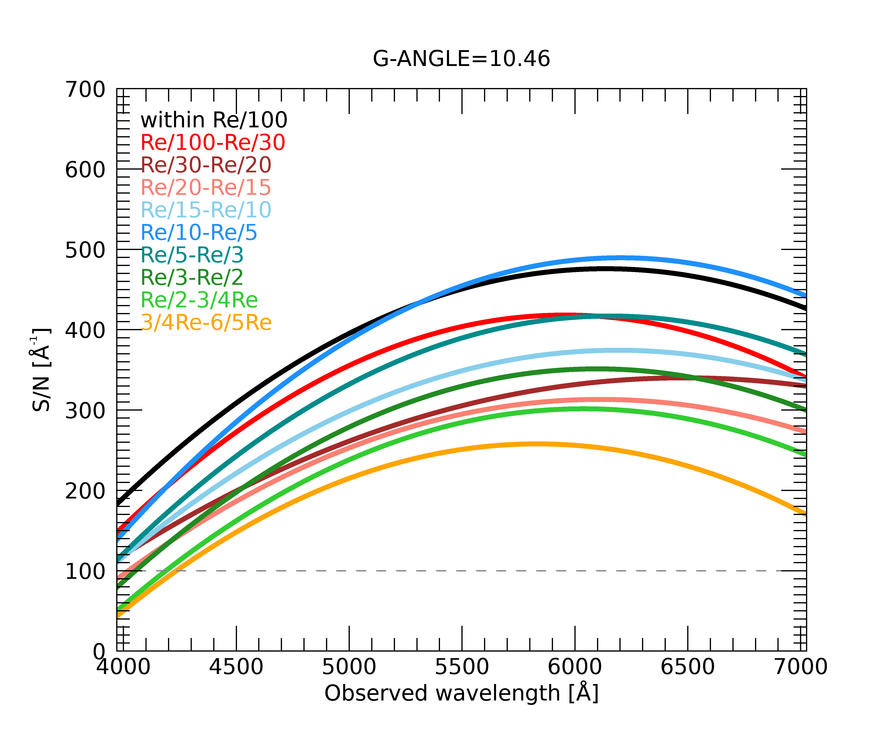}
\includegraphics[width=8.cm]{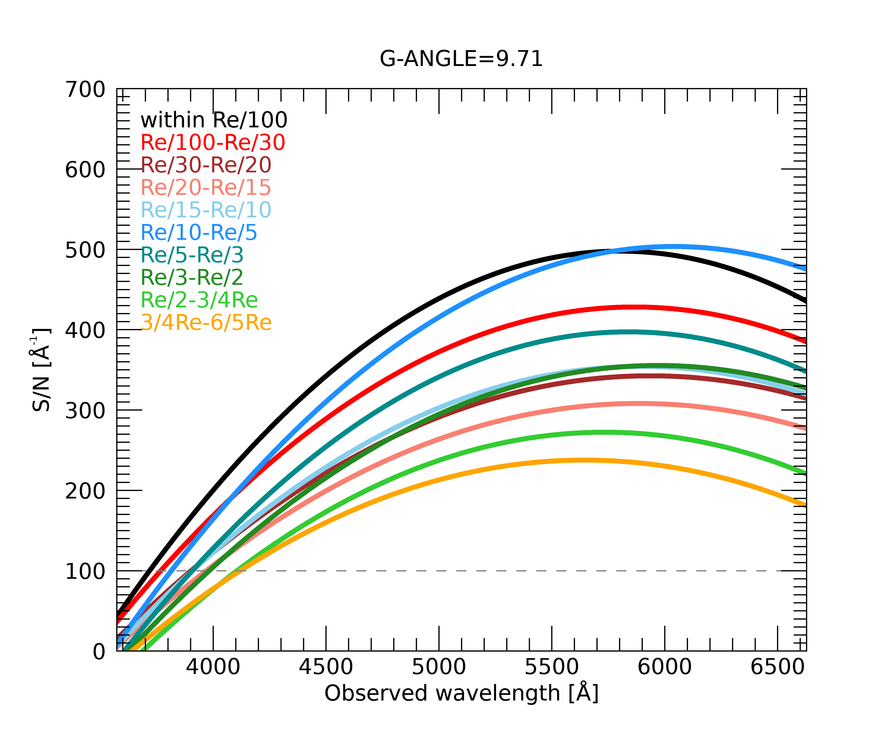}
\includegraphics[width=8.cm]{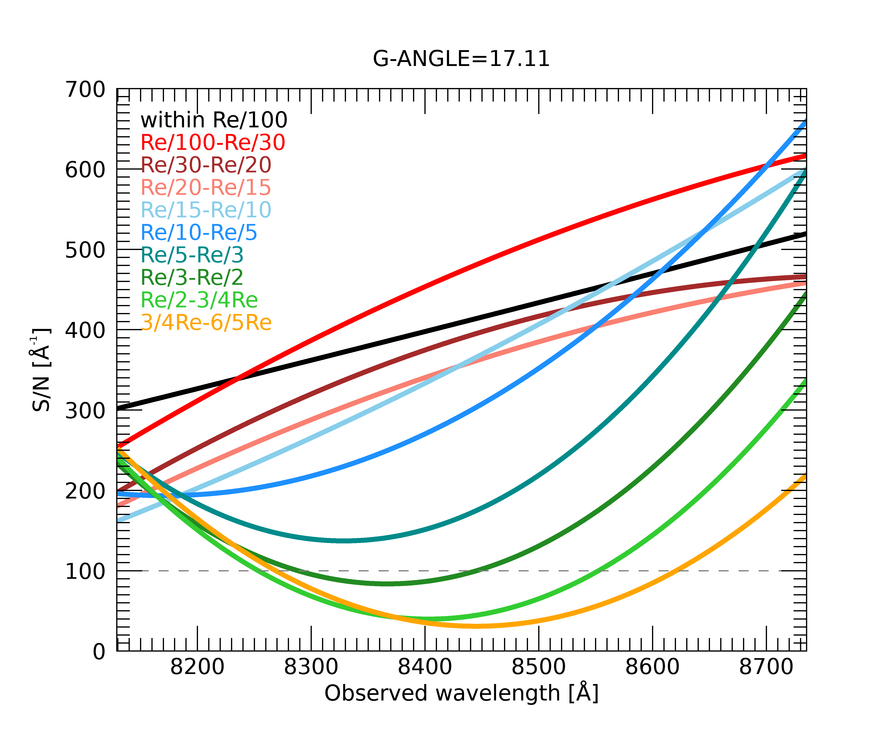}
\includegraphics[width=8.cm]{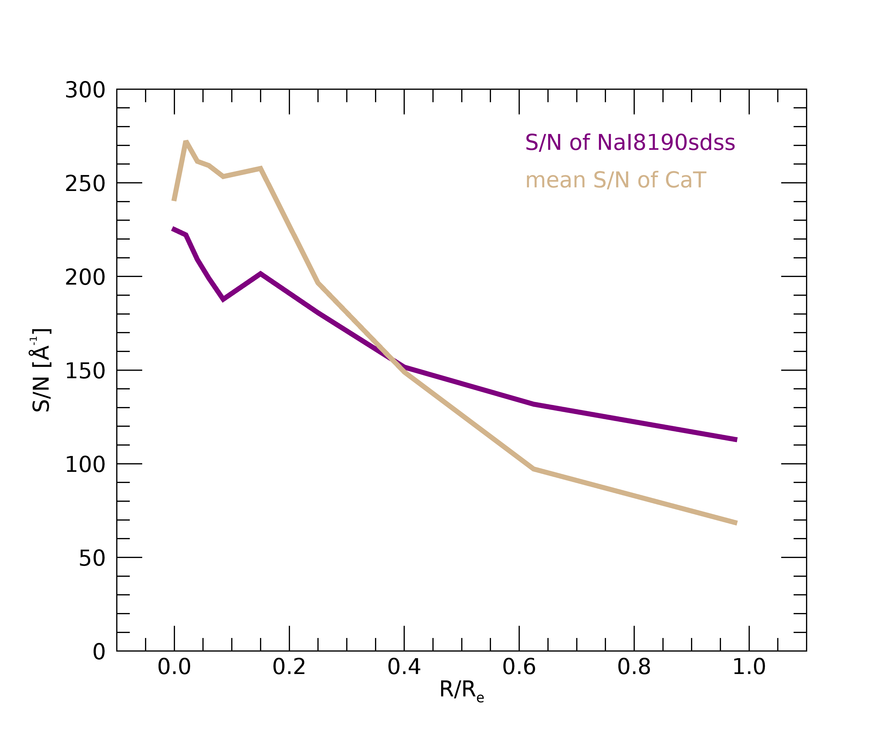}
\caption{\small{\textit{Top panels and bottom-left panel:} Final S/N (/\AA) for the three GA configurations. Only for the purpose of showing the overall trends over the three configuration wavelength regions, these curves have been smoothed. Unsmoothed S/N curves reflect the per-pixel noise and were used for the index error estimates. Colors show the different radial spectra as in Figure \ref{fig:spectra}. At $\lambda>8000$\AA\space sky residuals are important in the outer radial bins; however, in this region the measured spectral indices are only partially affected by residuals as can be seen by their unsmoothed S/N radial trends shown in the bottom-right panel. \textit{Bottom-right panel:} detailed S/N trends as a function of radius (/R$_e$) for NaI8190sdss (purple line) and CaT (tan line) as derived by the unsmoothed S/N curves. }}
\label{fig:SN}
\end{centering}
\end{figure*}

In Figure \ref{fig:SN}, the final S/N curves as a function of the wavelength for each GA are shown. We note that the curves shown in these plots are highly smoothed in order to present the mean trends over the wavelength ranges at different radii and that instead the per-pixel noise has been taken into account when deriving the error for each index (see Appendix \ref{app:errors}).
As already mentioned, the mean S/N over the spectra is well above $100$/\AA. Notice that the S/N of outer regions in the red GA is lower than $100$/\AA\space where the background residuals are high, as evident in the bottom panel of Figure \ref{fig:spectra}. Such residuals are due to the strong sky emission affecting this wavelength region. However, no useful spectral indices have been measured in those regions as detailed in Section \ref{sec:indices}. \\

We corrected the red GA data, which were affected by significant fringing that had to be carefully attenuated before proceeding with standard data reduction. Details regarding the data reduction steps of this region (above $8000$\AA) are presented in Appendix \ref{app:fringes}.


\section{Analysis}
\label{sec:analysis}

In this section we present the method we used to derive the radial SP properties of M89, as well as a summary of  the simulations we performed to test the bias effects on the derivation of the IMF.\\

We fit a large number of spectral indices in the optical and NIR region (see Section \ref{sec:indices}) in order to take into account all of their possible dependencies and to alleviate the degeneracies. 
Furthermore, the high S/N of all the radial bins up to $1$R$_e$ allowed us to analyze the radial variation of spectral features with small uncertainties (as shown in Section \ref{sec:indices}), thus providing us the radial profile for each SP property, e.g. age, metallicity, elemental abundances, and IMF slope. In particular, we used the spectral indices radial variation as a probe of the IMF gradient by means of the comparison with a wide set of SP models with a varying IMF slope and including many elemental abundances (as detailed in Section \ref{sec:fit}). In order to reduce the degeneracies, we required that all the stellar properties that each chosen index is sensitive to, was included as a free parameter in the fit, including elemental abundances. This naturally increased the computational effort but guarantees that all the variables at play are taken into account. 
However, also considering all of the parameters in the fit with a large number of measured indices, does not eliminate the presence of degeneracies. We then carefully checked all possible biases occurring at each step of the analysis performing parallel simulated fits, as fully described in Section \ref{sec:sim}. In this way, knowing the input SP properties values and applying the same procedures and using the same indices as on real data, it was possible to uncover the hidden biases which naturally arise in each particular situation. 

We will demonstrate at the end of the analysis that this series of steps performed to retrieve the IMF slopes from our data set, did not lead to any bias in our results.


\subsection{Spectral Indices}
\label{sec:indices}

As already stated, the wide wavelength range covered by our data offers us the opportunity to measure a large number ($\sim90$) of spectral indices from $3500$\AA\space to $9000$\AA\space with a high S/N ($>100$/\AA). Following the definition of both atomic and molecular indices with the respective band-passes ranges as defined in \citet{hamilton85}, \citet{gregg94}, \citet{worthey97}, \citet{trager98}, \citet{cenarro01}, \citet{thomas03}, \citet{labarbera13} and \citet{spiniello14}, we measured all spectral indices in the available wavelength range. Where possible, residual bad pixels have been interpolated over; otherwise the index has not been measured. Indices with band-passes across chip gaps have not been measured. In the optical band where we have an overlap of wavelength regions from the two chosen GA configurations (see Figure \ref{fig:spectra}), we separately measured the index values in the two sets of spectra. Only after estimating their errors, in particular the systematic errors which depend considerably on their position on the CCD, we combined together those indices in common performing a weighted mean. 

The uncertainties associated with the index measurements have both a statistical and systematic component; thus they have been estimated separately and finally quadratically summed. In particular, index errors include the contribution of the sky line residuals. All the details on the error estimation are described in Appendix \ref{app:errors}.\\

From the $89$ measured indices in our optical+NIR data (including also multiple definitions of the same feature), we choose $32$ reliable indices for our analysis following these criteria:
    \begin{itemize}
        \item S/N $>100$/\AA \ for all radial bins.
        \item No sky and telluric residuals.
        \item Not sensitive to abundances that are not taken into account in the analysis, like [N/H], [Cr/H] etc...
        \item Not strongly affected by gas emission as, for example, H$\alpha$.
        \item Excluding multiple definitions of the same feature.
    \end{itemize}

For the last radial bin TiO$1$ and Ca$2$ have not been included in the fit since sky residuals prevented a reliable index measurement. Before using the final chosen indices in the analysis, they have been further corrected to take into account the possible presence of gas emission lines and to bring all radial bin values to the same velocity dispersion value, as described below in Section \ref{sec:kin}.

\subsubsection{Kinematics and Gas Emission Removal}
\label{sec:kin}

\begin{figure}[ht!]
\begin{centering}
\includegraphics[width=8.5cm]{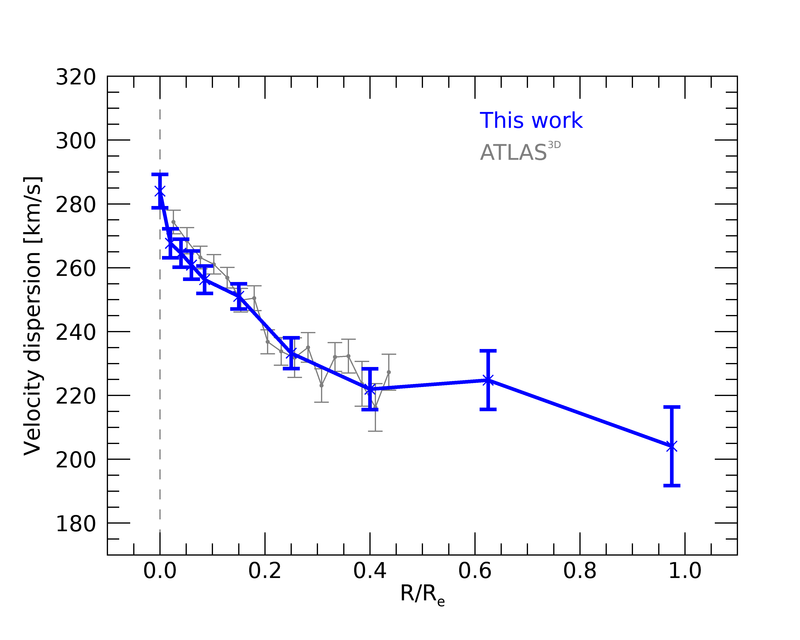}
\caption{\small{Velocity dispersion profile for M89. The gray represents line shows data from ATLAS$^{3D}$ (\citealt{emsellem04, cappellari11}) available up to R/R$_e\sim0.4$.}}
\label{fig:kin}
\end{centering}
\end{figure}

Before summing spectra from opposite sides of the galaxy, the velocity dispersion profile was found to be very symmetric. We repeated the measurement with PPXF over the whole spectral range from $4000$\AA\space to around $6500$\AA\space with the same models, masking chip gaps and residuals, and found consistent trends. The profiles derived from both optical GA were consistent within the uncertainties. For these reasons, to derive the final velocity dispersion profile we computed a mean of both GA profiles and then a mean between the north and south profiles. This final velocity dispersion profile is shown in Figure \ref{fig:kin} together with a comparison of the values from ATLAS$^{3D}$ (\citealt{emsellem04, cappellari11}) which shows consistent results.  

\begin{figure*}[ht!]
\begin{centering}
\includegraphics[width=18cm]{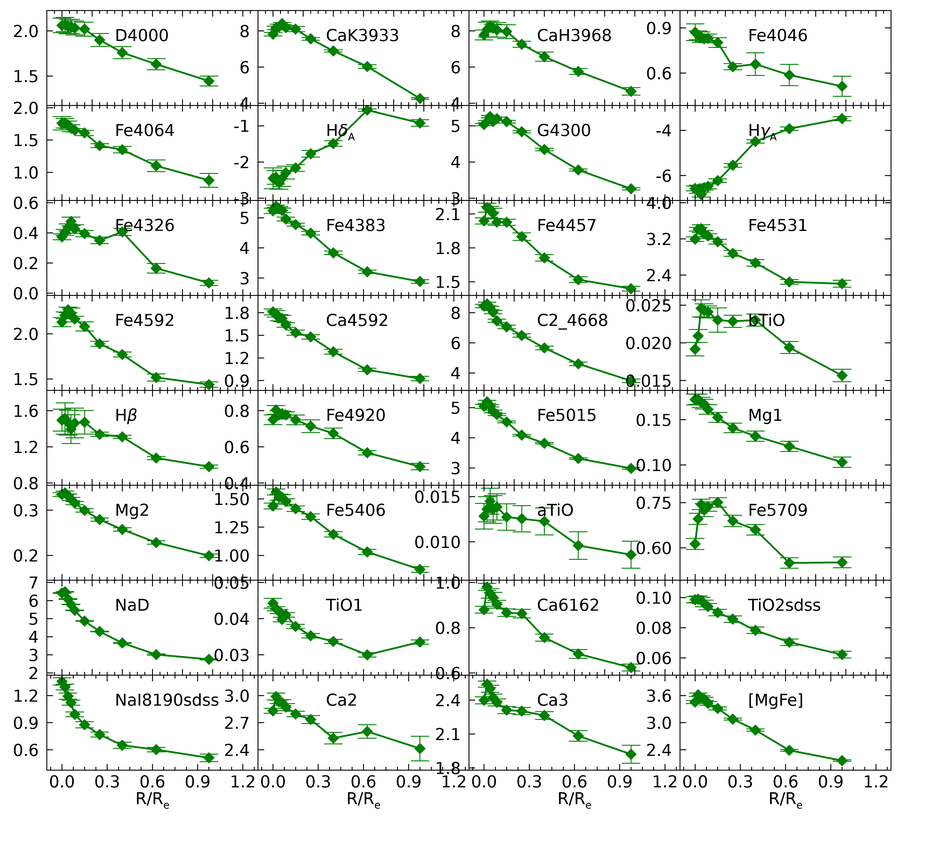}
\caption{\small{Radial trends for all of the spectral indices included in the fit. All index values are corrected to match the center value of the velocity dispersion, i.e. $284$ km/s (see Section \ref{sec:kin} for details). Those indices affected by gas emission infilling (i.e. H$\delta_A$, H$\gamma_A$, H$\beta$ and Fe$5015$) have been corrected up to R/R$_e\sim0.2$ as described in Section \ref{sec:kin}. Error bars include both statistical and systematic errors. }}
\label{fig:indices}
\end{centering}
\end{figure*}

Deriving a robust velocity dispersion profile is important because spectral indices are sensitive to spectral resolution variations. The steep measured velocity dispersion profile means that each index radial profile is partly affected by a radially changing spectral resolution. To remove this effect, we applied a correction bringing all non-central radial bin index values to match the resolution of the central bin. We derived this correction for each spectral index by taking the output best-fit model spectrum obtained from the PPXF fits on each radial spectrum and then downgrading its spectral resolution to that of the central radial spectrum (i.e. $284$ km/s). We then compared the value of each index measured on the original model with the one of the downgraded version and corrected our index measurements. For atomic indices this correction was multiplicative, while additive for molecular indices. Radial index profiles shown in Figure \ref{fig:indices} are already corrected for this effect. 

The last correction we applied to the spectra was removal of the presence of gas emission infilling from the absorption lines. From measured gas emission fluxes of M89 in the ATLAS$^{3D}$ data \citep{cappellari11}, we expected to detect gas emission up to R/R$_e\sim0.2$, while in the outer regions the level of gas is consistent with zero. Gas infilling is indeed visible in our central spectra in the Balmer lines, mostly in H$\alpha$. We ran PPXF with the inclusion of gas components. In particular, we included Balmer lines (H$\alpha$, H$\beta$, H$\gamma$ and H$\delta$ imposing the Balmer decrement), OI[$6300.30$\AA, $6363.67$\AA], OII[$3726.03$\AA, $3728.82$\AA], OIII[$4958.92$\AA, $5006.84$\AA], SII[$6716.47$\AA, $6730.85$\AA] and NII[$6548.03$\AA, $6583.41$\AA]. In order to take into account all these elements together, we combined the optical spectra with different GA into one single spectrum, thus with a wider range and without chip gaps. The fitted wavelength range was $3600-7080$\AA. In PPXF fits we used MILES models \citep{vazdekis10} with a wide range of metallicity (from [Z/H]$=-0.66$ to [Z/H]$=+0.40$) and ages (from 3 to 14 Gyr), and with a fixed unimodal IMF slope$=2.3$. Differently from previous PPXF fits that were focused on the kinematic properties and used additive polynomials, in these SP fits we made use of multiplicative polynomials with degree$=10$. In agreement with ATLAS$^{3D}$ data, we found a decreasing amount of gas emission from the center to R/R$_e\sim0.2$. Regarding the kinematics of these gas components, both the radial velocity and the velocity dispersion show a clear decreasing trend from the center to R/R$_e\sim0.1$ and then a slightly increasing trend toward R/R$_e\sim0.2$. Given that the stellar spectrum has been already corrected to the rest-frame, this means that the kinematics of the gas is at some level decoupled from the stellar one. Beyond R/R$_e>0.2$ the fits did not produce any converged solution so we conclude that, if present, the gas component amount is so small that it is not detectable and thus negligible. We removed the gas component from the inner radial spectra and we measured again those spectral indices that are affected by gas infilling.
As far as the stellar velocity dispersion that resulted from these fits, we found good agreement with the profile shown in Figure \ref{fig:kin} out to R/R$_e=0.2$.\\

Figure \ref{fig:indices} shows the final radial trend of all the spectral indices we included in the analysis following the criteria stated above in Sec \ref{sec:indices}. Error bars include both statistical and systematic errors. It is easily seen that all trends present a gradient which is clearly negative with the exception of H$\delta$ and H$\gamma$ that have instead positive gradients. \\
It can also be noticed that almost all indices (excluding: Ca$4592$, Fe$4046$, Fe$4064$, NaI$8190$sdss and TiO1) have slightly lower values in the very center of the galaxy (first to second radial bins) that can probably be connected with the presence of the inner dust disk of M89 (\citealt{lauer05,bonfini18}) (more comments follow in Section \ref{sec:discussion}).

\subsection{IMF Retrieval}
\label{sec:fit}

When considering all of the SP parameters that can be acting on the measured spectral indices, the number of models that includes their entire combination is huge. Therefore, to limit the computational time, we inspected all the index elemental abundance dependencies looking at models (see below), and chose a subset of indices that are only sensitive to the following parameters (as hinted in Section \ref{sec:indices}): age, metallicity, IMF slope, [Ca/H], [Na/H], [Mg/H], [Ti/H], [Fe/H], [C/H] and [O+Ne+S/H] (hereafter [O/H]). The list of indices considered in the analysis has been already shown in Figure \ref{fig:indices}.
Moreover, in order to further decrease the number of models in the fit, we split our analysis in subsequent steps as detailed below in Section \ref{sec:steps}.

In each step of the analysis we based our SP properties retrieval on $\chi^2$ minimization processes, following 
\begin{equation}
\chi^2=\sum_i^N \left[\frac{\text{EW}_i - \left(\text{EW}_{\text{mod},i}+\sum_{\text{X}} \Delta_{\text{X},i} \cdot \text{[X/H]}\right)}{\sigma_{\text{EW}_i}}\right]^2
\label{eq:chi2}
\end{equation}
where EW$_i$ is the measured value of the equivalent width $i^{\text{th}}$ index, EW$_{\text{mod},i}$ is the corresponding index value expected by models, $\Delta_{\text{X},i}$ is a correction to the $i^{\text{th}}$ index for non-solar values of the [X/H] elemental abundance (details below, Eq. 2) and $\sigma_{\text{EW}_i}$ is the uncertainty on the $i^{\text{th}}$ measured index. 

We built the probability density functions (PDF) starting from the weights associated with each $\chi^2$ through $w\propto e^{-\chi^2/2}$, marginalizing over each SP parameter. Assuming a Gaussian profile, for each distribution we computed the weighted mean and the $16^{th}$ - $84^{th}$ percentiles, considered to be the $1\sigma$ error of each retrieved quantity.
To be conservative, whenever the retrieved error of a SP property was smaller than the model grid step used, it was imposed to be equal to the grid step itself.
This method is similar to the one used in F20 (see Equation 1) for the spectral indices analysis, but in this work we did not apply a prior on the age estimate or a proxy on the [$\alpha$/Fe] value. This is because of the availability of more measured indices which gave us the possibility to fit more free parameters simultaneously.  By means of simulations, we have verified the feasibility of retrieving each parameter robustly, as shown in Section \ref{sec:sim}.

SP models have been selected that include a wide range of parameters, in particular IMF slopes, and a wide wavelength range with high spectral resolution. These requirements are covered by both the \citet{conroy18} models and EMILES models by \citet{vazdekis16}. With respect to the F20 study, we used the same Conroy models but instead of adopting the MILES models (which cover up to $\sim7500$\AA), here we needed to used the EMILES models in order to consider also our observed NIR region. In particular, the \citet{conroy18} models provide $100$km/s resolution spectra over our covered spectral range $3500-9000$\AA, for a wide range of ages up to $13.5$ Gyr and metallicities from [Z/H]$=-1.50$ to $+0.20$. The IMF is parameterized in two power-law parts: X1 is the slope from $0.08-0.5M_{\odot}$ and X2 from $0.5-1.0M_{\odot}$ (following $dN/dm \propto m ^{-x_i}$), while above $1M_{\odot}$ it has a constant Salpeter slope \citep{salpeter}. EMILES models, with the BaSTI theoretical isochrones \citep{pietrinferni04}, have instead a constant FWHM$=2.51$\AA\space over our spectral range and provide ages up to $14$ Gyr and metallicities from [Z/H]$=-2.27$ to $+0.40$. We selected EMILES models with a bimodal IMF that has a logarithmic slope $\Gamma_b$ which is tapered toward low masses around $0.5M_{\odot}$ (following $dN/d $log$ m \propto m^{ -\Gamma_b}$).

Both models have been homogenized to have a constant velocity dispersion of $284$ km/s (as the central value of M89) and sampling at $0.37$\AA, similar to our data. On both model sets, we measured the index values and then linearly interpolated age and metallicity over the same parameter grids. Ages are considered from $1$ to $14$ Gyr with a step of $0.25$ Gyr and metallicities from [Z/H]$=-1.50$ to $+0.4$ with a step of $0.1$ dex. IMF slope values for both X1 and X2, and for $\Gamma_b$ run from $0.5$ to $3.9$ with a step of $0.2$. 

To include  the effects of non-solar values of the many elemental abundances on spectral indices in the modelling, we used the response functions described in \citet{conroy18}. The response functions are provided for a Kroupa IMF \citep{kroupa}, with a constant $\sigma=100$ km/s resolution and for selected age and metallicity values. As in the case of the models, we downgraded the resolution of the response functions in the wavelength range of our data to $284$ km/s, then interpolated them over the age and metallicity grids of the initial Conroy and EMILES models, and applied them on the model spectra. We then measured the spectral indices on these new non-solar elemental abundances spectra in order to estimate each index correction $\Delta_{\text{X},i}$ used in Eq. \ref{eq:chi2}, computed as follows: 

\begin{equation}
\Delta_{\text{X},i}=\frac{\text{EW}_{\text{mod},i,\text{[X/H]}\neq0} - \text{EW}_{\text{mod},i,\text{[X/H]}=0}}{\delta\left(\text{[X/H]}\right)}
\label{eq:correction}
\end{equation}

\noindent
where $\text{EW}_{\text{mod},i,\text{[X/H]}\neq0}$ is the index value measured on models with non-solar value of [X/H] to which the response functions have been applied, while $\text{EW}_{\text{mod},i,\text{[X/H]}=0}$ is the same index at its solar value. $\delta\left(\text{[X/H]}\right)$ is the variation of each [X/H] from the solar value, i.e. $0$.

These index corrections can be computed for any desired value of each elemental abundance. Our spanned values, i.e. range and step, for each elemental abundance are quoted in the analysis description at each step (Section \ref{sec:steps}). To improve the efficiency of the parameter space exploration, an optimizer or MCMC sampler would be more suitable (see for example: \citealt{alton17,alton18}); nevertheless we adopted this standard method since our purpose is to test our analysis following what has been common practice in the literature (e.g.: \citealt{spiniello14,martin-navarro15,labarbera16,parikh18,sarzi18}). For comparison, we later do a full spectral fit in Section \ref{sec:fsf}.

\subsection{Simulations}
\label{sec:sim}

In this section we describe our parallel analysis of the uncertainties and, particularly, the biases that occur when fitting many stellar parameters simultaneously. In F20, we have already shown with simulations of spectral indices and full spectral-fitting analysis, that changing the set of indices or wavelength range and the number and set of chosen parameters, can lead to very different results in terms of biases in the retrieved quantities. Every region of a spectrum includes different information in different proportions so that the degeneracy is peculiar to the particular problem being considered. This once again highlights the complexity of stellar population analyses in general, and the necessity of taking the very real biases into account when assessing the results.
  
In our simulations we repeated the same analysis with the same data quality and fit parameters as conducted on the real data (as described in Section \ref{sec:fit}). To mimic the observed spectrum we started with a model template (from \citealt{conroy18}), with a chosen age, metallicity and IMF slopes and then applied the response function to produce non-solar values for the chosen elemental abundances. In particular, we proceeded studying both a Kroupa IMF like input, with X1$=1.3$ and X2$=2.3$, and a bottom-heavy (BH) IMF with X1$=3.1$ and X2$=3.5$, since we have previously noted that the quality of the IMF retrieval itself differs for different choices of the input IMF. The chosen model was then re-sampled and downgraded to the spectral resolution of the observed data, i.e. $284$ km/s, the central velocity dispersion of M89. To apply realistic noise to the model spectra, which can take into account the S/N variation as a function of the wavelength, we made use of the real data unsmoothed S/N curves over the spanned wavelength range, which thus reflect the real per-pixel noise and eventual sky residuals. In detail, we combined the S/N curves obtained for M89 from $3500$ to $9000$\AA\space and normalized it to its maximum at around $\sim6000$\AA. 
This final curve increases from short wavelengths, reaches its maximum and slightly decreases up to $\sim7000$\AA; then increases again from $>8000$\AA\space and decreases after $8500$\AA, reproducing the sensitivity of the spectrograph. In each simulation run, a noise curve is built such that the final S/N curve peak of the simulated data reaches the desired S/N value, following the shape of the data S/N curve. The noise is then added randomly at each pixel of the model spectrum from a Gaussian distribution whose width is the noise level at that wavelength. $1000$ realizations of the same spectrum were performed each time and considered as independent spectra. The measurements of indices and the parameter retrieval have been performed on each spectrum following the same procedure as on real data. We then focus on the distributions of all the retrieved parameters over the $1000$ realizations to check their accuracy and precision.  \\ 

From our simulation analysis, we present two important tests: the first one is a set of general simulations where we took a generic input spectrum with non-solar elemental abundances and tested the retrieval of IMF slopes  with different levels of knowledge of the input chemical composition (e.g. assuming the right value of the total metallicity and of only a part of the non-solar elemental abundances); and the second one is a set of simulations tailored to the steps used in the M89 analysis. Both sets are fully described in Appendix \ref{app:gensim} and \ref{app:sim89}, here we just mention the main results. In particular, the former general simulations led to the important result that in order to avoid severe biases on the IMF retrieval, not only the age and metallicity but all the elemental abundances to which the particular spectral indices are sensitive to, must also be well constrained. As a consequence of this result, we built the M89 test case data analysis as a chain of subsequent steps where we used only the spectral indices that depend on the stellar parameters that can be retrieved. In the second set of simulations we then have systematically checked the particular case of the stellar population properties at each radius in M89 by simulating each step of the analysis and demonstrating that the final results are bias-free. Comments on each step of the parallel simulations are also present in Section \ref{sec:steps}.


\subsection{Steps of the Analysis and Results}
\label{sec:steps}
As mentioned previously, in deriving the radial variation of the IMF, we have proceeded with a step-by-step analysis, in parallel with the simulations to test the level of bias that can occur at each step (see Appendix \ref{app:sim89} for a full description). Indeed, simultaneously fitting all of the many parameters considered in this analysis would be possible only by adopting a MCMC sampler (as in the case of \citealt{alton18}); however, since our intention in this analysis is to test the commonly-used spectral index fitting method, as noted previously, we proceeded by splitting the analysis into a series of steps. In addition, we do a full spectral fit for comparison in Section \ref{sec:fsf}.
Our general method was to start from fitting only a few stellar parameters including in the fit only spectral indices whose values variation is solely or mainly due to the selected abundances, then reduce the parameter ranges in accordance with the previous step and simulation results, and repeat the fit with new abundances and new indices (and decreasing the grid step of each parameter). As learnt from our simulations, proceeding in this way guarantees that no bias that has been addressed in this analysis is affecting the results at each step.
Those indices that can be strongly affected by gas emissions (i.e. H$\delta$, H$\gamma$, H$\beta$ and Fe$5015$) were also removed from the same index set (hereafter ``emission-free" indices) during all of the fits and steps to see if there were differences in the results. 

We repeated the same procedure with both Conroy and EMILES models, but for brevity in the following description we focus only on the Conroy models results. Since the details of the procedure are very similar, the EMILES models analysis is reported in Appendix \ref{app:emiles}. \\

\begin{figure*}[ht!]
\begin{centering}
\includegraphics[width=19.5cm]{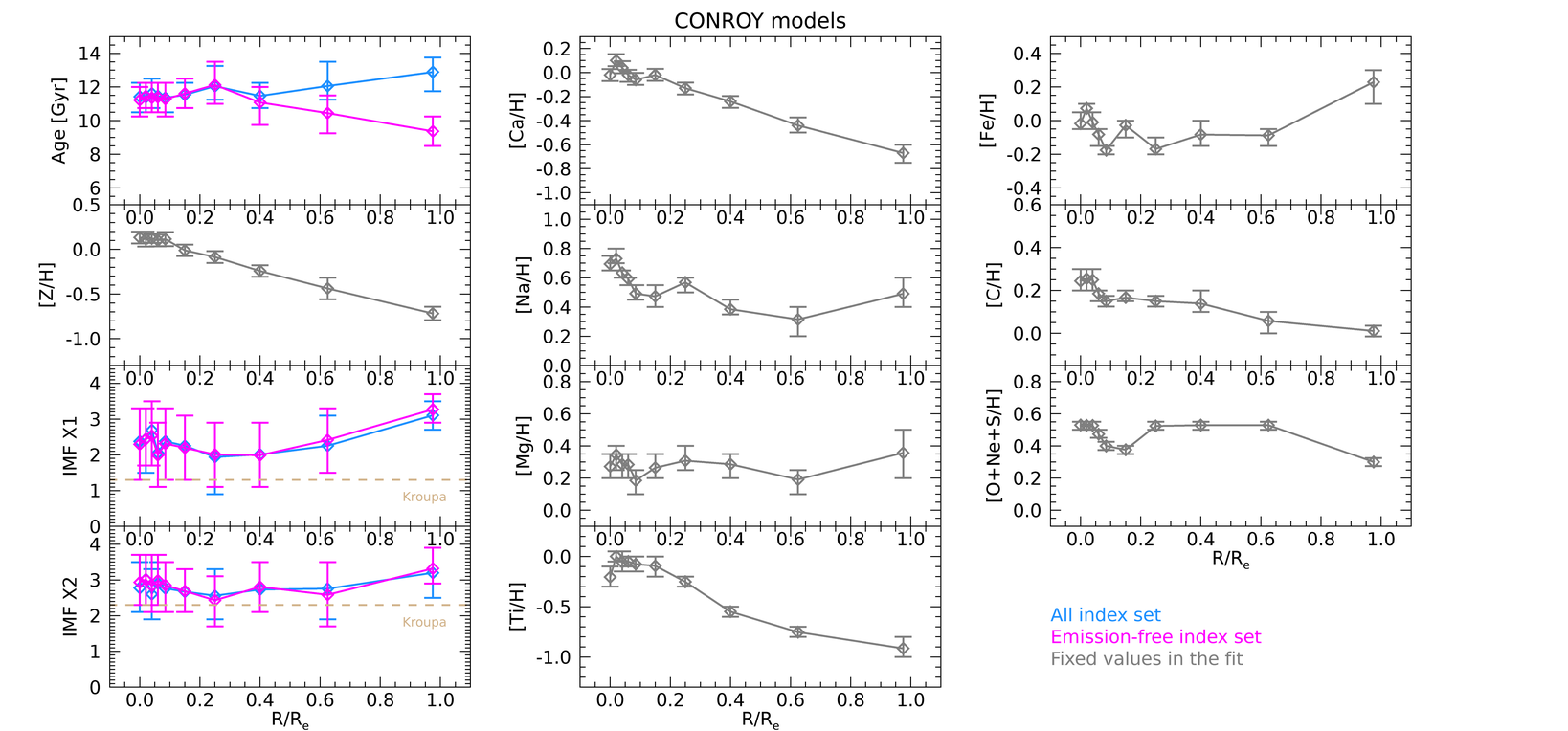} 
\caption{\small{Results of the SP retrieval for M89 from the final step five of the analysis using \citet{conroy18} models. All weighted means are plotted as a function of the radius (R/R$_e$). Gray lines indicate the fixed values during step five of the fit as retrieved from the previous steps. Magenta and cyan lines show the final values for age and IMF slopes for the all-indices and emission-free index sets, respectively.}}
\label{fig:results}
\end{centering}
\end{figure*}

The steps of the analysis consisted of:

\begin{enumerate}

    \item \textbf{Step one}. Free parameters: metallicity, IMF slopes, [Na/H], [Ca/H], [Mg/H] and [Fe/H]. The age is fixed. All other elemental abundances are assumed to be solar. This first step does not include most of the indices useful for deriving the IMF slopes, i.e. TiO1, TiO2, aTiO, bTiO: this is necessary to put constraints on some parameters ranges and thus to limit the number of models and the computational time. All parameters, except for the age, which is fixed, are free to run within the entire range offered by models and in some cases linearly extrapolated toward lower or higher values. This step is repeated fixing different values of ages ($5$, $10$, $11$, $12$, $13$ Gyr);    we checked the simulation results for possible bias due to a wrong age value. (INDICES: D4000, CaK$3933$, CaH$3968$, Fe$4046$, Fe$4064$, H$\delta_A$, H$\gamma_A$, Fe$4326$, Fe$4383$, H$\beta$, Fe$4920$, Fe$5406$, Ca$6162$, NaI$8190$, Ca$2$, Ca$3$, [MgFe]). 
    
RESULTS: Metallicity shows an overall decreasing radial trend from $0.3$ to $\sim-1.0$ dex. [Ca/H] is sub-solar, [Na/H] and [Mg/H] are super-solar and [Fe/H] goes from sub- to super-solar values. The IMF slopes are very poorly constrained. Agreement between all-indices and emission-free indices is good. Simulations show a very good retrieval of [Z/H], [Ca/H], [Mg/H] and [Fe/H]. [Na/H] is instead highly biased generally toward higher values and presents large uncertainties. Since [Na/H] is mostly constrained by NaI$8190$ at this step, and since this index depends mostly on the IMF slopes (and metallicity), which are still not well constrained, we attribute the offset of this first step to the degeneracy with the IMF slopes. 
Biases on the IMF slopes are more important for a Kroupa than for a bottom-heavy IMF. Since NaI$8190$ is the only index that has a significant sensitivity to the IMF as well as to Na abundance, we ascribe the observed biases to a degeneracy among IMF slopes and [Na/H] produced by that index. Imposing a wrong fixed age in the simulations results in adding an offset to the values of [Ca/H] and [Fe/H], with differences of $0.1$ dex when changing by $\pm2$ Gyr; the degeneracy with metallicity is also present with a bias of $0.2$ dex below the true value only in the case of overestimating the age. 

GAIN: From this first step we saved the information on the retrieved metallicity, which was the most solidly measured parameter constrained by the indices subset. Moreover, we used step one to rule out sub-solar values of [Na/H] and [Mg/H].

    \item \textbf{Step two}. Free parameters: metallicity (restricted), [Na/H], [Ca/H], [Mg/H], [Fe/H], [Ti/H], [C/H] and [O/H]. Fixed parameters: age and IMF slopes. The metallicity is constrained at each radius around $\pm0.2$ dex from the values found in step 1. The same fit is repeated for ages $= 10$, $11$, $12$, $13$ Gyr and IMF slopes put at the same fixed values $=1.5$, $2.5$, $3.1$. Ranges for elemental abundances: [Ca/H] from $-1$ to $0.2$, [Na/H] from $0.0$ to $1.0$, [Mg/H] from $0.0$ to $0.8$, [Fe/H] from $-0.4$ to $0.4$, [Ti/H] from $-1.0$ to $0.2$, [C/H] from $0.0$ to $0.4$ and [O/H] from $0.$ to $0.6$, all with a step of $0.2$ dex. (INDICES: same as step one,  and in addition, G$4300$, Fe$4457$, Fe$4531$, Fe$4592$, Ca$4592$, bTiO, Fe$5015$, Mg$1$, Mg$2$, aTiO, Fe$5709$, NaD, TiO$1$, TiO$2$).
    
RESULTS: The negative metallicity gradient is stable running from $\sim0.1$ to $\sim-0.7$ dex. Also [Ca/H] and [Ti/H] show a clear negative gradient from around solar values to rather extreme sub-solar values below $\sim-0.5$ dex. [Mg/H] and [Fe/H] have similar trends with a flat behavior in the first half of the galaxy and a mild increase in the second one.
[Na/H] is highly dependent on the choice of fixed IMF. [C/H] and [O/H] are poorly constrained but show a rather constant trend. The agreement between all-indices and emission-free indices is still very good. Simulations (see Figure \ref{fig:sim89}, orange, in Appendix \ref{app:sim89}) show an excellent recovery of [Z/H], [Ca/H], [Na/H], [Mg/H], [Fe/H] [C/H] and [O/H] for both Kroupa and bottom-heavy IMF inputs. Some small ($<0.2$ dex) biases are found for [Ti/H]. \\
 When imposing a wrong fixed value of the age ($\pm1$ Gyr) in the simulation fits, some biases arise mainly for the Kroupa inputs.  [Z/H] and [Ca/H] are completely unaffected by biases in all fits. \\
 When fixing wrong IMF slopes in simulations (both fixed to $2.5$ for bottom-heavy IMF input and $3.1$ for Kroupa), [Z/H] changes within $0.1$ dex (mostly for the Kroupa inputs), and most of the elemental abundances suffer biases up to $0.5$ dex.
 We took into account all of them in the allowed parameter ranges of the following step.

GAIN: From this step, for each radius and for each abundance, we looked at all the values spanned from all the choices of age and IMF and from simulations results, and keep these values for the allowed ranges in the next step three. Metallicity values are definitively fixed at this stage.

    \item \textbf{Step three}. Free parameters: IMF slopes, [Na/H], [Ca/H], [Mg/H], [Fe/H], [Ti/H], [C/H] and [O/H]. Fixed parameters: age and metallicity. IMF slopes are limited to three values, free in the fit: $1.90$, $2.70$ and $3.50$. The same fit is repeated for ages $=10$, $11$, $12$, $13$ Gyr. Metallicity values are fixed to those found in step 2 (see Figure \ref{fig:results} and Table \ref{tab:results} for their values). Steps for elemental abundances are reduced to $0.1$ dex. (INDICES: as for step two).
    
RESULTS: IMF slopes values lie from $\sim2$ to $\sim3$. [Ca/H] is very well constrained, confirming a clear decreasing trend from $\sim0$ down to $-0.6$. Also [Ti/H] confirms a negative gradient down to $-1.0$. [Na/H] shows a steep gradient within R/R$_e=0.1$ and then it is rather flat in the outer bins. Similar behavior is found for  [Mg/H], [Fe/H], [C/H] and [O/H]. Some difference, although within the error bars, is observed between all-indices and emission-free indices for the two outer radial bins. When changing the fixed age values, [Ca/H] values are found to be very stable while other elements show differences up to $0.1$ dex, although within their errors. In the simulations (see Figure \ref{fig:sim89}, green) all elemental abundances are retrieved well for both the bottom-heavy and Kroupa IMF input with only small biases for [Ti/H], [C/H] and [O/H]. [Ca/H] is very well constrained. The IMF slopes are biased for both the bottom-heavy and Kroupa IMF input, but no constraints on the IMF are yet adopted or placed based on this step. When simulating a fixed wrong age with $\pm1$ Gyr from the input value, all elemental abundances are generally still consistent with the input values, with some exceptions that are consistent to within $2\sigma$.

GAIN: We found very robust values of [Ca/H], and thus it was fixed in the next step. We also reduced the fitting ranges of the other abundances.

    \item \textbf{Step four}. Free parameters: age, IMF slopes, [Na/H], [Mg/H], [Fe/H], [Ti/H], [C/H], and [O/H]. Fixed parameters: metallicity and [Ca/H]. Age values run between $10$ to $13$ Gyr with a step of $0.25$ Gyr. IMF slopes are still limited to three values as for step 3. Steps for elemental abundances are further reduced to $0.05$ dex, with the exception of the last radial bin for which the step is still $0.1$ dex. (INDICES: as for step two).
    
RESULTS: The age is rather constant around $11$ Gyr. All the elemental abundances are well-defined and confirm the previous general results. Their trends can be seen in Figure \ref{fig:results}.

GAIN: From step four we can fix all the elemental abundance values and their errors as shown in Figure \ref{fig:results}.

    \item \textbf{Step five}. Free parameters: age and IMF slopes. Fixed parameters: metallicity and all elemental abundances. Age values run between $6$ to $14$ Gyr with a step of $0.25$ Gyr. IMF slopes are free to vary from $0.5$ to $3.9$ with a $0.2$ step. (INDICES: as for step two).
    
FINAL RESULTS: The final results for the IMF slopes retrieved with the Conroy models show a flat trend for both X1 and X2, with values $\sim2.5$ and $\sim3$ respectively. An apparent mild negative gradient for X1 can be seen in the inner half of the galaxy (up to R/R$_e\sim0.25$), but the large error bars prevent meaningful constraints. The age is rather constant in the inner part of the galaxy at $11.5$ Gyr, while starting from R/R$_e>0.4$ there is a different trend for all-indices results that start to increase up to $13$ Gyr; while fitting only emission-free indices the trend slightly decreases in the last radial bins. No differences are seen for X1 and X2 when fitting the two different sets of indices. The very outer bin values of [Na/H], [Mg/H], [Fe/H], and [O/H], at R/R$_e=1$, show a visible change of their trend that probably also affects the IMF values of both X1 and X2. Since the S/N of the bluer indices in this outer radial bin was at the limit of acceptability for an IMF slope retrieval, we are inclined to suppose that the results obtained at R/R$_e=1$ are not robust and presumably biased.
Simulations (see Figure \ref{fig:sim89}, blue) of the final step show excellent agreement for the age and both IMF slopes.

\end{enumerate}    

\begin{figure}[ht!]
\begin{centering}
\includegraphics[width=8.5cm]{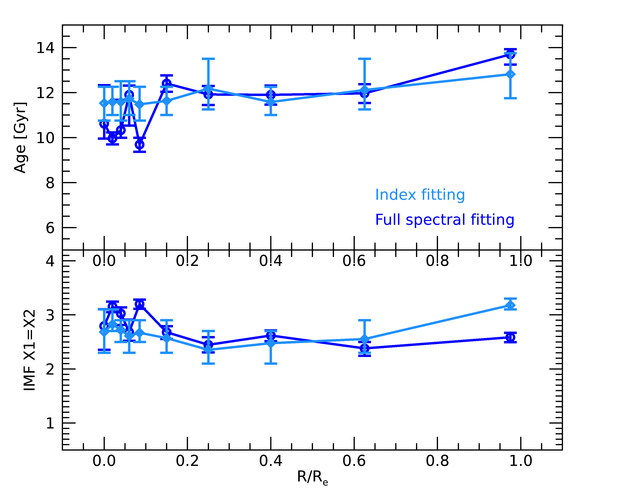} 
\caption{\small{Same results as in Fig. \ref{fig:results} (light blue lines) but in the case when X1 is fixed to be equal to X2. Dark blue lines refer to the results of the full spectral fitting as in Figure \ref{fig:indexFull}, always when imposing X1 equal to X2 (see Section \ref{sec:fsf}). Not shown here are the fixed values of [Z/H] and elemental abundances (same as in Fig. \ref{fig:results}). Being cancelled the mutual correlation between X1 and X2 (see text), error bars are smaller than for the case shown in Fig. \ref{fig:results} for both age and IMF.  }}
\label{fig:resultsX1eqX2}
\end{centering}
\end{figure}

\begin{table*}[ht!]
 \begin{centering}
 \caption{\small{Final Results for M89. Radial gradients for the retrieved SP parameters: age, metallicity, X1 and X2 IMF slopes, [Ca/H], [Na/H], [Mg/H], [Ti/H], [Fe/H], [C/H], and [O/H]. The quantities are the weighted means retrieved with the all-indices set (blue lines in Figure \ref{fig:results}).}} 
 \label{tab:results}
 \begin{tabular}{lcccc}
 \hline
 \hline
Radius [R/R$_{e}$] & Age [Gyr]            & [Z/H]          & X1                  & X2       \\           
 \hline
0.00   & 11.4$\pm0.9$         &  0.13$\pm0.07$             & 2.4$\pm0.9$         & 2.8$\pm0.7$ \\
0.02   & 11.4$\pm0.8$         & +0.11$\pm0.08$             & 2.4$\pm0.9$         & 3.0$\pm0.7$ \\
0.04   & 11.6$\pm0.9$         & +0.12$\pm0.05$             & 2.7$^{+0.8}_{-0.9}$ & 2.6$\pm0.7$ \\
0.06   & 11.5$^{+0.8}_{-0.9}$ & +0.10$\pm0.07$             & 2.0$^{+0.8}_{-0.9}$ & 2.9$\pm0.6$  \\
0.085  & 11.3$\pm0.9$         & +0.11$\pm0.08$             & 2.4$\pm0.9$         & 2.8$\pm0.7$  \\
0.15   & 11.5$\pm0.7$         & -0.01$\pm0.06$             & 2.2$\pm0.9$         & 2.7$\pm0.6$  \\
0.25   & 12.0$^{+1.2}_{-0.8}$ & -0.08$\pm0.07$             & 1.9$\pm1.0$         & 2.6$\pm0.7$  \\
0.4    & 11.5$^{+0.8}_{-0.7}$ & -0.24$\pm0.06$             & 2.0$\pm0.9$         & 2.7$\pm0.7$  \\
0.625  & 12.1$^{+1.4}_{-0.8}$ & -0.44$\pm0.12$             & 2.2$\pm0.8$         & 2.7$\pm0.8$  \\
0.975  & 12.9$^{+0.9}_{-1.1}$ & -0.72$\pm0.08$             & 3.1$\pm0.4$         & 3.2$\pm0.7$  \\
 \hline
 \end{tabular}

\vspace{0.5cm}

 \begin{tabular}{lccccccc}
 \hline
 \hline
Radius [R/R$_{e}$] & [Ca/H]      & [Na/H]                  & [Mg/H]                  & [Ti/H]                  & [Fe/H]                  & [C/H]                   & [O/H]  \\
 \hline
0.00   & -0.02$\pm0.05$          & +0.69$^{+0.06}_{-0.04}$ & +0.27$\pm0.07$          & -0.20$^{+0.10}_{-0.09}$ & -0.02$^{+0.07}_{-0.03}$ & +0.24$^{+0.06}_{-0.04}$ & +0.53$^{+0.02}_{-0.03}$ \\
0.02   & +0.10$\pm0.05$          & +0.73$^{+0.07}_{-0.03}$ & +0.34$^{+0.05}_{-0.09}$ & +0.00$\pm0.05$          & +0.07$\pm0.03$          & +0.25$\pm0.05$          & +0.53$\pm0.03$   \\
0.04   & +0.04$\pm0.05$          & +0.63$^{+0.02}_{-0.03}$ & +0.29$^{+0.06}_{-0.08}$ & -0.06$\pm0.10$          & -0.01$^{+0.06}_{-0.04}$ & +0.25$\pm0.05$          & +0.53$^{+0.02}_{-0.03}$ \\
0.06   & -0.02$^{+0.04}_{-0.06}$ & +0.59$^{+0.01}_{-0.04}$ & +0.28$^{+0.06}_{-0.08}$ & -0.05$\pm0.05$          & -0.08$^{+0.03}_{-0.07}$ & +0.18$^{+0.01}_{-0.03}$ & +0.47$\pm0.03$    \\
0.085  & -0.06$^{+0.06}_{-0.04}$ & +0.49$^{+0.06}_{-0.04}$ & +0.19$^{+0.06}_{-0.08}$ & -0.08$\pm0.07$          & -0.18$\pm0.03$          & +0.15$\pm0.03$          & +0.40$\pm0.02$    \\
0.15   & -0.02$\pm0.05$          & +0.47$^{+0.08}_{-0.07}$ & +0.26$^{+0.09}_{-0.06}$ & -0.09$^{+0.09}_{-0.11}$ & -0.03$\pm0.03$          & +0.17$^{+0.03}_{-0.02}$ & +0.38$\pm0.03$ \\
0.25   & -0.13$\pm0.05$          & +0.57$^{+0.03}_{-0.07}$ & +0.31$^{+0.09}_{-0.06}$ & -0.25$\pm0.05$          & -0.17$^{+0.07}_{-0.03}$ & +0.15$\pm0.03$          & +0.52$\pm0.03$ \\
0.4    & -0.24$\pm0.05$          & +0.38$^{+0.06}_{-0.04}$ & +0.29$^{+0.06}_{-0.09}$ & -0.55$\pm0.05$          & -0.08$^{+0.08}_{-0.07}$ & +0.14$^{+0.06}_{-0.04}$ & +0.53$^{+0.03}_{-0.02}$    \\
0.625  & -0.44$^{+0.07}_{-0.06}$ & +0.31$\pm0.1$           & +0.19$^{+0.06}_{-0.09}$ & -0.75$\pm0.05$          & +0.09$^{+0.04}_{-0.06}$ & +0.06$^{+0.04}_{-0.06}$ & +0.53$^{+0.03}_{-0.02}$    \\
0.975  & -0.67$^{+0.07}_{-0.08}$ & +0.49$^{+0.12}_{-0.09}$ & +0.36$^{+0.14}_{-0.16}$ & -0.91$^{+0.11}_{-0.08}$ & +0.23$^{+0.07}_{-0.13}$ & +0.01$\pm0.03$          & +0.30$\pm0.03$    \\
 \hline
 \end{tabular}

 \end{centering}
\end{table*}

From Figure \ref{fig:results} it can be seen that the error bars on X1 and X2 are still relatively large also after fixing the majority of parameters. Looking at the error ellipses showing the covariance distributions of X1 vs X2 in this last fit (not shown here), we noticed a strong mutual negative correlation at all radii. Only a weak positive correlation is instead found between age and X1 and a moderate negative one between age and X2. Thus, for a test, we repeated the fit fixing X1 equal to X2 and found that in this case the uncertainties on the IMF slope are visibly smaller as shown in Figure \ref{fig:resultsX1eqX2}, indicating that the errors on X1 and X2 in the previous fit were dominated by this degeneracy. Only a very mild correlation with age is left. \\

The final results will be discussed in Section \ref{sec:discussion}. In the next Subsection \ref{sec:fsf} we focus on the comparison of these results with those obtained from a full spectral fitting analysis.

\subsubsection{Comparison with Full Spectral Fitting Results}
\label{sec:fsf}
To further verify the robustness of the results obtained with the method described above (Section \ref{sec:steps}), we compare them with those derived when performing a full spectral fitting with the code PYSTAFF \citep{vaughan18}. As shown in F20 on a similar data set, the full spectral fitting is able to give accurate and precise results on many stellar parameters together.

As input, we prepared a merged version of the reduced spectra combining those of the two different GA configurations in order to fit the full spectra at each radius without chip gaps\, sampled on a $1.25$ \AA\space spaced grid. The wavelength regions considered in the fit are the same as those covered by the definitions of the indices used; that is $3750-6455$ \AA\space and $8143-8727$ \AA, with masked regions at $5738-5862$ \AA,   $8244-8474$ \AA, and $8486-8515$\AA. Bad pixels were also masked. 

We used the \cite{conroy18} models with a constant spectral resolution of $100$ km/s as templates. Since our data have a constant FWHM $= 5.5$ \AA\space over the wavelength, in order to match the spectral resolution of data and models, we convolved the observations to the template resolution at $\lambda> 7500$ \AA\space (where data resolution reaches the value of $100$ km/s), and the models to the data resolution at shorter wavelengths. 

We fit the following SP parameters:  age, metallicity, IMF slopes, [Na/H], [Ca/H], [Mg/H], [Fe/H], [Ti/H], [C/H], [O/H], and in addition, [N/H] and [Si/H]. Ranges are $1$ to $14$ Gyr for age, $-1.5$ to $0.3$ dex for [Z/H], $0.3$ to $3.5$ for both IMF slopes, and for the  nine element abundances $-0.45$ to $1.0$ dex for [Na/H], $-0.2$ to  $0.2$ dex for [C/H], $0$ to $0.45$ dex for [O/H], and $-0.45$ to $0.45$ dex for [Ca/H], [Fe/H], [N/H], [Ti/H], [Mg/H], [Si/H]. We simultaneously fit the stellar kinematics ($V$ and $\sigma$).  To explore the large parameter space, PYSTAFF employs the emcee package \citep{foreman-mackey13}. We used $100$ walkers and $5000$ steps for most spectra; only for the outermost bin we increased the steps to $8000$ as the fit did not converge within $5000$ steps.

\begin{figure*}[ht!]
\begin{centering}
\includegraphics[width=19.5cm]{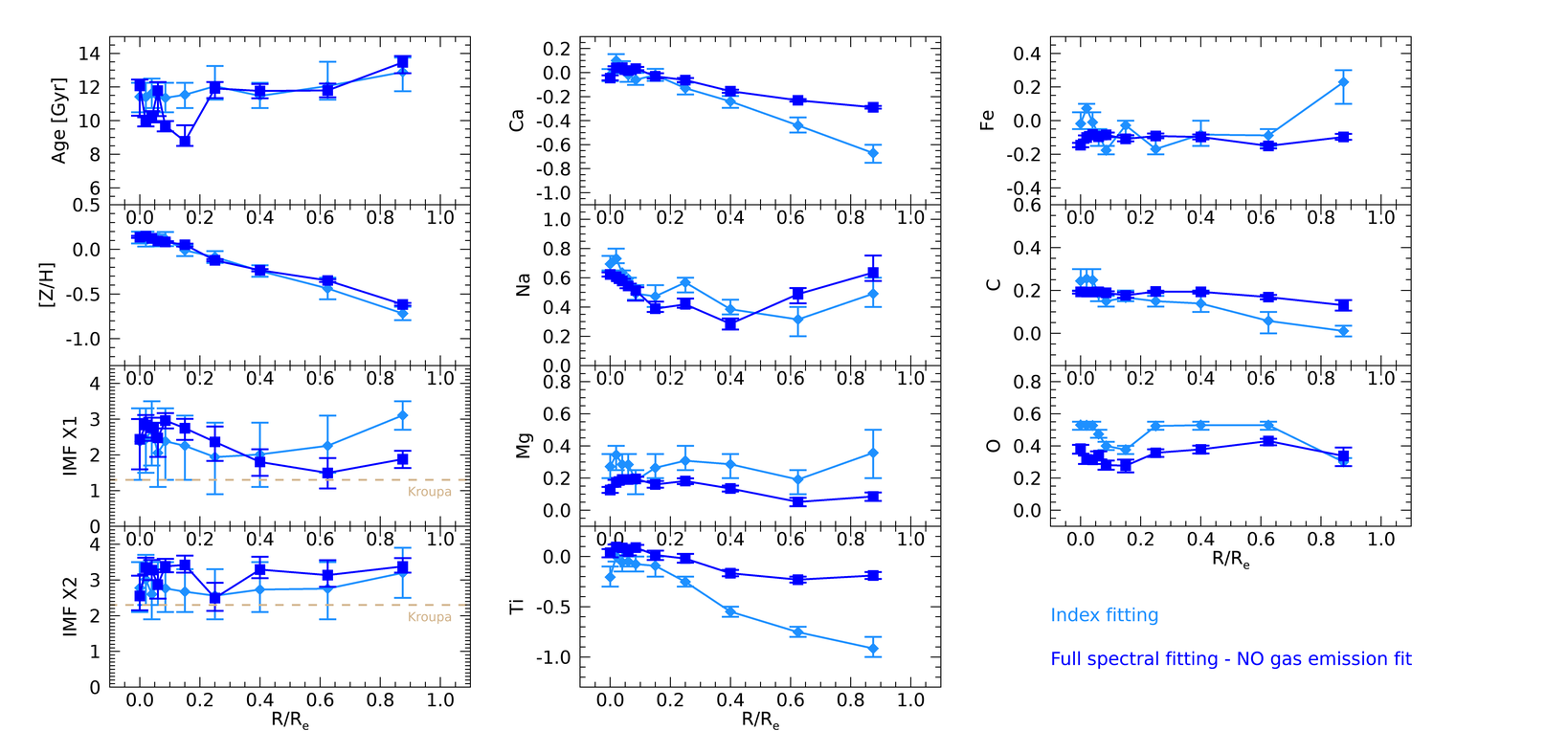} 
\caption{\small{Comparison of the results from index fitting and full spectral fitting,  both performed using Conroy models. In particular, light blue symbols and lines show the all-indices set final results as shown in Figure \ref{fig:results}, and the blue ones the results from full spectral fitting with PYSTAFF without including gas emission in the fit. The comparison shows good consistency between the two methods for all the retrieved parameters, including the IMF slopes. As detailed in the text, inconsistency arises only in those cases where the two methods adopted different extrapolation algorithms, e.g. for very low values of [Ca/H] and [Ti/H], and for higher values of [O/H].}}
\label{fig:indexFull}
\end{centering}
\end{figure*}

The comparison with our final results obtained from spectral index fitting in Section \ref{sec:steps} is shown in Figure \ref{fig:indexFull} where, similar to Figure \ref{fig:results}, each stellar parameter is plotted as a function of the radius. In particular, light blue lines refer to the index fitting results and dark blue lines to the full spectral-fitting ones. The overall comparison shows very good consistency of age, metallicity, IMF slopes and many elemental abundances. As expected, full spectral-fitting retrieved quantities have smaller error bars. This allows us to confirm the presence of a negative gradient in X1, which was only partially constrained in the index-fitting analysis. Moreover, again as a result of the more precise retrieval, X2 is confirmed to be bottom-heavy (with only two low-significance exceptions). When repeating the test of imposing X1 equal to X2 (see Section \ref{sec:steps}) with the full spectral fitting, as shown in Figure \ref{fig:resultsX1eqX2} as dark blue lines, we again find good consistency with the index-fitting results (light blue lines) and confirm that the error bars in this case are smaller due to the removal of the mutual correlation between the two IMF slopes.

Some important differences are present when comparing some elemental abundances like [Ca/H], [Ti/H] and [O/H]. In particular, these arise when extreme values of each parameter are found in the index fitting (e.g. the outer bins of [Ca/H] and [Ti/H]). This is due to a different method adopted to extrapolate the elemental abundances out of the values provided by the response functions (generally below $-0.3$ dex and above $+0.3$). That is,  in the case of the index fitting, the extrapolation is performed linearly, whereas the PYSTAFF code adopts a Taylor expansion which at distant values produces evident distortions of the spectrum that are consistently different from a more simple linear projection. The result is that values above $\sim0.5$ or below $\sim-0.5$ dex are unlikely to be preferred during the full spectral fitting process. We thus observe a discrepancy in the results from the two methods whenever the index fitting prefers extrapolated values, as in the cases of outer bins of [Ca/H] and [Ti/H], and for [O/H] at all radii.\\

The full spectral fitting has been also repeated allowing the presence of gas emission as a free parameter for all the radial bins. We remind the reader that the spectra (and thus all indices) have been already cleaned for gas emission according to the results obtained with PPXF (with MILES models), as described in Section \ref{sec:kin}, up to R/R$_e=0.2$. For the outer 4 radial bins PPXF could not converge to reliable results because of insufficient S/N. To test if there are any residuals of  gas emission in the spectra, and how it influences the results, we  fit the Balmer gas emission flux and kinematics with PYSTAFF. The fluxes of the different Balmer lines were tied to each other (H$5$=0.037 H$\alpha$, H$\epsilon$=0.056 H$\alpha$, H$\delta$=0.091 H$\alpha$, H$\gamma$=0.164 H$\alpha$, H$\beta$=0.35 H$\alpha$).

\begin{figure}[h]
\begin{centering}
\includegraphics[width=8cm]{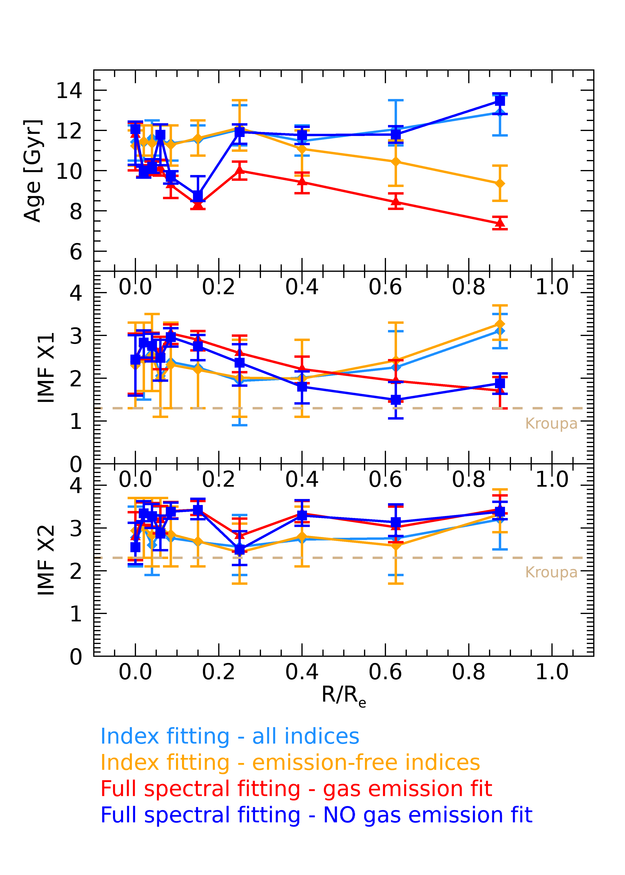} 
\caption{\small{Comparison of the results from index fitting and full spectral fitting analysis as in Figure \ref{fig:indexFull}, but for age and IMF slopes only. The full spectral fit with the inclusion of gas emission as a free parameter is added (red lines), as well as the results of the index fitting performed with the emission-free index set (orange lines). The fact that the full spectral fitting results with gas emission (red lines) are very close to the emission-free index set results for the age in the outer bins, suggests that some amounts of gas should be removed from emission-sensitive features in these outer regions also. In any case, the retrieved IMF slopes are not affected by this effect as demonstrated in the lower panels. In addition, other parameters, not shown here, do not show differences.}}
\label{fig:indexFullgas}
\end{centering}
\end{figure}

In Figure \ref{fig:indexFullgas}, we show the results of adding the gas emission in the fit (red lines) for age and the IMF slopes alone. For R/R$_e<0.02$ the results for the age are consistent with those retrieved without correction for gas emission (blue), while in the outer bins evident differences arise shown by the red line, which follows a decreasing trend toward $7$ Gyr. The IMF slopes are fully consistent. We also included the results from the index analysis obtained when fitting the emission-free index set (orange lines). Although not consistent to within $1\sigma$, the red and orange lines have the same trend at R/R$_e>0.02$, likely meaning that a small amount of gas emission is actually present in the outer bins, causing the emission-sensitive indices to force the age values toward older ages in the fit (blue line). As shown in the lower panels (Figure \ref{fig:indexFullgas}), however, the IMF slopes are not affected by the difference in the age values, in addition to all of the other parameters not shown here.

\section{Discussion}
\label{sec:discussion}

Our final results on the retrieval of IMF X1 and X2 slopes derived with the Conroy models from the step-by-step analysis described in Section \ref{sec:analysis}, are shown in Figure \ref{fig:results} and detailed in Table \ref{tab:results}. Both slopes show  consistency with a flat trend over the entire radial profile. The associated probability density functions (not shown) at each radius for both X1 and X2 are well-shaped with very few exceptions, confirming the good retrieval also for values close to the limit of the models. As earlier stated, the error ellipses for X1 and X2 show a clear correlation between the two slopes that explains the large error bars in this case.

The EMILES models results (described in Appendix \ref{app:emiles}  and shown in Figure \ref{fig:resultsEMILES}) have smaller error bars, due to a one-slope parameterization of the IMF. With these models we obtain a negative gradient of $\Gamma_b$  up to half R$_e$." 

Although it is not possible to directly compare the results from different models due to the different definitions of the slopes, the negative gradient seen for X1 with the Conroy models is likely consistent despite the larger error bars. 
Moreover, in the comparison with the results of the full spectral fitting (as described in Section \ref{sec:fsf} and shown in Figure \ref{fig:indexFull})
X1 clearly shows a negative gradient from $\sim3$ to $1.5$, being at the same time fully consistent with the index fitting analysis.
Nevertheless, the common result for these different analyses is that the IMF of M89 is consistent with being slightly more bottom-heavy than a Kroupa-like IMF (within $1\sigma$).

The uncertainties that we found from the simulations (see Figure \ref{fig:sim89}) due to the degeneracies should be added to the overall uncertainty of X1 and X2. As detailed in Section \ref{sec:sim}, the amount and the direction of the bias on the IMF results can be specific for each given situation, and depends primarily on the IMF true value, which is, of course, unknown. For this reason, we did not include it quantitatively, but just qualitatively discuss it here. Looking at blue points in the upper panels of Figure \ref{fig:sim89} (which refer to the final fit, where the IMF determination is finalized), it can be seen, as mentioned previously, that if the input is a bottom-heavy IMF, its retrieval is simpler, with no evident bias. In the other case (lower panels) where the input is Kroupa-like, we have a small bias for the 2nd and 3rd simulated input values, which slightly decreases the retrieved values of X2. In our results we observe values around $2.0-2.5$ for X1 and $\sim3$ for X2. Regarding X1, we do not find any bias even if the true input is a Kroupa IMF, so we are confident that X1 is robustly constrained. For X2, if its true nature in the $0.1<R/R_e<0.6$ galaxy region were truly BH, its determination would be unbiased, or, at least with biases well within the error bars.

\begin{figure*}[ht!]
\begin{centering}
\includegraphics[width=7.2cm]{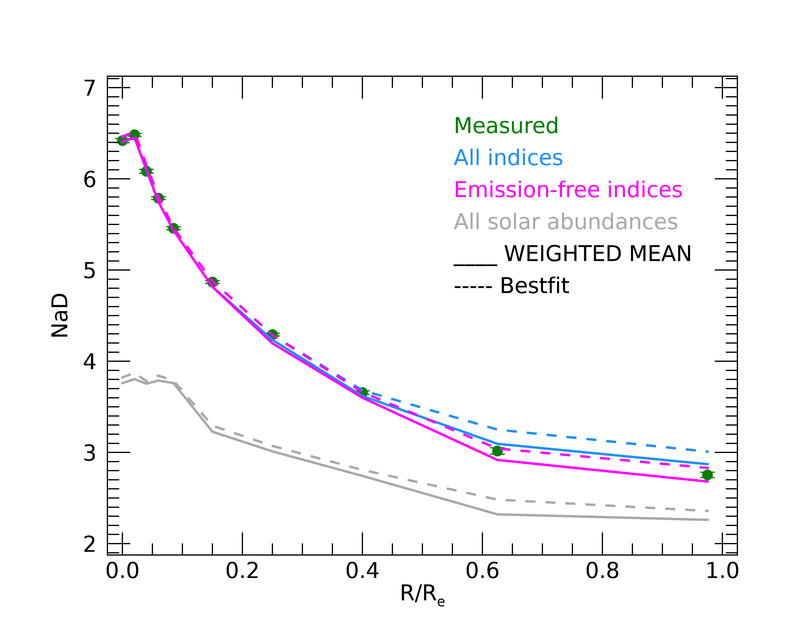}
\includegraphics[width=7.2cm]{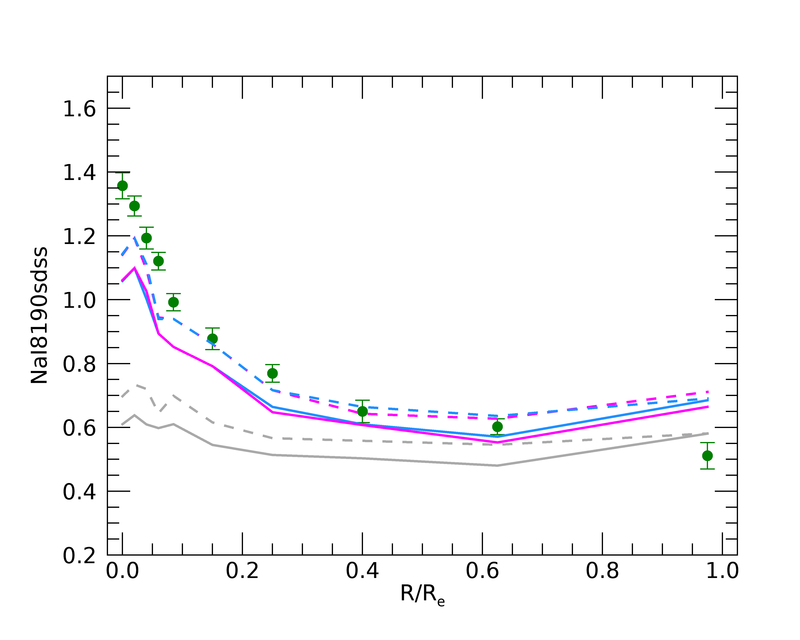} \includegraphics[width=7.2cm]{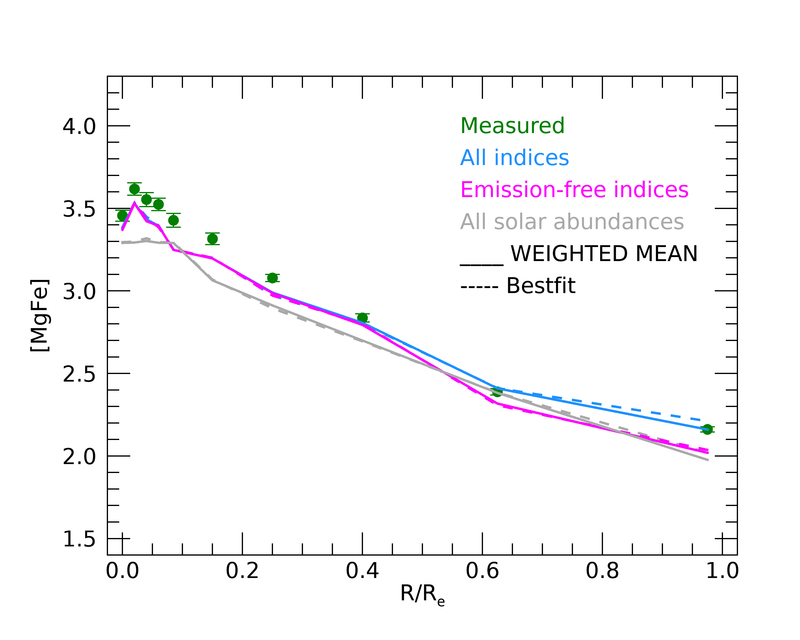}
\includegraphics[width=7.2cm]{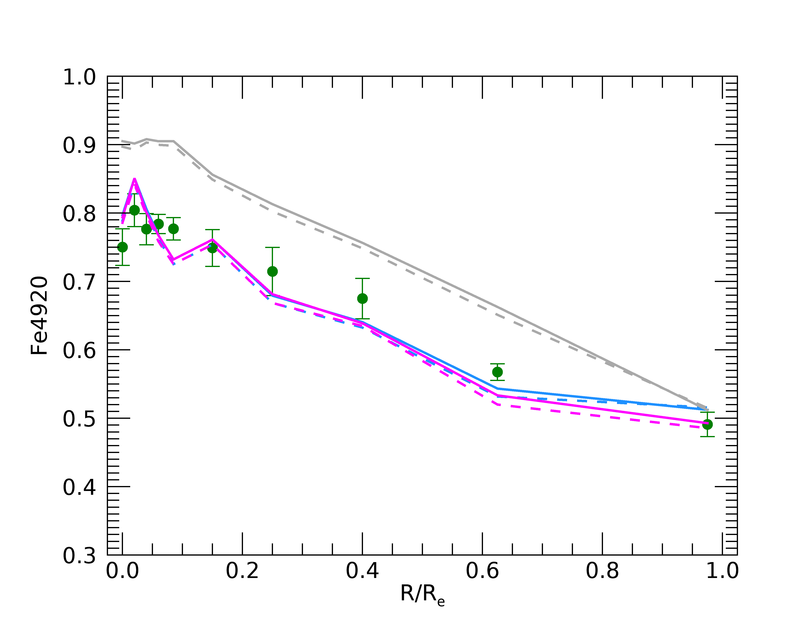} 
\includegraphics[width=7.2cm]{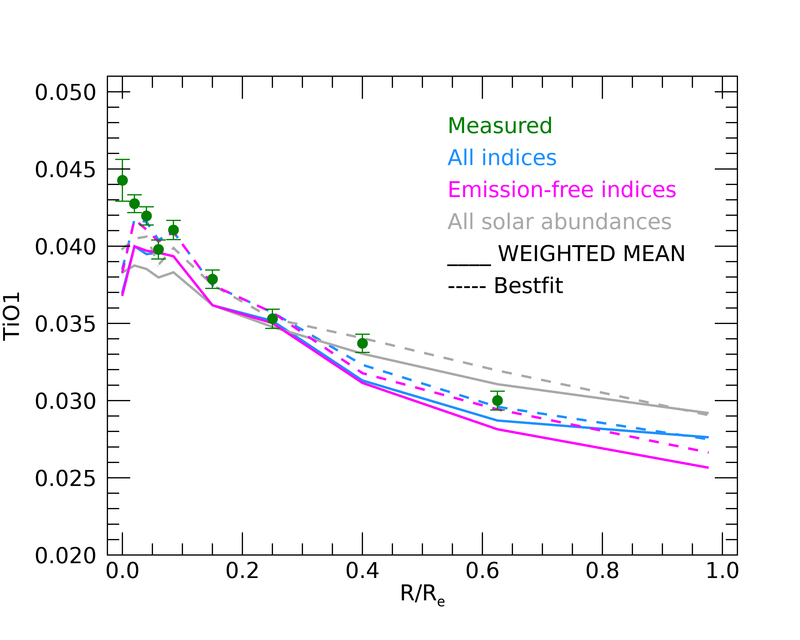}
\includegraphics[width=7.2cm]{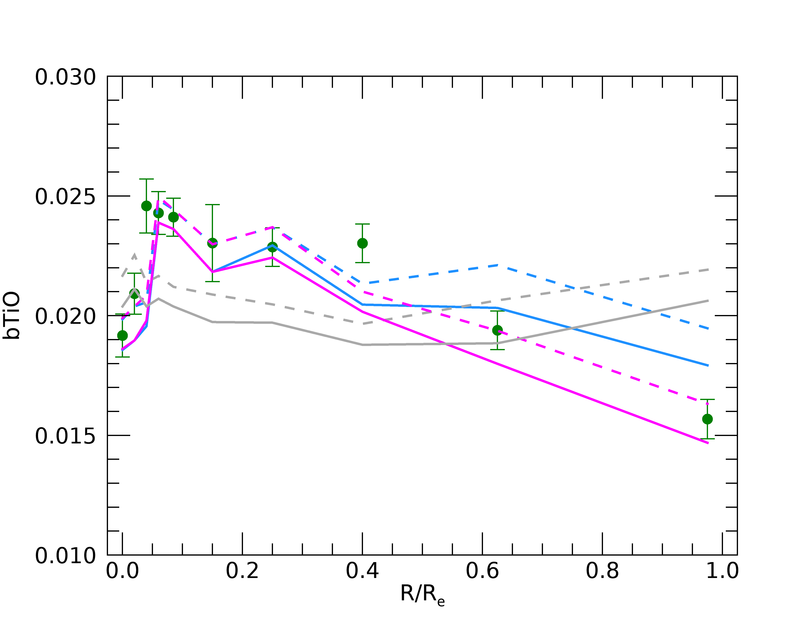} 
\includegraphics[width=7.2cm]{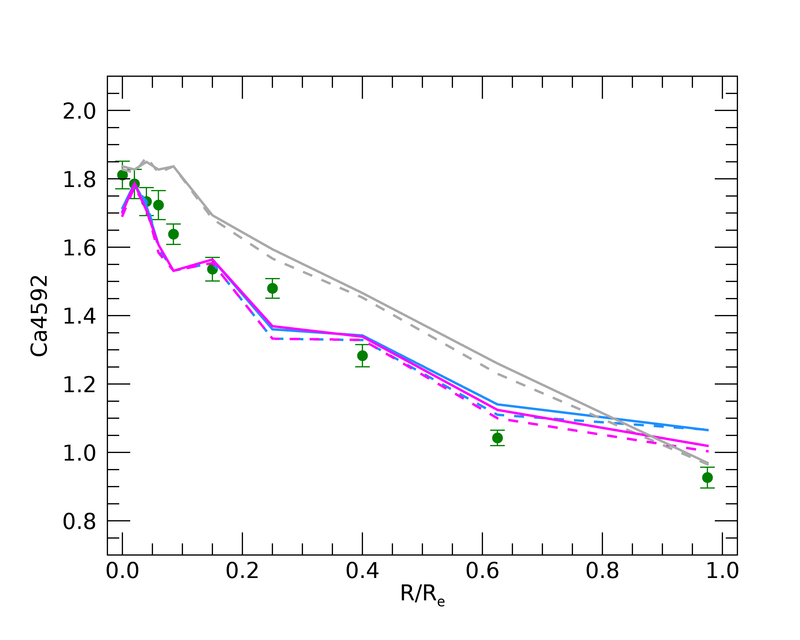} 
\caption{\small{Comparison of a subset of measured index values (green) with those of best-fit models (dashed lines) as a function of radius. Solid lines indicate the weighted mean values of models corresponding to Figure \ref{fig:results}. Gray lines show the same results, but without the corrections for elemental abundances, while blue and magenta lines include all of the abundances and are referred to the results obtained with the all-indices and the emission-free indices sets respectively (analogous to Figure \ref{fig:results}).}} 
\label{fig:indexbest}
\end{centering}
\end{figure*}

A final test of the robustness of the results was to compare the measured index values with those of the best-fit models as a function of radius. For each fitted index, as shown for a subset in Figure \ref{fig:indexbest}, we then compared its measured radial variation (green points) with the corresponding best-fit values (dashed lines) and weighted mean values (solid lines). The measured values are in overall good agreement, with a number of exceptions where the best-fit profile values are rigidly shifted up/down by $1$ or $2\sigma$. Other cases do not show consistency with the central bins or the outskirts. In Figure \ref{fig:indexbest} we show only some significant cases to illustrate, for example, the good retrieval of [Na/H] (with NaD and NaI8190sdss),  [Mg/H] and [Fe/H] (with [MgFe] and Fe$4920$) and [Ca/H] (with Ca$4592$). Regarding the IMF indicators, we show as examples TiO1 and bTiO (and NaI8190sdss which is an IMF indicator as well), where we have good consistency with the weighted mean or the best fit depending on each radial bin. TiO1 and NaI8190sdss,  in particular, show a $>2\sigma$ inconsistency in the first central bins. Not shown, aTiO, shows good agreement with the trend of the solar abundances best-fit/mean (although shifted), but not when including the non-solar elemental abundances in the first half of the galaxy; TiO2sdss has a deviation from the measured values of $<2\sigma$ in the first half of the galaxy. This comparison also illustrates the overall ability of models to reproduce the data. \\

Figure \ref{fig:results} illustrates the successful retrieval of most of the elemental abundances, in particular in the central regions. Moreover, because of our detailed  simulations, we have confidence that these values are not affected by significant degeneracies. 
Indeed, during each step of the analysis up to step four, where we definitively constrained the abundances, we carefully simultaneously fit, together with age, the metallicity, the IMF slopes, and all of the elemental abundances that influence the used spectral indices.
From the figure, looking at the plots it can be seen that there are two different trends in the radial profiles; the first is for the center out to $0.1$ R/R$_e$ and the second for the outer region ($>0.1$ R/R$_e$). For most of the chemical species (i.e. [Na/H], [Mg/H], [Fe/H], [C/H], and [O/H]), in the first radial bins there is a rapid and systematic decrease of their values, while after $0.1$ R/R$_e$ there is a change of trend where the values rise or flatten off. For [Ca/H] and [Ti/H] we instead see flat values in the central region of the galaxy and then a constant decreasing gradient. Remarkably, in the outskirts we observe rather extreme sub-solar values of [Ca/H] and [Ti/H], which are not common in early-type galaxies. High super-solar values of [Na/H] found in the center (up to $0.8$ dex), are instead usually observed in local galaxies (F20, \citealt{conroy12,jeong13,labarbera17}). 

The trend of elemental abundances with radius may be connected with the kinematics of M89 and in particular with its velocity dispersion profile (shown in Figure \ref{fig:kin}), which interestingly follows the same pattern, i.e. a steep decrease and then a smoother negative gradient. From studies on the kinematics of local galaxies with different masses (e.g.: \citealt{conroy14, worthey14, parikh18}), we know that there is an observed correlation between the velocity dispersion of a galaxy and its chemical characteristics, as well as correlations amongst different species. For example, in \citet{conroy14}, the authors have analysed stacked spectra of local early-type galaxies in different velocity dispersion bins and found that in the three highest $\sigma$-bins (those similar to the velocity dispersion values of M89) [C/H], [O/H], [Mg/H], [Fe/H] and [Na/H] are significantly correlated with $\sigma$, while [Ca/H] ad [Ti/H] (the heaviest $\alpha$-elements) have no correlation. Although those findings apply to entire galaxies and not within a single galaxy, our results are consistent, in particular for the inner regions, with the correlations found by \citet{conroy14}. With only one galaxy, this argument lacks statistical significance; however, in Feldmeier-Krause et al., (in preparation), we focus on the correlation among radial gradients and stellar parameters (including velocity dispersion) for a sample of 9 galaxies.\\

This qualitative correlation amongst the velocity dispersion profile and the profiles of many elemental abundances and stellar metallicity could be a sign that the star formation history of M89 is also connected with its kinematic history. A possible second stellar population with different chemical composition could have been accreted at a later time in the outskirts of an already present central bulge (within R/R$_e\sim0.2$). However, this merging event could not have happened too long after the formation of the central region since the ages of the two stellar populations are roughly the same (see the flat trend of the age in Figure \ref{fig:results}). The complexity of the accretion history of M89 has been already suggested by \citet{Janowiecki2010} who found the presence of a variety of tidal structures such as shells and plumes, in its nearby vicinity. These features can be connected with both multiple accretion events or a major merger that happened in the past. 

Another peculiarity of M89 that reaffirms its complicated assembly history is the presence of the inner dust disk. As hinted in Section \ref{sec:indices}, looking at the measured index radial variations (Figure \ref{fig:indices}), a small deviation from the general trend in the first inner radial bin is spotted for most of them. This effect is reflected in the results shown in Figure \ref{fig:results} in the first radial bin of many elemental abundances that show slightly lower values, particularly [Ca/H], [Ti/H] and [Fe/H]. The few exceptions (i.e. Ca$4592$, Fe$4046$, Fe$4064$, NaI$8190$sdss and TiO1) have index values that could be consistent with lower values given their large error bars (larger than the difference with the subsequent radial bin), or indices that are more sensitive to [Z/H] and IMF (and age) like the NaI$8190$sdss. 
We speculate that this difference, though small, in the SP properties at the very center of the galaxy may be connected to the presence of the dust disk in the center of M89. Indeed, as measured in \citet{bonfini18}, the radial trend of the color $F475W - F850LP$ shows a ``red bump", a signature of the dust disk, within a radius of $\sim0.4"$, similar to the scale of our inner radial bin (i.e. $\sim0.3"$). However, while the seeing in the optical region was excellent ($\sim0.5"-0.6"$), the $2.5"$ slit width causes a partial light dilution. Thus, we expect that this effect is only mildly observed in our data. The origin of the inner dust disk in local early-type galaxies is still debated and attributed both to external (e.g. \citealt{xu10}) and internal causes (e.g. \citealt{lauer05,richtler20}). Since M89 exhibits shells and stellar streams (\citealt{Schweizer92,bonfini18}), the presence of dust (and gas, that is observed at larger radii) can be a signature of an accreted small stellar component that can have different characteristics (although diluted by the bulk of stars of M89).

We further considered the possibility that the central lower values observed in many measured indices could be connected with the presence of an active galactic nucleus (AGN). Due to its weak emission lines compared to the underlying strong stellar emission, M89 has been classified as LINER (\citealt{ho97,maoz05}), as well as a low-luminosity AGN \citep{gonzalez09}. In a Hubble Space Telescope (HST) spectrum of the $<0".2$ inner region of M89, \citet{cappellari99} detected broad emission lines ($\sim3000$ km/s) of both forbidden and permitted transitions, confirming the presence of an AGN. This central region is also known to show significant emission variability (\citealt{cappellari99,maoz05}) on the timescale of years, which supports its AGN nature. 
From our IMACS spectra we find the presence of the same emission lines seen in the HST spectrum (\citealt{cappellari99}, fig.6), although diluted. Indeed, our measured velocity dispersion trends of emission line fluxes with radius (as derived from PPXF fits, see Section \ref{sec:kin}) show a decrease from $\sim300$ km/s (around the stellar value) in the center to below $\sim100$ km/s at R/R$_e\sim0.1$, for both permitted and forbidden lines, thus revealing only poor details of the nuclear activity. As described in Section \ref{sec:kin}, we removed the fitted non-stellar components from each radial spectrum before measuring indices. As a result, the observed dip in the central radial bin of many indices is not likely connected with the presence of nuclear emission. Moreover, the fact that several of these indices fall in spectral regions free from emission lines corroborates this idea. Aside from the emission lines, there is also a contribution to the shape of the continuum observed in the HST spectrum shown in \citet{cappellari99}; however, it only affects the UV region below $3000$\AA. We thus conclude that the presence of the AGN is not likely the cause of the observed dip in the central spectral index values, and that it is more plausibly an actual change in the stellar population content.

\subsection{Comparison with Literature}
\label{sec:literature}

The radial trend of the IMF slope, as well as other SP properties of M89 have been already analyzed by \citet{martin-navarro15} and \citet{vandokkum17}. 
Both works relied on high S/N spectroscopic data and observed a clear negative gradient of the IMF despite applying two very different methods.

\begin{figure}[ht]
\begin{centering}
\includegraphics[width=9.5cm]{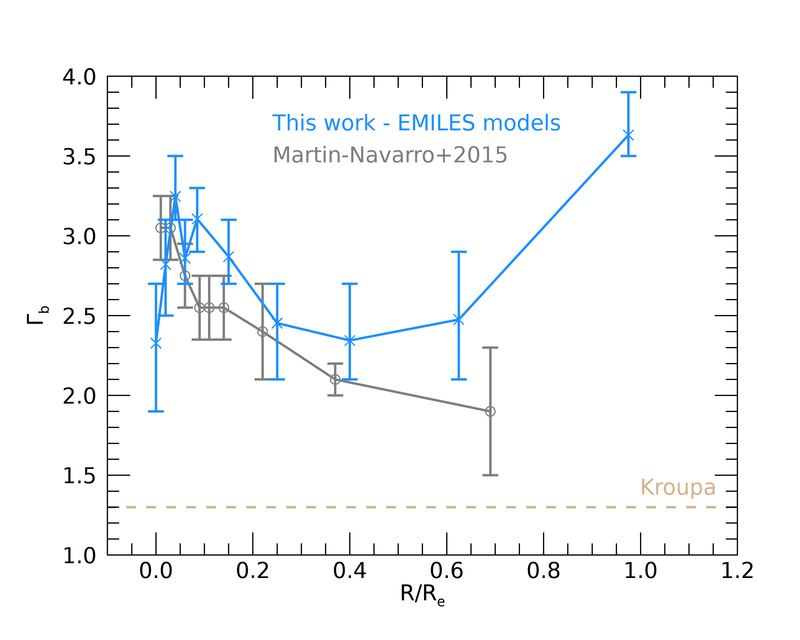}
\includegraphics[width=8cm]{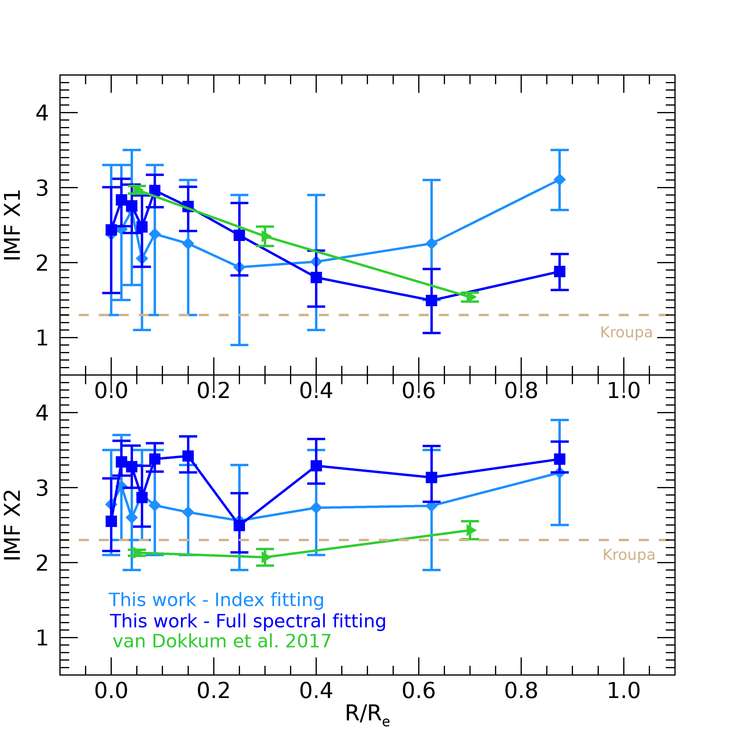}
\caption{\small{Left: comparison of the radial trend of the IMF slope $\Gamma_b$ retrieved with EMILES models (light blue line), with the results shown in \citet[gray line]{martin-navarro15}. The data show good consistency though a small offset is present. Right: comparison of the radial trends of the IMF slopes X1 and X2 retrieved with \citet{conroy18} models from spectral index (light blue lines) and full spectral fitting (blue lines), with the mean results of \citet{vandokkum17} over their galaxy sample (which includes M89, green lines). Excellent agreement is found for X1, and moderate agreement (always within $1\sigma$ of the index fitting) for X2.}}
\label{fig:MNcomparison}
\end{centering}
\end{figure}

In the former work,  the authors retrieved the IMF slope $\Gamma_b$ with the MILES models \citep{vazdekis12}. With a prior on the age estimate obtained from a parallel full spectral fit, they exploited the measures of TiO$1$, TiO$2$, Mg$4780$ (similar definition as bTiO), Ca$2$ and [MgFe] to constrain $\Gamma_b$. In the fit, [Mg/Fe] was previously determined (as described in Appendix \ref{app:alpha_est}), while age, $\Gamma_b$, [Z/H], [$\alpha$/Fe] and [Ti/Fe] were set as free parameters. They observed a clear negative gradient of $\Gamma_b$, and the comparison with our results is shown in Figure \ref{fig:MNcomparison}, left panel. Light blue lines refer to this work using the EMILES models (Appendix \ref{app:emiles}) while gray lines refer to to those from \citet{martin-navarro15}. We find consistent results for $\Gamma_b$ although with a shift of $\sim0.3$ to higher values.

Regarding the comparison with other available parameters from \citet{vazdekis12}, on one hand we find full consistency with the age to within $1\sigma$; on the other hand, the metallicity and $\alpha$-enhancement (shown in Figure \ref{fig:alphaproxy}) exhibit significant inconsistencies with only few exceptions (in the center for the age and around $0.1$R/R$_e$ for [$\alpha$/Fe]). We suspect this is a result of the different characterization of the chemical abundances. We have many more indices in our analysis, and as a result, we have been able to constrain more elemental abundances and therefore give a more detailed (and less biased) description of the chemical balance at each radius.\\

In \citet{vandokkum17}, the authors performed a full spectral fitting analysis on a wide spectral range with the \citet{conroy17} models, retrieving 36 free parameters including IMF slopes and many elemental abundances. Due to lower S/N, the \citet{vandokkum17} data extend only up to around $0.4$ R/R$_e$. Nevertheless, for X1 and X2, the authors report the mean values found for their whole galaxy sample that extend to higher radii. The comparison is shown in Figure \ref{fig:MNcomparison}, right panel. Blue and light blue lines are the trends of our results using the \citet{conroy18} models with full spectral and index fitting respectively, while green lines are the \citet{vandokkum17} results.

The agreement with X1 is excellent, being fully consistent with both the index and full spectral fitting analysis of our data. X2 is lower for  \citet{vandokkum17}, but as a result of our rather large error bars, still consistent with our findings to within $1\sigma$. Regarding the other parameters, not shown in Figure \ref{fig:MNcomparison}, we find that age is largely not consistent in the central $<0.2$ R/R$_e$, probably  due to the secondary stellar population component fit in the \citet{vandokkum17} analysis. Elemental abundances (the 6 out of 17 that are shown in the paper) show a moderate consistency, with \citet{vandokkum17} results having in general steeper gradients. Due to the unavailability of the remaining 11 elemental abundances not shown in \citet{vandokkum17}, little more can be said regarding the reason for the discrepancies.

\section{Summary and Conclusions}
\label{sec:conclusions}

The present work follows the analysis described in F20 where we started the investigation of the  retrieval of the IMF, exploiting spectroscopic data and undertaking analysis comparing different methods and testing a variety of different assumptions.  In that work we showed the complexity of the problem and the different results and biases that different approaches can bring. With this second work we have focused on the IMF slope retrieval using only spectral indices, shedding light on the possible biases that affect such analyses, and suggesting a method to alleviate the degeneracy. 

Our analysis is based on high quality ($100<$ S/N $<500$) spectroscopic data in the optical and NIR wavelength range,  obtained with IMACS at the Magellan-Baade telescope of the local early-type galaxy M89. These data allowed us to divide the light profile of the target into 10 radial bins, while preserving the spectral quality out to $\sim1 R_e$. We measured many spectral features, collecting radial information on the SP properties from $3500-9000$\AA. As a result of the comparison with different families of models (Conroy and EMILES), which offer spectral information for a variety of stellar population parameters including elemental abundances and IMF slopes, we derived such quantities by means of a $\chi^2$ minimization process. 
Due to the large number of models and to the intrinsic dependence of each spectral index on different parameters, we divided the analysis in subsequent steps in order to derive quantities fitting indices that only depend on those parameters. To quantify the biases that arise at each step, we performed parallel simulations aimed at highlighting any possible degeneracy among the retrieved parameters. In this way, both the uncertainties connected with the quality of the data and those resulting from the degenerate nature of the problem are taken into account.
Indeed, with the extended analysis of simulated fits, we demonstrated that unless all of the abundances upon which the chosen set of indices are sensitive to are taken into account, then severe biases affect the results on the IMF slopes. The only way to accurately constrain the IMF slopes is then to have tools to robustly derive the elemental abundances values. In our analysis this was possible due to the many spectral indices included in the fits which, as demonstrated again by means of simulations, are solid indicators of the explored elemental abundances. 

With this method, we derived the radial profiles of the following stellar properties: age, metallicity, IMF slopes (X1 and X2), [Ca/H], [Na/H], [Mg/H], [Ti/H], [Fe/H], [C/H] and [O/H], all shown in Figure \ref{fig:results} and detailed in Table \ref{tab:results}). The age is found to be rather constant with values around $11$ Gyr as found in \citet{martin-navarro15}. Only in the outermost radial bins the age derived with the all-indices set is not consistent to within 1$\sigma$ errors with the emission-free indices set prediction. The metallicity presents a negative gradient from slightly super-solar values toward sub-solar values ($\sim-0.7$ dex). 
The derived elemental abundances generally show decreasing trends from the center to the outer region. For some of them, we speculate that a correlation with the velocity dispersion trend could be a sign of a past merging event.

Our results on the retrieval of the IMF slopes X1 and X2, as parametrized in the Conroy models, are consistent with constant trends of both slopes with bottom-heavy values. X1 is also consistent with a negative gradient although not well constrained due to large error bars, a result of the correlation between the two slopes. For X2, many points are also consistent to within $1\sigma$ with a Kroupa IMF. Both IMF slopes are consistent with the findings of \citet{vandokkum17} for the mean values trends of their galaxy sample.
Adopting the EMILES models, the bimodal IMF slope shows a negative gradient (in agreement with the findings of \citealt{martin-navarro15}) in the first half of the galaxy, but values are bottom-heavy across the overall radial region.
Also when accounting for systematic errors, as suggested from the simulations tailored to M89, the results are clearly showing a bottom-heavy like IMF. 
We also compared the results with those obtained when performing a full spectral fitting analysis on the same spectra. They are fully consistent with the index fitting results on the IMF, as well as many other stellar parameters, and confirmed the negative gradient of X1 from $\sim3$ to $\sim1.5$ along the radius.  

With this work, we again note the complexity of the IMF determination in local early-type galaxies, even when high quality data are available. The level of degeneracy is high and the methods applied to address the degeneracies can fail in different ways, thus leading to different results. With our analysis we have taken a different approach to try and understand the biases affecting different SP properties that arise when fitting several spectral indices. Our findings clearly show that to avoid severe degeneracy in the IMF slopes' retrieval, one must take into account in the fit all the elemental abundances that the set of chosen indices is sensitive to.

\acknowledgments
We thank the anonymous referee for the helpful comments and suggestions which improved the manuscript. I.L. thanks B. Madore for his helpful suggestions on fringing removal during observations and data reduction.

%

\vspace{5mm}
\facility{Magellan: Baade (IMACS)}
\software{IRAF (Tody 1986, Tody 1993), IDL, MOLECFIT (Smette et al. 2015; Kausch et al. 2015), PPXF (Cappellari 2017), PYSTAFF (Vaughan et al. 2018),  emcee (Foreman-Mackey et al. 2013)}





\appendix

\section{Fringing in the Red Grating Spectra}
\label{app:fringes}

\begin{figure*}[ht!]
\begin{centering}
\includegraphics[width=18cm]{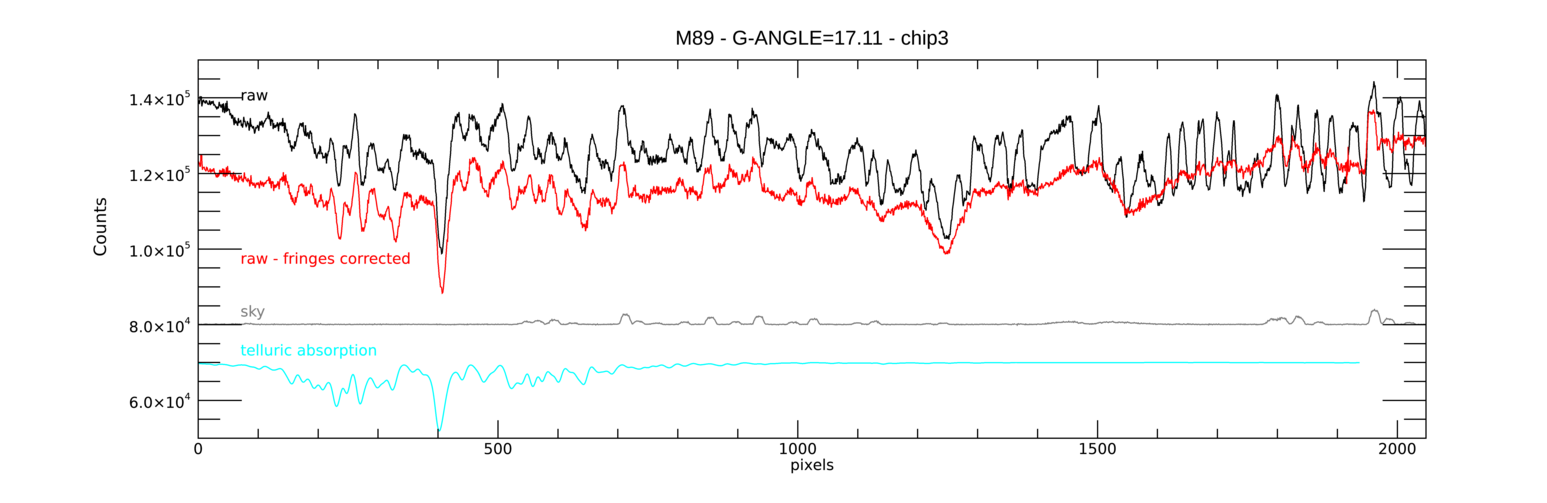}
\caption{\small{Fringing removal for the GA=17.11 spectra. The black line is the raw spectrum extracted from the center of the galaxy; the red line is the same raw spectrum extracted after the removal of fringes; the gray line is a pure sky spectrum for reference, extracted from the same frame but away from the source, and the cyan line a typical telluric absorption spectrum in the same spectral range. The x-axis is in units of pixels.}}
\label{fig:fringes}
\end{centering}
\end{figure*}

During the observation run in May 2018, red grating data were taken with an exposure time of 1h. However, no spectroscopic flat has been taken in the same night and with the same grating configuration. Those data are not useful due to the presence of a high level of fringing which only spectroscopic flats can remove. The shape and amplitude of fringes change with different disperser and grating angles. We noticed also some level of time variation. For this reason, we observed M89 again in May 2019 for one hour with the new red grating configuration (see Table \ref{tab:infoM89}), together with spectroscopic flats taken closely before and after the scientific exposure. The fringing distortion is, in first approximation, proportional to the flux, so the flat field, being constant over all the chip, is the best tool to remove it. After removing bad pixels and cosmic rays, we normalized, slightly smoothed, and applied the flat field to the raw scientific frame. In particular, for each of the three scientific exposures we created an interpolated version of the flat field at the central time of the actual observation time range. This provides the closest version of the fringing effect during each exposure. The level of fringing of the chip (called chip $3$) around $8500$\AA\space is $\sim15-18$\%. Another process that improved the fringing removal was adding the spectra below and above the galaxy center, which have different fringing patterns, but summed together they result in a smoother spectrum. 
Figure \ref{fig:fringes} shows the comparison of raw data (black line) with respect to the flat fielded ones (red lines) where the amplitude of the fringing is highly removed. Both spectra are extracted from the center of the galaxy within $R_e/100$. For reference, a pure sky spectrum and a telluric absorption spectrum are shown with gray and cyan lines respectively. 

We also reduced the $10000$\AA\space region in order to measure the FeH spectral features, but the level of fringing in that region (around $30$\%) prevented us from obtaining a reliable measurement of FeH, even after fringing removal. This is mostly due to the presence of a residual non-linear proportionality of the fringing left in the sky emission lines that lie within the FeH features and that are much stronger than the galaxy flux. Just applying the flat field to the raw frame in this case caused fringes residuals on emission lines that thus cannot be well removed during the background subtraction. On the other hand, other measured features like the CaT and NaI8190\AA\space are fortunately free from strong emission lines and the fringes are sufficiently well removed. 

We compared our data to similar IMACS data taken before 2018 and found that the fringing effect in the red grating of IMACS became worse after the beginning of 2018. Previous data show an almost negligible level of fringing (around $5$\%). This change may be due to the Mosaic2 dewar mounting on the f/4 Channel in IMACS that happened in December 20, 2017.

\section{Statistical and Systematic Errors of Spectral Indices}
\label{app:errors}

Statistical errors have been computed by means of a Monte Carlo approach, repeating the same index measurement $5000$ times on the same observed spectrum perturbed at each pixel with a Gaussian random generated noise. In particular, the noise has been locally estimated as the standard variation calculated over a $60$ pixel wide portion of the spectrum. The standard deviation of the distribution of values of the same index obtained from the $5000$ measurements gives the statistic Gaussian error of that index.
The width of the distributions of percentage statistic errors of all indices runs from $\sim2\%$ for the central radial bins to $\sim5\%$ for the outermost bin.

Systematic errors are more complex since they include different kinds of biases. One source of systematics is the wavelength calibration, causing shifts and distortions of the wavelength solution. Connected to this point is the uncertainty on the radial velocity measurement that could contribute to a residual wrong rigid shifts of the wavelengths happened during the process of bringing spectra to restframe. Both effects can impact index measurements since they slightly change the band-pass ranges of the index definition and alter their value. We took both effects into account by comparing the peak positions of the emission lines of the arc frames, calibrated with the same solution as scientific frames, with the reference values. The difference of each emission line wavelength has been fitted with a polynomial curve of second or third order. The resultant curves for each CCD chip have been applied to the corresponding wavelength arrays. We repeated the index measurements using these corrected wavelengths. The discrepancy with the original values gives an estimate of the systematic uncertainty. As expected, the error values are bigger in regions where the number of arc lines is smaller and the wavelength calibration is less precise. In any case, the percentage errors on index values due to this bias are small, the majority being well below $1\%$. Similarly, we estimated the errors due to rigid shifts of the wavelengths starting from the uncertainty obtained from PPXF fits. The errors on the radial velocity estimates have been transformed in shifts and applied in both positive and negative direction to the wavelengths. As before, index measurements have been repeated on shifted spectra and then used to compute the systematic error of each index. Also in this case, percentage error levels are below $1\%$ for optical GA and around $1\%$ for the red region.\\

The second source of systematic is the relative flux calibration. Distortions of the spectrum shape can alter, in a non-proportional way, the mean flux values of the indices and create biases. This is not the case for the absolute flux calibration, since multiplying the spectrum by any constant value does not change any index value as indices are defined as flux ratios. In our data, shape distortions are present at the edges of each chip. To estimate these distortions in the optical range, we exploited the wavelength range difference between the two GA optical configurations: the GA values have been chosen in such a way that the same wavelength pixel that appears in one configuration close to the edge of the chip, is instead in the middle of the chip in the second configuration. This allows us to compare the shapes of the same spectrum observed in the two different configurations and to estimate their discrepancy. As expected, most differences are found to be close to the edges of the chips. Obviously, in the remaining regions without overlap we could not apply this method, however we did not use indices below $3900$\AA\space or above $6700$\AA\space for our analysis due to low S/N in the first case and for telluric residuals in the second one. For each GA spectrum, we then built the straight lines that connected the edge and the center of the chip following the distortion as it appears in the comparison with the respective other GA. We applied the correction and measured again the index values to estimate the systematic error. As expected, those indices measured close to the chips edges are more affected by this bias. However, for each index that can be combined with its counterpart in the other GA spectrum, this error is minimized. Percentage values of these systematic errors are very small being all below $0.5\%$. For the red region we estimated this systematic component comparing the index values with those measured on spectra that were not flux calibrated. We estimate that  the percentage error in this case is overestimated by $1.5\%$.\\


\section{General Simulations} 
\label{app:gensim}

To systematically examine the level of biases on the IMF retrieval induced by fitting many parameters simultaneously, we took a generic input spectrum with non-solar elemental abundances. We tested the effect of considering a number of abundances from zero to seven, in each of their possible combinations. In particular, as input we chose one of the templates described below in Appendix \ref{app:sim89}, i.e. number 2 in Table \ref{tab:inputsim}, which has an age of $11$ Gyr, solar metallicity, and moderate values of the chemical abundances; we considered both its Kroupa-like and BH-like IMF input values. 
These tests are focused on analysing the effects of using IMF-sensitive indices to retrieve the IMF slopes: thus, for simplicity, we have always assumed that the age, metallicity and all elemental abundances values are already perfectly retrieved (by means, for example, of the presence in the fits of other indices). This means that in all performed fits, we have fixed these quantities to their true input values instead of trying to retrieve them as free parameters. In this way, we are able to isolate the effects of biases on the IMF retrieval only due to the choice of abundances in the fit, thus avoiding those caused by failing to constrain the abundances themselves.
We considered in parallel two sets of IMF-sensitive indices: the first set includes aTiO, bTiO, TiO1 and TiO2, and the second set is the same with the addition of Na$8190$\AA, Ca$1$, Ca$2$ and Ca$3$ from the NIR region. 
Knowing the set of indices, in the analysis we evaluated only the elemental abundances that these indices are sensitive to: these are [Na/H], [Ca/H], [Mg/H], [Ti/H], [Fe/H], [C/H], and [O/H]. 

We started from the simplest case of retrieving the IMF slopes X1 and X2, fixing the age and metallicity at their true input values (as stated before, assuming that they are known by means of other spectral features), and fixing all of the elemental abundances explored to solar values (which are not their true values). 

In the next step, we changed in turn the value of each elemental abundance to its true value in order to see the changes in the X1 and X2 retrieved values. In the following set of fits we then fixed two elemental abundances at a time to their true values, considering all possible combinations. Step by step we increased the number of elemental abundances that have true values until we considered all 7 elemental abundances. Our results are shown in Figs. \ref{fig:simgen1}-\ref{fig:simgen6} (fully described in Appendix \ref{app:subgensim}) 
and \ref{fig:simgenall}. 

The detailed comments on each step of these general simulations are presented in Appendix \ref{app:subgensim}.  Those can be useful to the reader to see the level of bias that is encountered if approaching each particular analysis. 

The upper panel of Figure \ref{fig:simgenall} shows a comparison of the two extreme cases,  one with all of the abundances fixed to solar values (blue points) and the other where all of the abundances are fixed to their true input values (red points). For both kinds of IMF input (Kroupa-like and BH), it is clear that the biases observed when the abundances are not set to their true values completely disappear when all abundances are set to their input values. 

We expected to see a gradual decrease of the observed biases when taking into account the range of abundances from zero to the maximum adopted abundances, but this is not what we found (see Appendix \ref{app:subgensim}). Indeed, in these simulations the source of bias can be not only a result of the elemental abundances that are not set to their true values,  but also the degeneracy between X1 and X2 themselves. Both effects contribute each time in a different way to create severe biases on both X1 and X2. The only case that does not show any of the two effects is the last one where we have fixed all of the abundances at their input values.	\\

To disentangle this double degeneracy, as for the last simulation set shown in Appendix \ref{app:subgensim}, we also repeated the very first simulation when all chemical abundances were solar (Figure \ref{fig:simgenall}, upper panel, blue points), fixing in turn X1 or X2 and thus isolating the biases due only to wrong abundance values. The results are shown in the lower panel of Figure \ref{fig:simgenall} where the same set of blue points from the upper panel plot is shown as green points when X1 is further fixed and as orange points when X2 is fixed. As expected, when fixing X1 and leaving only X2 free (green), there is an improvement in accuracy with respect to fitting both X1 and X2 simultaneously (always leaving all abundances to solar values), meaning that a degeneracy between X1 and X2 is effectively present. Less expected is the case where X1 is set as the only free parameter (orange): while for the smaller set of indices (diamonds) the results are again improving, for the larger set (squares), the retrieval of X1 for both inputs (Kroupa and BH) is considerably more biased. This suggests that in the case where both X1 and X2 are free parameters (blue), the biased X2 values are in some way compensating for the effects of other degeneracies due to elemental abundances, and mimicking a good retrieval of X1. 

A general and evident consideration, looking at all the simulations results (Figs. \ref{fig:simgen1}-\ref{fig:simgen6}, \ref{fig:simgenall}) is that retrieving the IMF slopes is always more difficult and results in biases when the input is a Kroupa-like IMF. This was already observed in F20 when undertaking similar simulations. Indeed, BH spectra have more pronounced features and stronger absorption lines that help to break the degeneracy with other parameters. \\

We also find that adding the $4$ indices of the second set (squared points) in the fit, does not always improve the X1-X2 retrieval. This happens because the  indices added also have  other elemental sensitivities that must be taken into account, so the potential gain in constraining the IMF is masked by the bias introduced by new elemental abundance sensitivity. In this case, for example, the sensitivity of the Calcium Triplet on [Ca/H] and of Na$8190$\AA\space on [Na/H] are further increasing the complexity of the problem, while at the same time contributing to the IMF determination.
More general comments are detailed in Appendix \ref{app:subgensim}.

\begin{figure}[ht!]
\begin{centering}
\includegraphics[width=8cm]{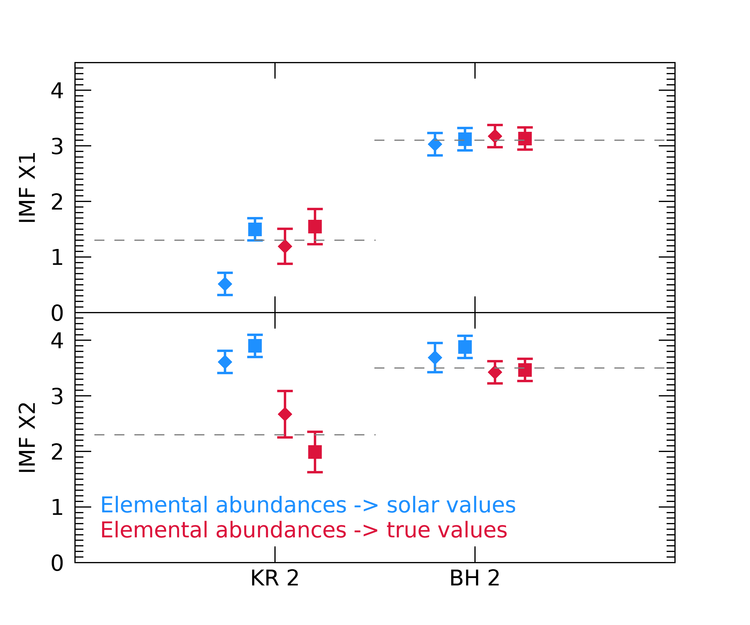}
\includegraphics[width=8cm]{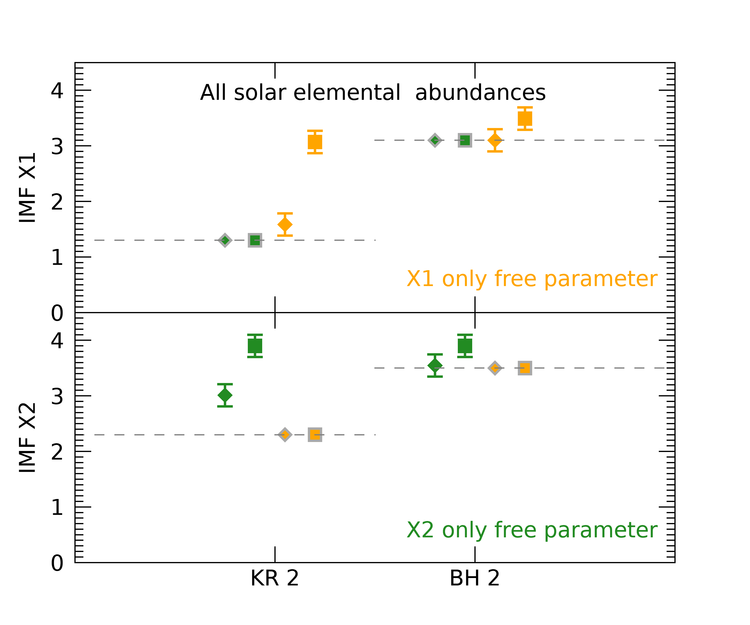}
\caption{\small{Upper panel: simulation results when fitting the input spectra number $2$ (see Table \ref{tab:inputsim}), for both a Kroupa-like IMF (left set of points) and a BH IMF (right set of points). In the fit, age and [Z/H] are fixed to their input values, while X1 and X2 are the only free parameters. For blue points all elemental abundances have solar values, while for red points all of them are fixed to their true input values. Diamonds are the results for the smaller set of indices (aTiO, bTiO, TiO1, TiO2) and squares for the larger set (adding Na$8190$\AA, Ca$1$, Ca$2$ and Ca$3$). Dashed gray horizontal lines indicate the input values. Lower panel: as for the blue set of the upper panel (all solar elemental abundances), but either X1 is fixed (green points) or X2 (orange points).}}
\label{fig:simgenall}
\end{centering}
\end{figure}

We conclude from this general analysis that for a chosen set of available indices, it is necessary to constrain all abundances that the indices are sensitive to, to reliably retrieve values for X1 and X2. Otherwise, severe biases can affect the measurement of the IMF in directions that are particular to each individual analysis.

\subsection{Comments on General Simulations}
\label{app:subgensim}

We devote this appendix section to the detailed comments on the general simulations as presented above, where step by step we have simulated the IMF slopes retrieval with an increasing number of elemental abundances on their true input values. As already mentioned, all the other SP properties such as age and metallicity are fixed to their input values (see Table \ref{tab:inputsim}).

\begin{itemize}
    \item \textbf{First set: One elemental abundance at a time set to its true value.} Focusing on this first set of simulations shown in Figure \ref{fig:simgen1}, it can be noticed that for the Kroupa-like input, X2 is severely biased toward high values at the edge of the modelling for almost all the simulations, without dependency on the used set of indices; For the BH input instead, biases for X2 are observed toward both higher and lower values. X1 is better retrieved with the exceptions of few cases. The best choice to retrieve both X1 and X2 for a Kroupa-like input is to well constrain [Mg/H] to its true value (orange points). 
    
\begin{figure*}[ht!]
\begin{centering}
\includegraphics[width=15.5cm]{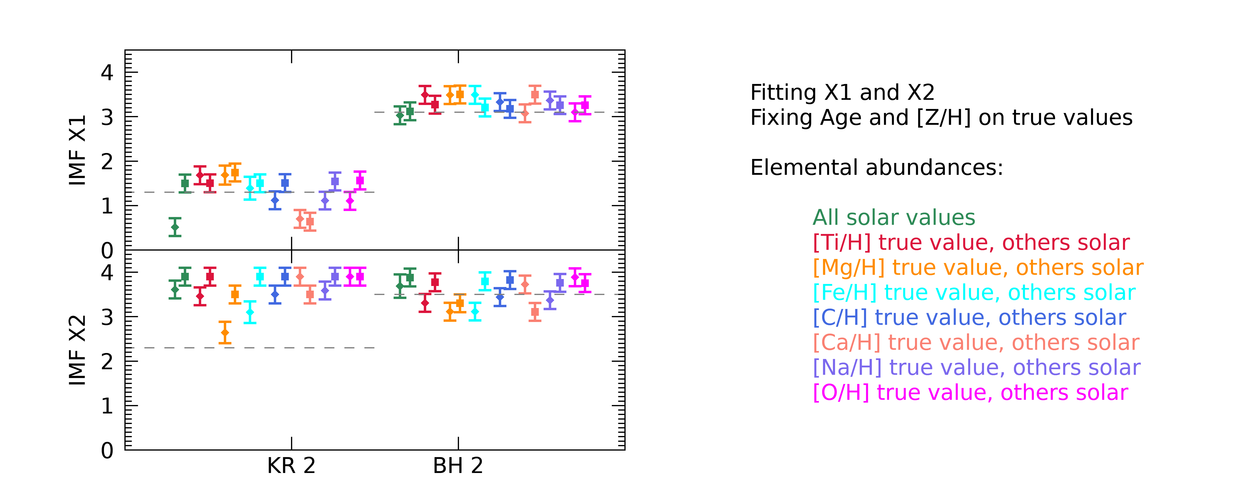}
\caption{\small{Simulation results when fitting the input spectra number 2 (see Table \ref{tab:inputsim}), for both Kroupa-like IMF (left set of points) and BH IMF (right set of points). In the fit, age and [Z/H] are fixed to their input values, while X1 and X2 are the only free parameters. For each colored point, one elemental abundance at a time is fixed to its true input value, while all the others are solar. Diamonds are the results for the smaller set of indices (aTiO, bTiO, TiO1, TiO2) and squares for the larger set (adding Na$8190$\AA, Ca$1$, Ca$2$ and Ca$3$). Dashed gray horizontal lines indicate the input values. The plot clearly shows that if the input IMF is Kroupa-like, severe biases occur on the retrieved IMF slopes values, in particular for X2. Mild degeneracy is observed when the input is BH. }}
\label{fig:simgen1}
\end{centering}
\end{figure*}

    \item \textbf{Second set: Two elemental abundances at a time set to their true value.} Adding a second elemental abundance set to its true value, as shown in Figure \ref{fig:simgen2}, does not much improve the IMF retrieval, still showing a strong bias for X2 toward high values for Kroupa-like input. In this second case, the best choice for a Kroupa-like input is with [Mg/H]-[Ca/H]. The biases for the BH input are again clearly less evident than Kroupa-like input, but still present for most cases. 
    
\begin{figure*}[ht!]
\begin{centering}
\includegraphics[width=15.5cm]{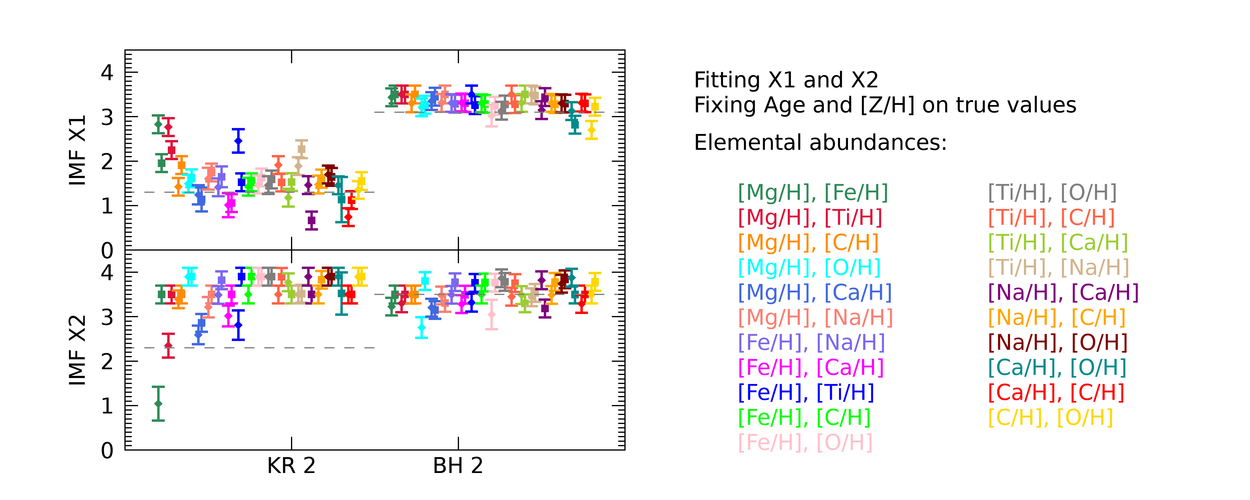}
\caption{\small{As Figure \ref{fig:simgen1} but with 2 elemental abundances fixed each time to their true input values. Again, the X1 and X2 retrieval when the input is Kroupa-like is significantly biased, with X2 still toward higher values. For BH input the biases are less evident but present.}}
\label{fig:simgen2}
\end{centering}
\end{figure*}

    \item \textbf{Third set: Three elemental abundances at a time set to their true value.} With 3 elemental abundances, see Figure \ref{fig:simgen3}, more cases appear to be consistent with the X2 input values, while for X1 more degeneracies are observed pushing values to both lower (at the limit of models) and higher levels. Especially for X1, important differences are found for values retrieved with different index sets. The best choice for a Kroupa-like input is [Ca/H]-[Mg/H]-[C/H], thus improving the previous choice. 
    In these sets of simulations it is possible to spot a regularity for the lower biases of X1 that happens anytime the [O/H] is set to its true value, causing X1 to be at the lowest values allowed by models. Not always this affects both sets of indices. This anomaly is generally confirmed also in the following sets of simulations. 

\begin{figure*}[ht!]
\begin{centering}
\includegraphics[width=15.5cm]{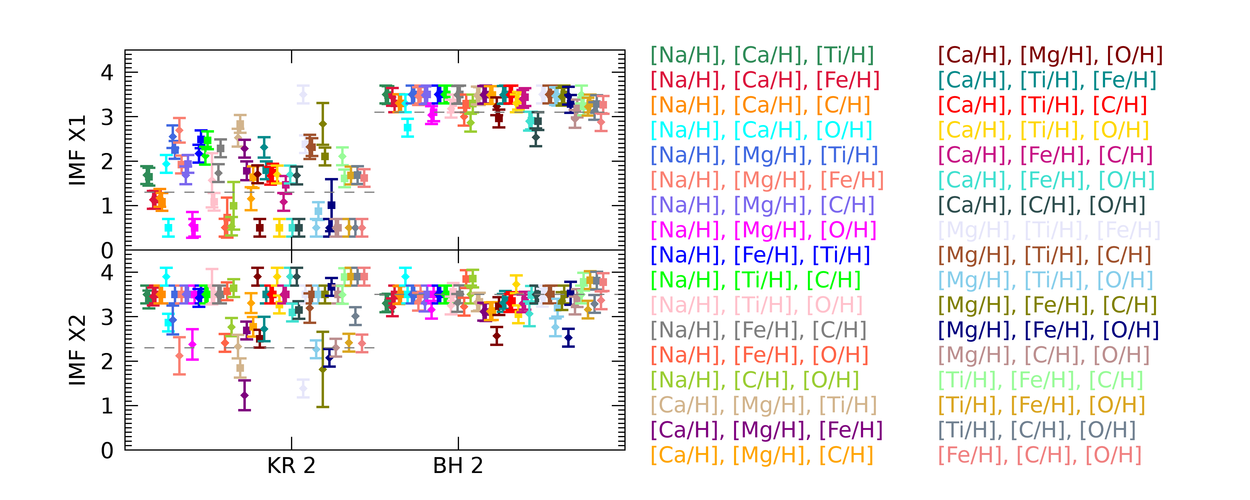}
\caption{\small{As Figure \ref{fig:simgen1} but with 3 elemental abundances fixed each time to their true input values. With respects to previous sets, more cases are less biased for X2, while for X1 more degeneracies appear.
Generally, when one slope is correctly retrieved, the other is severely biased.}}
\label{fig:simgen3}
\end{centering}
\end{figure*}
    
    \item \textbf{Fourth set: Four elemental abundances at a time set to their true value.} From 3 to 4 abundances set to their true values, see Figure \ref{fig:simgen4}, the level of bias of X1 is still high; it improves for X2 only for BH input. As noticed before, those abundance sets which include [O/H] have X1 biased toward the lowest values. 
    The previous good triplet [Ca/H]-[Mg/H]-[C/H] for Kroupa-like input with the addition of the fourth abundance set to its correct value is now generally good only for the retrieval of X2. No choice of 4 abundances is able to give consistent results for both slopes for the Kroupa-like case.

\begin{figure*}[ht!]
\begin{centering}
\includegraphics[width=16cm]{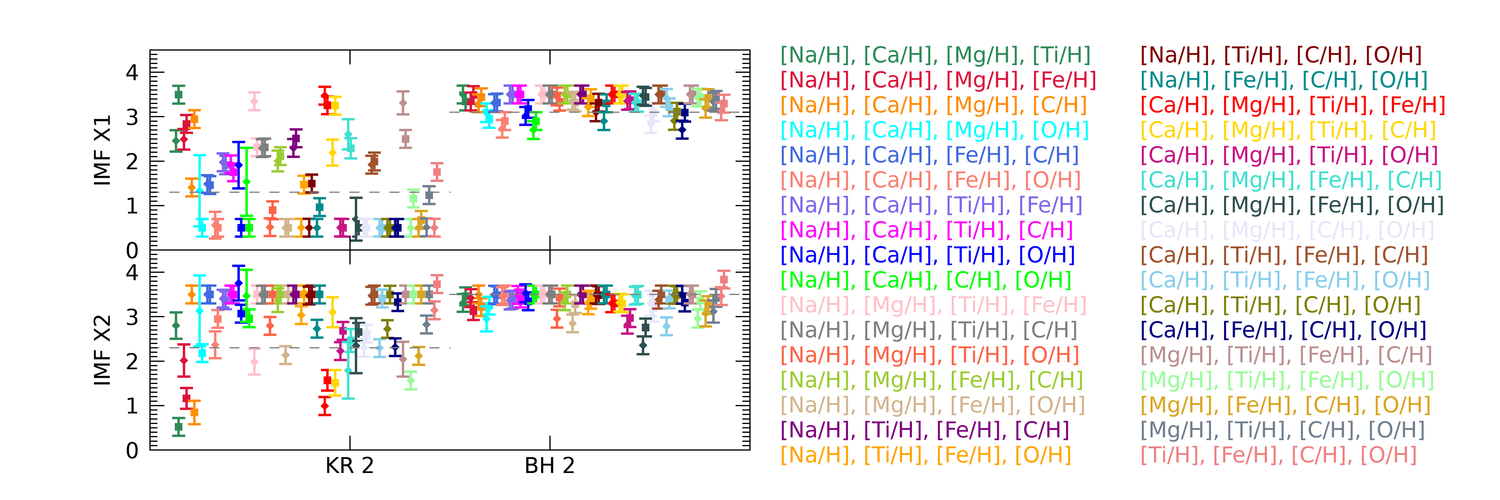}
\caption{\small{As Figure \ref{fig:simgen1} but with 4 elemental abundances fixed each time to their true input values. The level of bias for both X1 and X2 is still high with the only exception of X2 when the input is BH.}}
\label{fig:simgen4}
\end{centering}
\end{figure*}
    
    \item \textbf{Fifth set: Five elemental abundances at a time set to their true value.} With 5 abundances, see Figure \ref{fig:simgen5}, the level of biases of X1 further increases for both kinds of input. Again, the presence of [O/H] creates an underestimation of X1 for Kroupa-like input, with only 2 exceptions for the sets [Na/H]-[Mg/H]-[Ti/H]-[Fe/H]-[O/H] and [Na/H]-[Mg/H]-[Ti/H]-[C/H]-[O/H]. Indeed, these two combinations are the best choices for a Kroupa-like input.

\begin{figure*}[ht!]
\begin{centering}
\includegraphics[width=15.5cm]{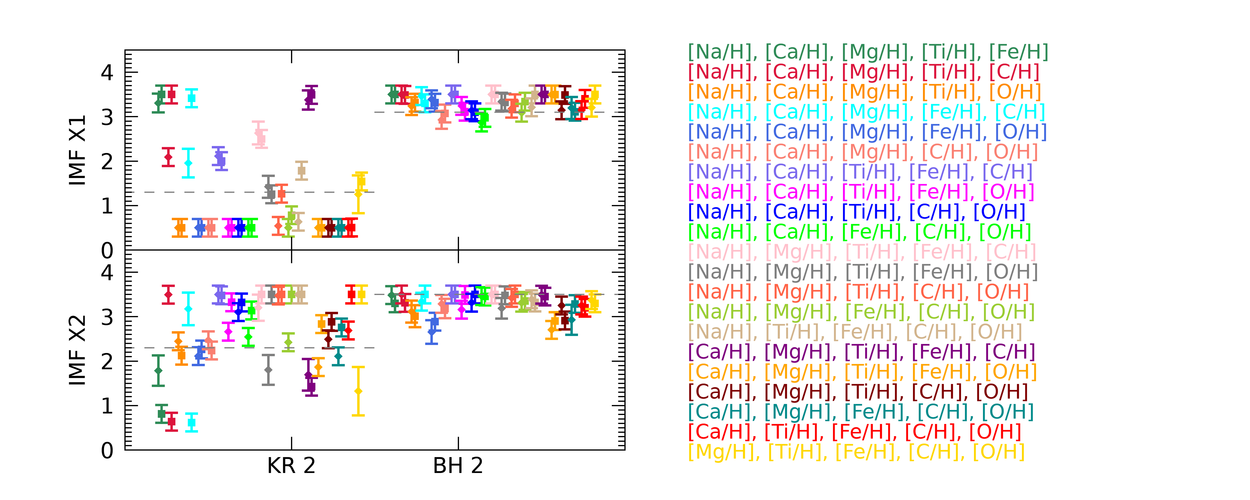}
\caption{\small{As Figure \ref{fig:simgen1} but with 5 elemental abundances fixed each time to their true input values. In this set the level of bias is further increased for both Kroupa-like and BH input.}}
\label{fig:simgen5}
\end{centering}
\end{figure*}
    
    \item \textbf{Sixth set: Six elemental abundances at a time set to their true value.} Finally, with 6 correct elemental abundances over 7 (Figure \ref{fig:simgen6}, upper panel) it is possible to check the effect of missing each single abundance's true value, and, against the expectations, the level of bias is still high for both Kroupa-like and BH input. For the Kroupa-like input results it can be noticed that the presence of [O/H] in almost all the sets, as before, produces biases for X1 toward lower values. The best choice of abundances for the Kroupa-like input is when excluding [Ca/H] (red) while for a BH input the best ones are when excluding [Mg/H] (orange) and excluding [Ti/H] (cyan).
    
\begin{figure*}[ht!]
\begin{centering}
\includegraphics[width=17cm]{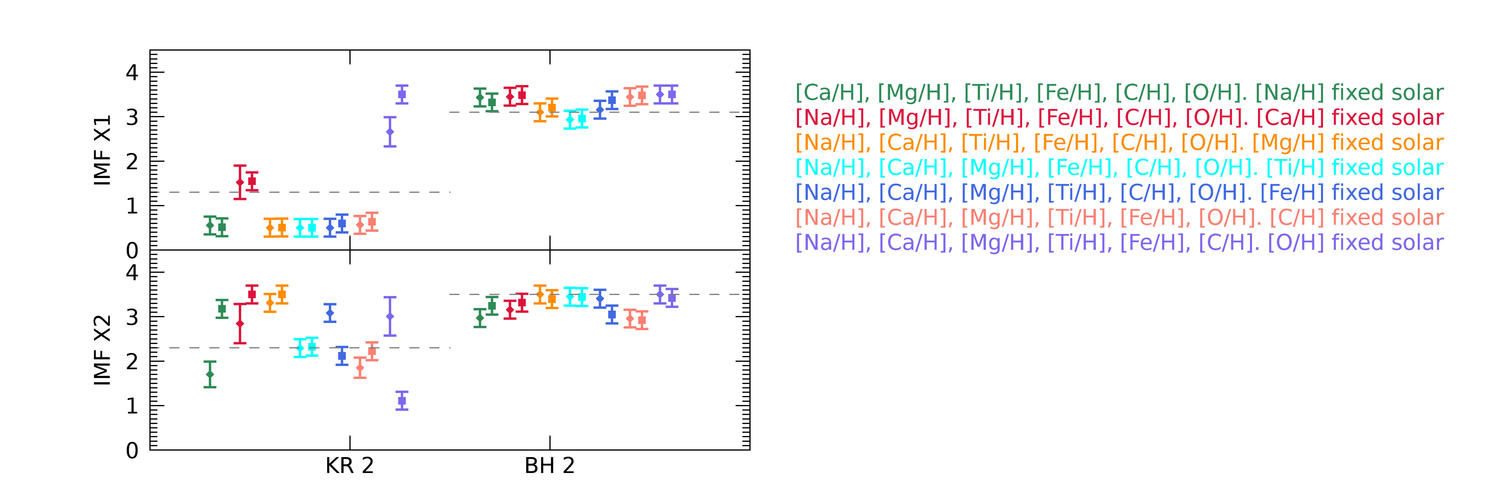}
\includegraphics[width=17cm]{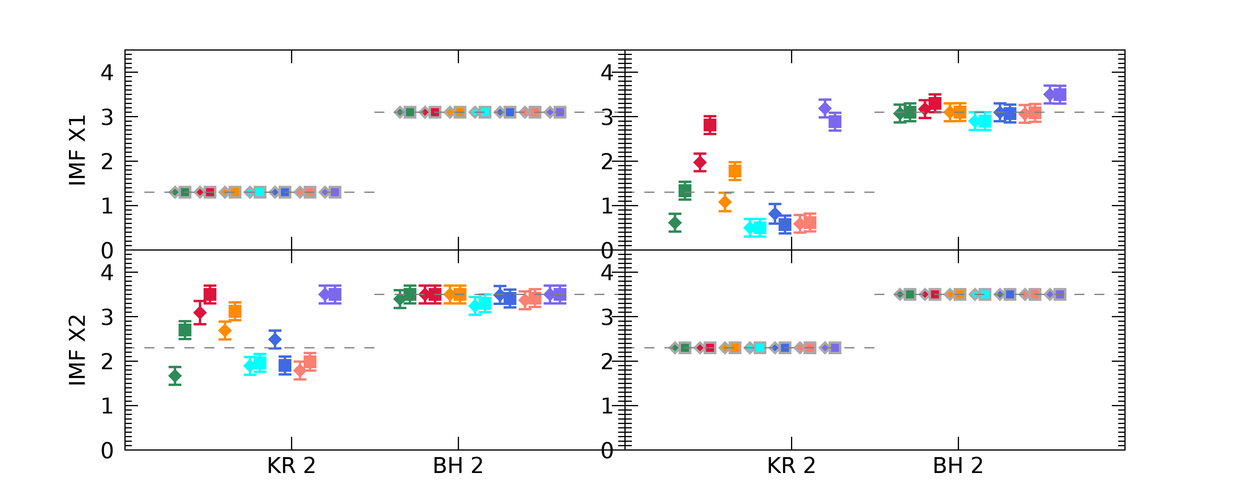}
\caption{\small{Upper panel: as Figure \ref{fig:simgen1} but with 6 (over 7) elemental abundances fixed each time to their true input values. Most of the retrievals are still significantly biased for both Kroupa-like and BH input. The best elemental abundances choice is when excluding [Ca/H] (red) for the Kroupa-like input, and when excluding or [Mg/H] (orange) or [Ti/H] (cyan) for a BH input.
Lower panels: either X1 is also fixed (left panel) or X2 (right panel), same colors as upper panel. Thanks to this plot it is possible to see the biases only due to elemental abundances, since the degeneracy between X1 and X2 is deleted. }}
\label{fig:simgen6}
\end{centering}
\end{figure*}

\end{itemize}

It is worth stressing that in these biases shown so far, there is both the effect of wrong values of the elemental abundances and the degeneracy between X1 and X2. To disentangle this duplicity, we then repeated the last set of simulations (with 6/7 elemental abundances set to their true values) fixing either X1 or X2, thus deleting the X1-X2 degeneracy and obtaining biases that are only due to wrong elemental abundances values. The results are shown in the lower panels of Figure \ref{fig:simgen6}, 
and from them we can observe that: (1) When missing [Na/H] (green) for Kroupa-like input the difference with respect to the previous case (see upper panel) is very small and we can conclude that the biases are due only to the solar value of [Na/H]. On the contrary, for BH input the bias disappeared thus it was only due to the X1-X2 degeneracy. (2) When only [Ca/H] is solar (red), for Kroupa-like, as before, the differences are small and the bias is still present meaning that it is caused by the wrong [Ca/H]. Again for BH, the previous bias was due to the X1-X2 degeneracy. (3) When missing [Mg/H] (orange), for Kroupa-like input the bias is smaller particularly for X1, therefore the previous degeneracy was also due to the X1-X2 degeneracy and not only to the solar [Mg/H]. For BH in the upper panel there were no biases and the result is confirmed. (4) When only [Ti/H] is solar (cyan), X1-X2 values are slightly more biased for both Kroupa and BH inputs, in particular for X2. It can be that the two sources of degeneracy (wrong [Ti/H] value and the X1-X2 degeneracy) produce biases in opposite directions and thus adding them together (upper panel of Figure \ref{fig:simgen6}) brings apparently consistent values. (5) When missing [Fe/H] (blue), for the Kroupa-like input the values are slightly better retrieved, while for BH the bias completely disappeared. (6) Pink set, Kroupa-like input: the difference with respect to the previous figure is very small, the biases are then due to the solar [C/H]. For BH input: again the bias disappeared thus it was only due to the X1-X2 degeneracy. (7) Finally, when only [O/H] is solar (purple), for Kroupa-like input biases are higher for both X1 and X2, with the larger set of indices (squares) going from a very low value to a very high value. Also for BH input, the bias for X1 is still present meaning that it is due to the solar [O/H].\\

In addition to the comments above, we remark some useful general considerations on the results:
\begin{itemize}
\item The  biases obtained toward lower (higher) values for a Kroupa-like input can be higher (lower) for the BH input, meaning that the degeneracy strongly depends on the value  of the indices in a non trivial way. 
\item Typically, for Kroupa-like input, the retrieved values are generally biased toward lower values for X1 and higher values for X2, often pushing to the models' limit values ($0.5$ lowest, $3.9$ highest). 
\item When one slope is biased toward higher values, the other slope tends toward lower values and vice versa, likely indicating the presence of the degeneracy between the two slopes. 
\item Overall, the error bars are very small indicating that the observed biases are well defined and not due to index uncertainties. We tested lowering the S/N of the simulated spectra. The uncertainty increases and can partially mask the effect of degeneracy, and, as a consequence, more fits are consistent with input values. 
\item Adding new indices to a previous set can improve the retrieval but only if all their abundance sensitivities are well taken into account. 
\item Considering all of the studied elemental abundances, [O/H] seems to be most responsible for resulting in biases.
\item It is hard to say which slope is more accurately retrieved since it depends on how many and which elemental abundances are set to their correct values. This effect is mostly correlated with the chosen sets of indices, which include indices sensitive both to X1 and X2 in a balanced way. 
\end{itemize}


\section{Simulations for M89 Analysis}
\label{app:sim89}
\begin{figure*}[ht!]
\begin{centering}
\includegraphics[width=19cm]{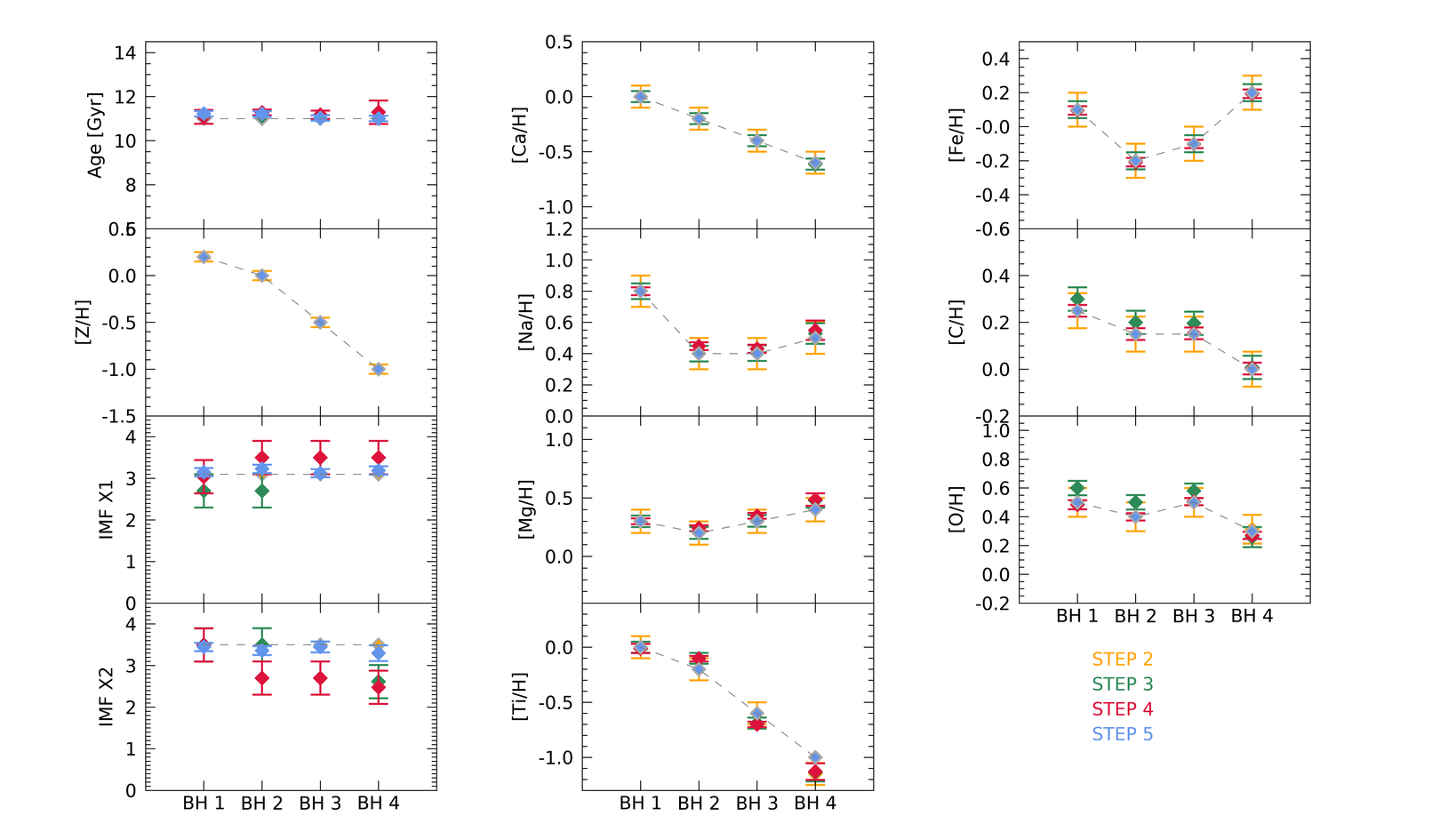}
\includegraphics[width=19cm]{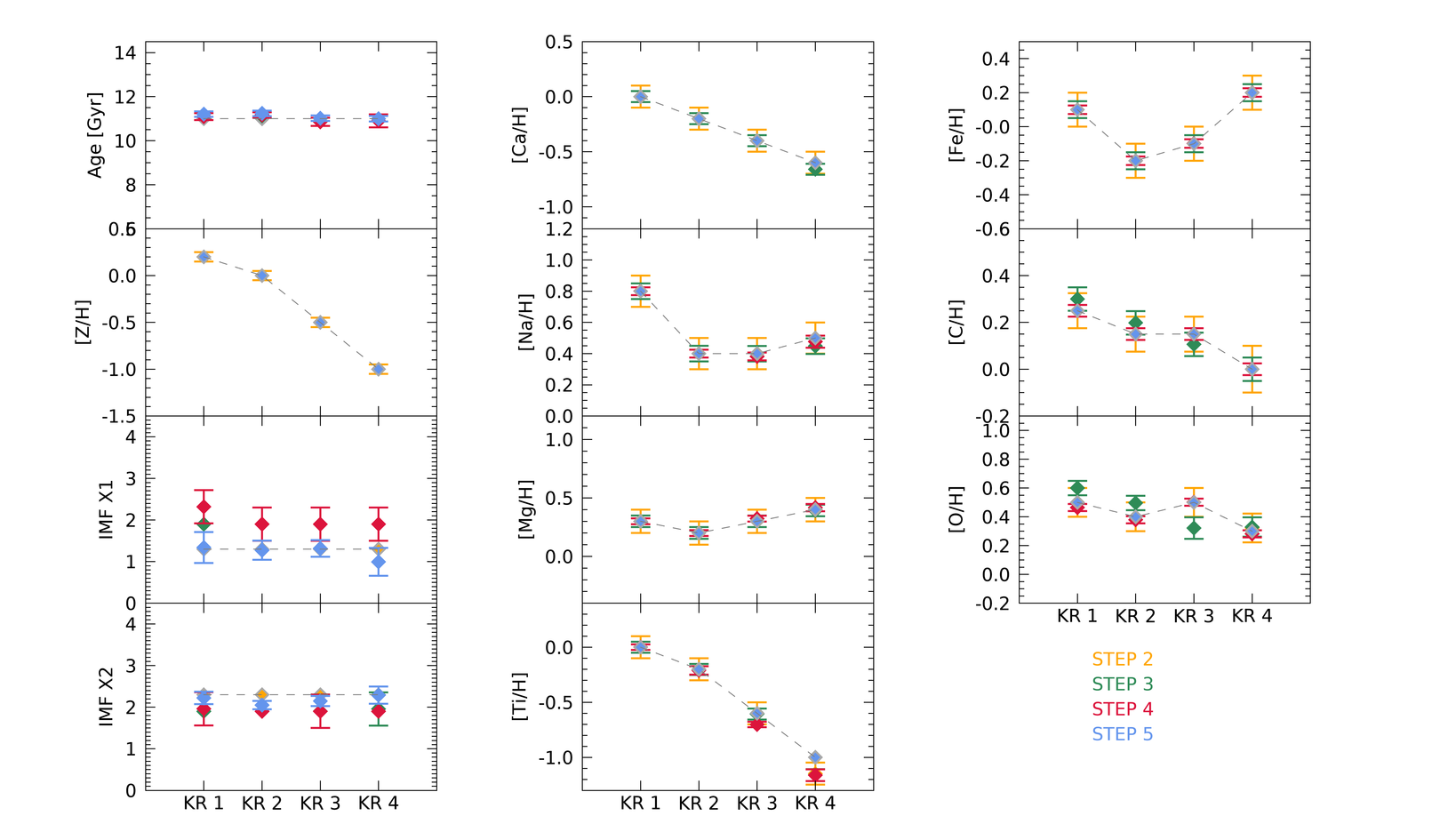} 
\caption{\small{Simulations tailored for the M89 analysis steps for the input values described in Table \ref{tab:inputsim}. Each color indicates a set of simulations performed in the setup of the five steps described in Section \ref{sec:steps}. Upper panels for BH inputs and lower panels for analogous with Kroupa-like inputs. Dashed gray lines show the values of the input parameters. When points of a parameter are encircled in gray, in that step that parameter has been fixed and the error bar is not present. Shown here are only the results when fitting the all-indices set described in Section \ref{sec:steps}; the corresponding ones with the emission-free index set are similar.}} 
\label{fig:sim89}
\end{centering}
\end{figure*}

In the previous section \ref{app:gensim} we have shown the complexity of trying to retrieve the IMF even when the set of indices includes many IMF indicators.
In our analysis we further exploited simulations to be better aware of the biases that can arise in the parameter retrieval we performed on M89 data, as described in Section \ref{sec:steps}.
In each step of the analysis on real M89 data we simultaneously fit different set of parameters, so for each of them, in parallel, we repeated the same fit on $1000$ mock realizations of $4$ input spectra chosen to resemble the derived SP radial profile. We tested the results starting from both BH and Kroupa-like IMF slopes. As for the real data analysis, each simulation has been performed with both the whole spectral index set and the emission-free index set. We chose a decreasing S/N from $450$/\AA\space for the center-like input out to $200$/\AA, following the real data quality. The details of the input spectra are listed in Table \ref{tab:inputsim} and shown in Figure \ref{fig:sim89}.

\begin{table*}[ht!]
 \begin{centering}
 \caption{\small{SP parameters and S/N for the simulated mock input spectra.}}
 \label{tab:inputsim}
 \makebox[\textwidth][r]{
 \begin{tabular}{lcccccccccccc}
 \hline
 \hline
          & Age [Gyr] & [Z/H]  & X1  & X2  & [Ca/H] & [Na/H] & [Mg/H] & [Ti/H] & [Fe/H] & [C/H] & [O/H] & S/N[/\AA] \\
 \hline
 BH 1     & 11.0      &  0.2   & 3.1 & 3.5 &  0.0    & +0.8    & +0.3    &  0.0    & +0.1   & +0.25  & +0.5   & 450 \\
 BH 2     & 11.0      &  0.0   & 3.1 & 3.5 & -0.2    & +0.4    & +0.2    & -0.2    & -0.2   & +0.15  & +0.4   & 400 \\
 BH 3     & 11.0      & -0.5   & 3.1 & 3.5 & -0.4    & +0.4    & +0.3    & -0.6    & -0.1   & +0.15  & +0.5   & 300 \\
 BH 4     & 11.0      & -1.0   & 3.1 & 3.5 & -0.6    & +0.5    & +0.4    & -1.0    & +0.2   & +0.00  & +0.3   & 200 \\
 Kroupa 1 & 11.0      &  0.2   & 1.3 & 2.3 &  0.0    & +0.8    & +0.3    &  0.0    & +0.1   & +0.25  & +0.5   & 450 \\
 Kroupa 2 & 11.0      &  0.0   & 1.3 & 2.3 & -0.2    & +0.4    & +0.2    & -0.2    & -0.2   & +0.15  & +0.4   & 400 \\
 Kroupa 3 & 11.0      & -0.5   & 1.3 & 2.3 & -0.4    & +0.4    & +0.3    & -0.6    & -0.1   & +0.15  & +0.5   & 300 \\
 Kroupa 4 & 11.0      & -1.0   & 1.3 & 2.3 & -0.6    & +0.5    & +0.4    & -1.0    & +0.2   & +0.00  & +0.3   & 200 \\
 \hline
 \end{tabular}
 }
 \end{centering}
\end{table*}

As already described in Section \ref{sec:steps}, we started our data analysis by fitting a small set of indices and considering only 4 elemental abundances ([Na/H], [Ca/H], [Mg/H] and [Fe/H]), as well as the metallicity and IMF slopes. The age was fixed and the same fit was repeated for different values of the age. The aim of this first step was to constrain the metallicity at each radius. For this very first step we did not use the input spectra described in Table \ref{tab:inputsim}, but rather, a larger set of 6 inputs with slightly different SP parameters. As detailed in the step 1 description, the results of these simulations show good retrieval of most of the elemental abundances (with the exception of [Na/H]) and an excellent retrieval of [Z/H]. 
No significant differences are found when using the whole index set or the emission-free index set. We repeated the simulated fits, deliberately setting a wrong fixed age value in order to see the size of the bias when retrieving the other parameters. The age input values for all 6 sets of input was $11$ Gyr, and as wrong values we considered $\pm2$ Gyr from that values, i.e. $9$ and $13$ Gyr. The $13$ Gyr case gives a lower result for [Z/H] by 0.2 dex for both IMF cases. For this reason in step 2 of the data analysis we did not completely fix the metallicity, but confined it to within a $\pm0.2$ dex range at each radius. The opposite case, with age $=9$ Gyr did not produce any significant bias in [Z/H]. 

From step 2 to the final fit we ran simulations with the input described in Table \ref{tab:inputsim}. We summarize the results in Figure \ref{fig:sim89}. 
The allowed models range and grid step for each parameter reflect those effectively used in the real data analysis. 
We also ran simulations with fixed wrong values as for the previous case of age in step 1. For example, since in step 2 (orange) age, X1 and X2 were fixed in the fit, we have explored the cases where we put in wrong values of these parameters in turn (X1 and X2 always coupled). The purpose was to take into account also these biases when deciding the allowed values range of each parameter in the following step. 

The step 2 results (orange) result in an accurate retrieval of metallicity and elemental abundances. Only a small level of bias of [Ti/H] values for the input 4 occurs. Step 3 and 4 (green and red) show an improvement in terms of precision in the elemental abundance retrieval although the IMF is still biased. The improvement is due to smaller allowed fitting ranges around the true values and a smaller grid step (from $0.1$ to $0.05$ dex). A small increased level of bias is observed in [C/H] and [O/H] values. In step 4, the age is not biased for both Kroupa-like and BH inputs and the retrieval of elemental abundances is excellent, with the only exception of [Ti/H] which shows  a small discrepancy in a few cases, which is, however, negligible when considering the  error bars for the real data.

Also in the case of these 4 inputs, as for the general simulations discussed in Appendix \ref{app:gensim}, when putting all the elemental abundances (chosen according to the sensitivity of the adopted spectral indices) at their true values, X1 and X2 are well retrieved in all cases; also together with the age as a free parameter (which is well retrieved as well). This important result is shown in the step 5 set in Figure \ref{fig:sim89} (blue). It is worth mentioning that in repeating these simulations with the smaller set of emission-free indices, the age retrieval is still consistently good. Thus the different behavior of age in the final step of the data analysis (Figure \ref{fig:results}) is not due to hidden biases. 

With this set of simulations we have thus demonstrated that the results obtained in the real data analysis are not affected by biases in the IMF slopes' retrieval. If a small bias is present, it is well below the estimated error bars for the actual data.

\section{EMILES Models Results}
\label{app:emiles}

\begin{figure*}[ht]
\begin{centering}
\includegraphics[width=19.5cm]{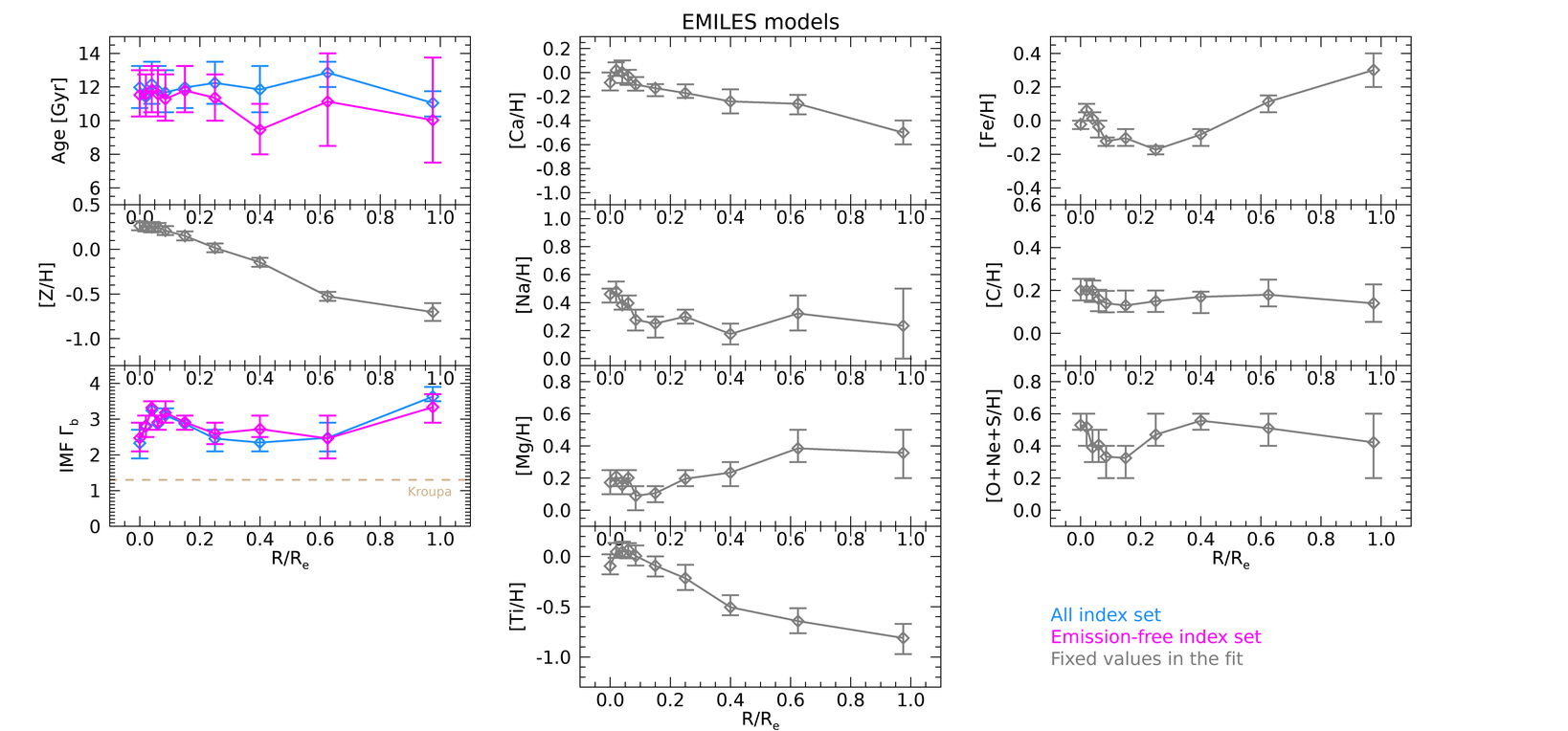}
\caption{\small{EMILES models. Results of the SP retrieval for M89 from the final step 5 of the analysis. All weighted means are plotted as a function of the radius (R/R$_e$). Gray lines indicate the fixed values during step 5. Magenta and cyan lines show the final retrieved values for age and IMF slope for all-indices and for only emission-free indices, respectively.}}
\label{fig:resultsEMILES}
\end{centering}
\end{figure*}

In this Appendix we show the results of the stellar parameters retrieval obtained for M89 that are the analogous to those described in Section \ref{sec:fit} but performed with the EMILES models \citep{vazdekis16}. As already detailed in Section \ref{sec:fit}, the EMILES models have been homogenized to the Conroy models used in the previous analysis in terms of sampling, spectral resolution, interpolation over the same parameters grid and with the application of the same response functions to obtain non-solar values of the elemental abundances. We also highlight that the two model families share the same stellar library (i.e. MILES library) over a part of the spectral range, so they are not fully independent. The only difference with respect to the Conroy models is the parametrization of the IMF slope. In this case, only one parameter is used to shape the IMF trend, i.e. the bimodal logarithmic slope $\Gamma_b$. One parameter less means more precision in the IMF slope retrieval, as observed in the results shown in Figure \ref{fig:resultsEMILES}. As for Figure \ref{fig:results}, the panels show the last step in the analysis where only age and IMF slope are free parameters in the fit, while metallicity and all the elemental abundances have been fixed to the values retrieved in the previous steps. 
Overall, the retrieved radial gradients for all parameters are very similar to the Conroy models results. 
[Ca/H] shows the same trend with the only exception of the two outer radial bins where with EMILES models values are $0.2$ dex higher. [Na/H] presents the same trend but with lower values in all regions of the galaxy, while [Mg/H] is lowered by $0.1$ dex only in the central regions.
[Fe/H], [C/H] and [O/H] have consistent trends with only [C/H] that shows a flatter profile.
Regarding the metallicity, its radial trend is similar but shifted by $\sim0.1$ dex toward higher values for EMILES models results. A systematically lower [Z/H] found with the Conroy models was also observed by F20, and it is due to the fact that generally for the super-solar metallicity and old age models the same index value for the Conroy models at a fixed [Z/H] is associated with instead a lower metallicity value in other models as EMILES and \citet{tmj}. Age has the same rather flat trend around $11.5$ Gyr although for EMILES models it has larger error bars and the results with only emission-free indices (magenta line in Figure \ref{fig:resultsEMILES}) is consistent also with low values down to $8$ Gyr. Due to the large error bars on the age values, on the contrary of what we found for Conroy models results, the all-indices and emission-free indices results are consistent. For other parameters and in particular for the IMF, as in the case of Conroy models, the two sets of indices give consistent results.

The bimodal IMF slope $\Gamma_b$, being the only IMF parameter, is better derived in terms of error bars and it shows a negative gradient from the center (but from R/R$_e$=0.04) out to R/R$_e=0.6$. Only in the outest radial bin, the slope increases again to high values, similarly to the X1/X2 behavior of the Conroy models results. As already hinted, the reliability of this radial bin is compromised by the low S/N of many fitted indices. Nevertheless, the IMF slope values are always well above the Kroupa IMF one, confirming the results.

\subsection{Estimation of [$\alpha$/Fe] with MILES Models}
\label{app:alpha_est}
\begin{figure}
\begin{centering}
\includegraphics[width=16.cm]{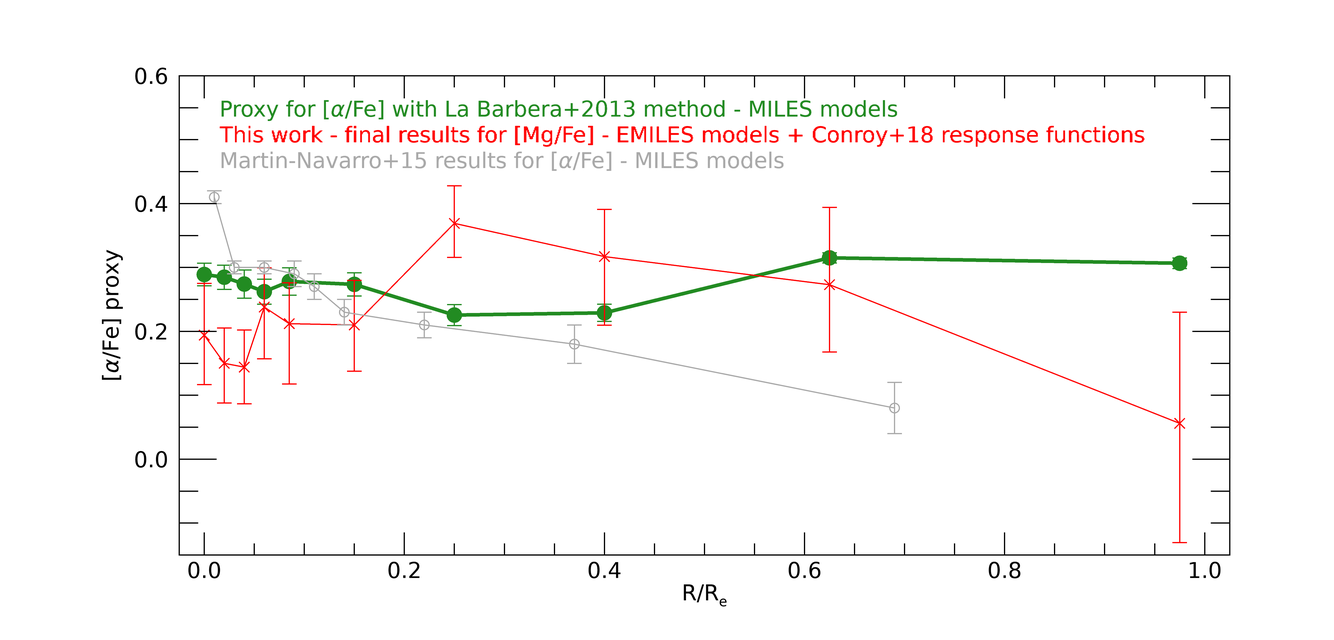}
\caption{\small{[$\alpha$/Fe] trends as a function of radius obtained with \citet{labarbera13} method with MILES models (green). As a comparison, the final [Mg/Fe] values obtained in this work with EMILES models (together with the \citealt{conroy18} response functions for the non-solar values of Mg) and the results of the work of \citet{martin-navarro15} are shown, respectively with red and gray lines and symbols.}}
\label{fig:alphaproxy}
\end{centering}
\end{figure}

As a final comparison, we briefly show the results of an alternative way to determine the [$\alpha$/Fe] value with spectral indices which is the method discussed in \citet{labarbera13}. This proxy makes use of the two metallicity indicators Mg$_b$ and Fe3 (defined as Fe3 $=($Fe$4383+$Fe$5270+$Fe$5335)/3$) to determine the amount of $\alpha$-elements that are needed. \citet{martin-navarro15} used the same method to derive the radial variation of [$\alpha$/Fe] for our same galaxy M89. In F20 we have shown that the Mgb-Fe3 grid is able to give consistent results on the $\alpha$-enhancement comparing with those of the full spectral fitting. We have thus checked the consistency of our results on [Mg/H] (see Figure \ref{fig:resultsEMILES}) with those extracted with this method and found rather consistent results (see Figure \ref{fig:alphaproxy}). To have non-solar [$\alpha$/Fe] values, we used the MILES models \citep{vazdekis15}, interpolating and extrapolating the metallicity up to $0.5$ dex and fixing the IMF bimodal slope to $3.0$ (mean value found from $\Gamma_b$ with EMILES models as described above). As age values in the chosen models we adopted those of our final results with EMILES models, comparing both those obtained with all indices and emission-free indices which gave almost equal results on [$\alpha$/Fe]. In particular, the results of this proxy show a rather constant value of [$\alpha$/Fe] around $\sim0.3$ dex (green line in Figure \ref{fig:alphaproxy}). A small inconsistency is seen in the inner radial bins for which our final results (red line) tend to be lower at around $\sim0.2-0.1$ dex. Comparing to \citet{martin-navarro15} findings (gray line), we observe a disagreement only in the very central bin (for which their result is around $0.4$ dex) and at the outer bins (for which their [$\alpha$/Fe] decreases gradually down to $0.1$ dex). With this test, we confirm the validity of this method to retrieve the [Mg/Fe] values. As we have shown with simulations (see Section \ref{sec:sim}), knowing the values of all the elemental abundances is of fundamental importance to retrieve the IMF slopes accurately, and determining the [Mg/Fe] value in a different way can be useful when the number of available measured indices prevents allowing this parameter to be free in the fit (as described in F20). 

\bibliography{LonoceIMF}{}

\begin{thebibliography}{}
\expandafter\ifx\csname natexlab\endcsname\relax\def\natexlab#1{#1}\fi
\providecommand{\url}[1]{\href{#1}{#1}}
\providecommand{\dodoi}[1]{doi:~\href{http://doi.org/#1}{\nolinkurl{#1}}}
\providecommand{\doeprint}[1]{\href{http://ascl.net/#1}{\nolinkurl{http://ascl.net/#1}}}
\providecommand{\doarXiv}[1]{\href{https://arxiv.org/abs/#1}{\nolinkurl{https://arxiv.org/abs/#1}}}

\bibitem[{{Alton} {et~al.}(2017{\natexlab{a}}){Alton}, {Smith}, \&
  {Lucey}}]{alton17}
{Alton}, P.~D., {Smith}, R.~J., \& {Lucey}, J.~R. 2017{\natexlab{a}}, \mnras,
  468, 1594, \dodoi{10.1093/mnras/stx464}

\bibitem[{{Alton} {et~al.}(2017{\natexlab{b}}){Alton}, {Smith}, \&
  {Lucey}}]{ASL17}
---. 2017{\natexlab{b}}, \mnras, 468, 1594, \dodoi{10.1093/mnras/stx464}

\bibitem[{{Alton} {et~al.}(2018){Alton}, {Smith}, \& {Lucey}}]{alton18}
---. 2018, \mnras, 478, 4464, \dodoi{10.1093/mnras/sty1242}

\bibitem[{{Auger} {et~al.}(2010){Auger}, {Treu}, {Gavazzi}, {Bolton},
  {Koopmans}, \& {Marshall}}]{auger10}
{Auger}, M.~W., {Treu}, T., {Gavazzi}, R., {et~al.} 2010, \apjl, 721, L163,
  \dodoi{10.1088/2041-8205/721/2/L163}

\bibitem[{{Barbosa} {et~al.}(2020){Barbosa}, {Spiniello}, {Arnaboldi},
  {Coccato}, {Hilker}, \& {Richtler}}]{barbosa20}
{Barbosa}, C.~E., {Spiniello}, C., {Arnaboldi}, M., {et~al.} 2020, arXiv
  e-prints, arXiv:2012.11609.
\newblock \doarXiv{2012.11609}

\bibitem[{{Barbosa} {et~al.}(2021){Barbosa}, {Spiniello}, {Arnaboldi},
  {Coccato}, {Hilker}, \& {Richtler}}]{barbosa21}
---. 2021, \aap, 645, L1, \dodoi{10.1051/0004-6361/202039810}

\bibitem[{{Bonfini} {et~al.}(2018){Bonfini}, {Gonz{\'a}lez-Mart{\'\i}n},
  {Fritz}, {Bitsakis}, {Bruzual}, \& {Cervantes Sodi}}]{bonfini18}
{Bonfini}, P., {Gonz{\'a}lez-Mart{\'\i}n}, O., {Fritz}, J., {et~al.} 2018,
  \mnras, 478, 1161, \dodoi{10.1093/mnras/sty1087}

\bibitem[{{Cappellari}(2017)}]{cappellari17}
{Cappellari}, M. 2017, \mnras, 466, 798, \dodoi{10.1093/mnras/stw3020}

\bibitem[{{Cappellari} {et~al.}(1999){Cappellari}, {Renzini}, {Greggio}, {di
  Serego Alighieri}, {Buson}, {Burstein}, \& {Bertola}}]{cappellari99}
{Cappellari}, M., {Renzini}, A., {Greggio}, L., {et~al.} 1999, \apj, 519, 117,
  \dodoi{10.1086/307342}

\bibitem[{{Cappellari} {et~al.}(2011){Cappellari}, {Emsellem}, {Krajnovi{\'c}},
  {McDermid}, {Scott}, {Verdoes Kleijn}, {Young}, {Alatalo}, {Bacon}, {Blitz},
  {Bois}, {Bournaud}, {Bureau}, {Davies}, {Davis}, {de Zeeuw}, {Duc},
  {Khochfar}, {Kuntschner}, {Lablanche}, {Morganti}, {Naab}, {Oosterloo},
  {Sarzi}, {Serra}, \& {Weijmans}}]{cappellari11}
{Cappellari}, M., {Emsellem}, E., {Krajnovi{\'c}}, D., {et~al.} 2011, \mnras,
  413, 813, \dodoi{10.1111/j.1365-2966.2010.18174.x}

\bibitem[{{Cappellari} {et~al.}(2012){Cappellari}, {McDermid}, {Alatalo},
  {Blitz}, {Bois}, {Bournaud}, {Bureau}, {Crocker}, {Davies}, {Davis}, {de
  Zeeuw}, {Duc}, {Emsellem}, {Khochfar}, {Krajnovi{\'c}}, {Kuntschner},
  {Lablanche}, {Morganti}, {Naab}, {Oosterloo}, {Sarzi}, {Scott}, {Serra},
  {Weijmans}, \& {Young}}]{cappellari12}
{Cappellari}, M., {McDermid}, R.~M., {Alatalo}, K., {et~al.} 2012, \nat, 484,
  485, \dodoi{10.1038/nature10972}

\bibitem[{{Cappellari} {et~al.}(2013){Cappellari}, {McDermid}, {Alatalo},
  {Blitz}, {Bois}, {Bournaud}, {Bureau}, {Crocker}, {Davies}, {Davis}, {de
  Zeeuw}, {Duc}, {Emsellem}, {Khochfar}, {Krajnovi{\'c}}, {Kuntschner},
  {Morganti}, {Naab}, {Oosterloo}, {Sarzi}, {Scott}, {Serra}, {Weijmans}, \&
  {Young}}]{cappellari13}
---. 2013, \mnras, 432, 1862, \dodoi{10.1093/mnras/stt644}

\bibitem[{{Cenarro} {et~al.}(2001){Cenarro}, {Cardiel}, {Gorgas}, {Peletier},
  {Vazdekis}, \& {Prada}}]{cenarro01}
{Cenarro}, A.~J., {Cardiel}, N., {Gorgas}, J., {et~al.} 2001, \mnras, 326, 959,
  \dodoi{10.1046/j.1365-8711.2001.04688.x}

\bibitem[{{Cenarro} {et~al.}(2003){Cenarro}, {Gorgas}, {Vazdekis}, {Cardiel},
  \& {Peletier}}]{cenarro03}
{Cenarro}, A.~J., {Gorgas}, J., {Vazdekis}, A., {Cardiel}, N., \& {Peletier},
  R.~F. 2003, \mnras, 339, L12, \dodoi{10.1046/j.1365-8711.2003.06360.x}

\bibitem[{{Chabrier}(2003)}]{chabrier03}
{Chabrier}, G. 2003, \pasp, 115, 763, \dodoi{10.1086/376392}

\bibitem[{{Chabrier} {et~al.}(2014){Chabrier}, {Hennebelle}, \&
  {Charlot}}]{chabrier14}
{Chabrier}, G., {Hennebelle}, P., \& {Charlot}, S. 2014, \apj, 796, 75,
  \dodoi{10.1088/0004-637X/796/2/75}

\bibitem[{{Conroy} {et~al.}(2014){Conroy}, {Graves}, \& {van
  Dokkum}}]{conroy14}
{Conroy}, C., {Graves}, G.~J., \& {van Dokkum}, P.~G. 2014, \apj, 780, 33,
  \dodoi{10.1088/0004-637X/780/1/33}

\bibitem[{{Conroy} \& {van Dokkum}(2012)}]{conroy12}
{Conroy}, C., \& {van Dokkum}, P.~G. 2012, \apj, 760, 71,
  \dodoi{10.1088/0004-637X/760/1/71}

\bibitem[{{Conroy} {et~al.}(2017){Conroy}, {van Dokkum}, \&
  {Villaume}}]{conroy17}
{Conroy}, C., {van Dokkum}, P.~G., \& {Villaume}, A. 2017, \apj, 837, 166,
  \dodoi{10.3847/1538-4357/aa6190}

\bibitem[{{Conroy} {et~al.}(2018){Conroy}, {Villaume}, {van Dokkum}, \&
  {Lind}}]{conroy18}
{Conroy}, C., {Villaume}, A., {van Dokkum}, P.~G., \& {Lind}, K. 2018, \apj,
  854, 139, \dodoi{10.3847/1538-4357/aaab49}

\bibitem[{{Dom{\'\i}nguez S{\'a}nchez} {et~al.}(2019){Dom{\'\i}nguez
  S{\'a}nchez}, {Bernardi}, {Brownstein}, {Drory}, \& {Sheth}}]{dominguez19}
{Dom{\'\i}nguez S{\'a}nchez}, H., {Bernardi}, M., {Brownstein}, J.~R., {Drory},
  N., \& {Sheth}, R.~K. 2019, \mnras, 489, 5612, \dodoi{10.1093/mnras/stz2414}

\bibitem[{{Dressler} {et~al.}(2006){Dressler}, {Hare}, {Bigelow}, \&
  {Osip}}]{dressler06}
{Dressler}, A., {Hare}, T., {Bigelow}, B.~C., \& {Osip}, D.~J. 2006, Society of
  Photo-Optical Instrumentation Engineers (SPIE) Conference Series, Vol. 6269,
  {IMACS: the wide-field imaging spectrograph on Magellan-Baade}, 62690F,
  \dodoi{10.1117/12.670573}

\bibitem[{{Eftekhari} {et~al.}(2019){Eftekhari}, {Mosleh}, {Vazdekis}, \&
  {Tavasoli}}]{eftekhari19}
{Eftekhari}, E., {Mosleh}, M., {Vazdekis}, A., \& {Tavasoli}, S. 2019, \mnras,
  486, 3788, \dodoi{10.1093/mnras/stz1113}

\bibitem[{{Emsellem} {et~al.}(2004){Emsellem}, {Cappellari}, {Peletier},
  {McDermid}, {Bacon}, {Bureau}, {Copin}, {Davies}, {Krajnovi{\'c}},
  {Kuntschner}, {Miller}, \& {de Zeeuw}}]{emsellem04}
{Emsellem}, E., {Cappellari}, M., {Peletier}, R.~F., {et~al.} 2004, \mnras,
  352, 721, \dodoi{10.1111/j.1365-2966.2004.07948.x}

\bibitem[{{Ferreras} {et~al.}(2013){Ferreras}, {La Barbera}, {de La Rosa},
  {Vazdekis}, {de Carvalho}, {Falcon-Barroso}, \& {Ricciardelli}}]{ferreras13}
{Ferreras}, I., {La Barbera}, F., {de La Rosa}, I.~G., {et~al.} 2013, \mnras,
  429, L15, \dodoi{10.1093/mnrasl/sls014}

\bibitem[{{Foreman-Mackey} {et~al.}(2013){Foreman-Mackey}, {Hogg}, {Lang}, \&
  {Goodman}}]{foreman-mackey13}
{Foreman-Mackey}, D., {Hogg}, D.~W., {Lang}, D., \& {Goodman}, J. 2013, \pasp,
  125, 306, \dodoi{10.1086/670067}

\bibitem[{{Gonz{\'a}lez-Mart{\'\i}n} {et~al.}(2009){Gonz{\'a}lez-Mart{\'\i}n},
  {Masegosa}, {M{\'a}rquez}, {Guainazzi}, \&
  {Jim{\'e}nez-Bail{\'o}n}}]{gonzalez09}
{Gonz{\'a}lez-Mart{\'\i}n}, O., {Masegosa}, J., {M{\'a}rquez}, I., {Guainazzi},
  M., \& {Jim{\'e}nez-Bail{\'o}n}, E. 2009, \aap, 506, 1107,
  \dodoi{10.1051/0004-6361/200912288}

\bibitem[{{Gregg}(1994)}]{gregg94}
{Gregg}, M.~D. 1994, \aj, 108, 2164, \dodoi{10.1086/117228}

\bibitem[{{Hamilton}(1985)}]{hamilton85}
{Hamilton}, D. 1985, \apj, 297, 371, \dodoi{10.1086/163537}

\bibitem[{{Ho} {et~al.}(1997){Ho}, {Filippenko}, \& {Sargent}}]{ho97}
{Ho}, L.~C., {Filippenko}, A.~V., \& {Sargent}, W. L.~W. 1997, \apjs, 112, 315,
  \dodoi{10.1086/313041}

\bibitem[{{Hopkins}(2013)}]{hopkins13}
{Hopkins}, P.~F. 2013, \mnras, 433, 170, \dodoi{10.1093/mnras/stt713}

\bibitem[{{Janowiecki} {et~al.}(2010){Janowiecki}, {Mihos}, {Harding},
  {Feldmeier}, {Rudick}, \& {Morrison}}]{Janowiecki2010}
{Janowiecki}, S., {Mihos}, J.~C., {Harding}, P., {et~al.} 2010, \apj, 715, 972,
  \dodoi{10.1088/0004-637X/715/2/972}

\bibitem[{{Jeong} {et~al.}(2013){Jeong}, {Yi}, {Kyeong}, {Sarzi}, {Sung}, \&
  {Oh}}]{jeong13}
{Jeong}, H., {Yi}, S.~K., {Kyeong}, J., {et~al.} 2013, \apjs, 208, 7,
  \dodoi{10.1088/0067-0049/208/1/7}

\bibitem[{{Kausch} {et~al.}(2015){Kausch}, {Noll}, {Smette}, {Kimeswenger},
  {Barden}, {Szyszka}, {Jones}, {Sana}, {Horst}, \& {Kerber}}]{molecfit2}
{Kausch}, W., {Noll}, S., {Smette}, A., {et~al.} 2015, \aap, 576, A78,
  \dodoi{10.1051/0004-6361/201423909}

\bibitem[{{Krajnovi{\'c}} {et~al.}(2011){Krajnovi{\'c}}, {Emsellem},
  {Cappellari}, {Alatalo}, {Blitz}, {Bois}, {Bournaud}, {Bureau}, {Davies},
  {Davis}, {de Zeeuw}, {Khochfar}, {Kuntschner}, {Lablanche}, {McDermid},
  {Morganti}, {Naab}, {Oosterloo}, {Sarzi}, {Scott}, {Serra}, {Weijmans}, \&
  {Young}}]{krajnovic11}
{Krajnovi{\'c}}, D., {Emsellem}, E., {Cappellari}, M., {et~al.} 2011, \mnras,
  414, 2923, \dodoi{10.1111/j.1365-2966.2011.18560.x}

\bibitem[{{Kroupa}(2001)}]{kroupa}
{Kroupa}, P. 2001, \mnras, 322, 231, \dodoi{10.1046/j.1365-8711.2001.04022.x}

\bibitem[{{La Barbera} {et~al.}(2013){La Barbera}, {Ferreras}, {Vazdekis}, {de
  la Rosa}, {de Carvalho}, {Trevisan}, {Falc{\'o}n-Barroso}, \&
  {Ricciardelli}}]{labarbera13}
{La Barbera}, F., {Ferreras}, I., {Vazdekis}, A., {et~al.} 2013, \mnras, 433,
  3017, \dodoi{10.1093/mnras/stt943}

\bibitem[{{La Barbera} {et~al.}(2017){La Barbera}, {Vazdekis}, {Ferreras},
  {Pasquali}, {Allende Prieto}, {R{\"o}ck}, {Aguado}, \&
  {Peletier}}]{labarbera17}
{La Barbera}, F., {Vazdekis}, A., {Ferreras}, I., {et~al.} 2017, \mnras, 464,
  3597, \dodoi{10.1093/mnras/stw2407}

\bibitem[{{La Barbera} {et~al.}(2016){La Barbera}, {Vazdekis}, {Ferreras},
  {Pasquali}, {Cappellari}, {Mart{\'\i}n-Navarro}, {Sch{\"o}nebeck}, \&
  {Falc{\'o}n-Barroso}}]{labarbera16}
---. 2016, \mnras, 457, 1468, \dodoi{10.1093/mnras/stv2996}

\bibitem[{{La Barbera} {et~al.}(2019){La Barbera}, {Vazdekis}, {Ferreras},
  {Pasquali}, {Allende Prieto}, {Mart{\'\i}n-Navarro}, {Aguado}, {de Carvalho},
  {Rembold}, {Falc{\'o}n-Barroso}, \& {van de Ven}}]{labarbera19}
---. 2019, \mnras, 489, 4090, \dodoi{10.1093/mnras/stz2192}

\bibitem[{{Larson}(1998)}]{larson98}
{Larson}, R.~B. 1998, \mnras, 301, 569,
  \dodoi{10.1046/j.1365-8711.1998.02045.x}

\bibitem[{{Larson}(2005)}]{larson05}
---. 2005, \mnras, 359, 211, \dodoi{10.1111/j.1365-2966.2005.08881.x}

\bibitem[{{Lauer} {et~al.}(2005){Lauer}, {Faber}, {Gebhardt}, {Richstone},
  {Tremaine}, {Ajhar}, {Aller}, {Bender}, {Dressler}, {Filippenko}, {Green},
  {Grillmair}, {Ho}, {Kormendy}, {Magorrian}, {Pinkney}, \& {Siopis}}]{lauer05}
{Lauer}, T.~R., {Faber}, S.~M., {Gebhardt}, K., {et~al.} 2005, \aj, 129, 2138,
  \dodoi{10.1086/429565}

\bibitem[{{Leier} {et~al.}(2016){Leier}, {Ferreras}, {Saha}, {Charlot},
  {Bruzual}, \& {La Barbera}}]{leier16}
{Leier}, D., {Ferreras}, I., {Saha}, P., {et~al.} 2016, \mnras, 459, 3677,
  \dodoi{10.1093/mnras/stw885}

\bibitem[{{Maoz} {et~al.}(2005){Maoz}, {Nagar}, {Falcke}, \& {Wilson}}]{maoz05}
{Maoz}, D., {Nagar}, N.~M., {Falcke}, H., \& {Wilson}, A.~S. 2005, \apj, 625,
  699, \dodoi{10.1086/429795}

\bibitem[{{Mart{\'\i}n-Navarro} {et~al.}(2015){Mart{\'\i}n-Navarro}, {La
  Barbera}, {Vazdekis}, {Falc{\'o}n-Barroso}, \& {Ferreras}}]{martin-navarro15}
{Mart{\'\i}n-Navarro}, I., {La Barbera}, F., {Vazdekis}, A.,
  {Falc{\'o}n-Barroso}, J., \& {Ferreras}, I. 2015, \mnras, 447, 1033,
  \dodoi{10.1093/mnras/stu2480}

\bibitem[{{McConnell} {et~al.}(2016){McConnell}, {Lu}, \& {Mann}}]{mcconnell16}
{McConnell}, N.~J., {Lu}, J.~R., \& {Mann}, A.~W. 2016, \apj, 821, 39,
  \dodoi{10.3847/0004-637X/821/1/39}

\bibitem[{{Parikh} {et~al.}(2018){Parikh}, {Thomas}, {Maraston}, {Westfall},
  {Goddard}, {Lian}, {Meneses-Goytia}, {Jones}, {Vaughan}, {Andrews},
  {Bershady}, {Bizyaev}, {Brinkmann}, {Brownstein}, {Bundy}, {Drory},
  {Emsellem}, {Law}, {Newman}, {Roman-Lopes}, {Wake}, {Yan}, \&
  {Zheng}}]{parikh18}
{Parikh}, T., {Thomas}, D., {Maraston}, C., {et~al.} 2018, \mnras, 477, 3954,
  \dodoi{10.1093/mnras/sty785}

\bibitem[{{Pietrinferni} {et~al.}(2004){Pietrinferni}, {Cassisi}, {Salaris}, \&
  {Castelli}}]{pietrinferni04}
{Pietrinferni}, A., {Cassisi}, S., {Salaris}, M., \& {Castelli}, F. 2004, \apj,
  612, 168, \dodoi{10.1086/422498}

\bibitem[{{Richtler} {et~al.}(2020){Richtler}, {Hilker}, \&
  {Iodice}}]{richtler20}
{Richtler}, T., {Hilker}, M., \& {Iodice}, E. 2020, \aap, 643, A120,
  \dodoi{10.1051/0004-6361/202038150}

\bibitem[{{Salpeter}(1955)}]{salpeter}
{Salpeter}, E.~E. 1955, \apj, 121, 161, \dodoi{10.1086/145971}

\bibitem[{{Sarzi} {et~al.}(2018){Sarzi}, {Spiniello}, {La Barbera},
  {Krajnovi{\'c}}, \& {van den Bosch}}]{sarzi18}
{Sarzi}, M., {Spiniello}, C., {La Barbera}, F., {Krajnovi{\'c}}, D., \& {van
  den Bosch}, R. 2018, \mnras, 478, 4084, \dodoi{10.1093/mnras/sty1092}

\bibitem[{{Schweizer} \& {Seitzer}(1992)}]{Schweizer92}
{Schweizer}, F., \& {Seitzer}, P. 1992, \aj, 104, 1039, \dodoi{10.1086/116296}

\bibitem[{{Smette} {et~al.}(2015){Smette}, {Sana}, {Noll}, {Horst}, {Kausch},
  {Kimeswenger}, {Barden}, {Szyszka}, {Jones}, {Gallenne}, {Vinther},
  {Ballester}, \& {Taylor}}]{molecfit1}
{Smette}, A., {Sana}, H., {Noll}, S., {et~al.} 2015, \aap, 576, A77,
  \dodoi{10.1051/0004-6361/201423932}

\bibitem[{{Smith} {et~al.}(2012){Smith}, {Lucey}, \& {Carter}}]{smith12}
{Smith}, R.~J., {Lucey}, J.~R., \& {Carter}, D. 2012, \mnras, 426, 2994,
  \dodoi{10.1111/j.1365-2966.2012.21922.x}

\bibitem[{{Spiniello} {et~al.}(2014){Spiniello}, {Trager}, {Koopmans}, \&
  {Conroy}}]{spiniello14}
{Spiniello}, C., {Trager}, S., {Koopmans}, L. V.~E., \& {Conroy}, C. 2014,
  \mnras, 438, 1483, \dodoi{10.1093/mnras/stt2282}

\bibitem[{{Thomas} {et~al.}(2003){Thomas}, {Maraston}, \& {Bender}}]{thomas03}
{Thomas}, D., {Maraston}, C., \& {Bender}, R. 2003, \mnras, 339, 897,
  \dodoi{10.1046/j.1365-8711.2003.06248.x}

\bibitem[{{Thomas} {et~al.}(2011{\natexlab{a}}){Thomas}, {Maraston}, \&
  {Johansson}}]{tmj}
{Thomas}, D., {Maraston}, C., \& {Johansson}, J. 2011{\natexlab{a}}, \mnras,
  412, 2183, \dodoi{10.1111/j.1365-2966.2010.18049.x}

\bibitem[{{Thomas} {et~al.}(2011{\natexlab{b}}){Thomas}, {Saglia}, {Bender},
  {Thomas}, {Gebhardt}, {Magorrian}, {Corsini}, {Wegner}, \&
  {Seitz}}]{thomasj11c}
{Thomas}, J., {Saglia}, R.~P., {Bender}, R., {et~al.} 2011{\natexlab{b}},
  \mnras, 415, 545, \dodoi{10.1111/j.1365-2966.2011.18725.x}

\bibitem[{{Tody}(1993)}]{tody93}
{Tody}, D. 1993, in Astronomical Society of the Pacific Conference Series,
  Vol.~52, Astronomical Data Analysis Software and Systems II, ed. R.~J.
  {Hanisch}, R.~J.~V. {Brissenden}, \& J.~{Barnes}, 173

\bibitem[{{Tortora} {et~al.}(2013){Tortora}, {Romanowsky}, \&
  {Napolitano}}]{tortora13}
{Tortora}, C., {Romanowsky}, A.~J., \& {Napolitano}, N.~R. 2013, \apj, 765, 8,
  \dodoi{10.1088/0004-637X/765/1/8}

\bibitem[{{Trager} {et~al.}(1998){Trager}, {Worthey}, {Faber}, {Burstein}, \&
  {Gonz{\'a}lez}}]{trager98}
{Trager}, S.~C., {Worthey}, G., {Faber}, S.~M., {Burstein}, D., \&
  {Gonz{\'a}lez}, J.~J. 1998, \apjs, 116, 1, \dodoi{10.1086/313099}

\bibitem[{{Treu} {et~al.}(2010){Treu}, {Auger}, {Koopmans}, {Gavazzi},
  {Marshall}, \& {Bolton}}]{treu10}
{Treu}, T., {Auger}, M.~W., {Koopmans}, L. V.~E., {et~al.} 2010, \apj, 709,
  1195, \dodoi{10.1088/0004-637X/709/2/1195}

\bibitem[{{van Dokkum} {et~al.}(2017){van Dokkum}, {Conroy}, {Villaume},
  {Brodie}, \& {Romanowsky}}]{vandokkum17}
{van Dokkum}, P., {Conroy}, C., {Villaume}, A., {Brodie}, J., \& {Romanowsky},
  A.~J. 2017, \apj, 841, 68, \dodoi{10.3847/1538-4357/aa7135}

\bibitem[{{van Dokkum} \& {Conroy}(2010)}]{vandokkum10}
{van Dokkum}, P.~G., \& {Conroy}, C. 2010, \nat, 468, 940,
  \dodoi{10.1038/nature09578}

\bibitem[{{van Dokkum} \& {Conroy}(2011)}]{vandokkum11}
---. 2011, \apjl, 735, L13, \dodoi{10.1088/2041-8205/735/1/L13}

\bibitem[{{van Dokkum} \& {Conroy}(2012)}]{vandokkum12}
---. 2012, \apj, 760, 70, \dodoi{10.1088/0004-637X/760/1/70}

\bibitem[{{Vaughan} {et~al.}(2018){Vaughan}, {Davies}, {Zieleniewski}, \&
  {Houghton}}]{vaughan18}
{Vaughan}, S.~P., {Davies}, R.~L., {Zieleniewski}, S., \& {Houghton}, R. C.~W.
  2018, \mnras, 479, 2443, \dodoi{10.1093/mnras/sty1434}

\bibitem[{{Vazdekis} {et~al.}(2016){Vazdekis}, {Koleva}, {Ricciardelli},
  {R{\"o}ck}, \& {Falc{\'o}n-Barroso}}]{vazdekis16}
{Vazdekis}, A., {Koleva}, M., {Ricciardelli}, E., {R{\"o}ck}, B., \&
  {Falc{\'o}n-Barroso}, J. 2016, \mnras, 463, 3409,
  \dodoi{10.1093/mnras/stw2231}

\bibitem[{{Vazdekis} {et~al.}(2012){Vazdekis}, {Ricciardelli}, {Cenarro},
  {Rivero-Gonz{\'a}lez}, {D{\'\i}az-Garc{\'\i}a}, \&
  {Falc{\'o}n-Barroso}}]{vazdekis12}
{Vazdekis}, A., {Ricciardelli}, E., {Cenarro}, A.~J., {et~al.} 2012, \mnras,
  424, 157, \dodoi{10.1111/j.1365-2966.2012.21179.x}

\bibitem[{{Vazdekis} {et~al.}(2010){Vazdekis}, {S{\'a}nchez-Bl{\'a}zquez},
  {Falc{\'o}n-Barroso}, {Cenarro}, {Beasley}, {Cardiel}, {Gorgas}, \&
  {Peletier}}]{vazdekis10}
{Vazdekis}, A., {S{\'a}nchez-Bl{\'a}zquez}, P., {Falc{\'o}n-Barroso}, J.,
  {et~al.} 2010, \mnras, 404, 1639, \dodoi{10.1111/j.1365-2966.2010.16407.x}

\bibitem[{{Vazdekis} {et~al.}(2015){Vazdekis}, {Coelho}, {Cassisi},
  {Ricciardelli}, {Falc{\'o}n-Barroso}, {S{\'a}nchez-Bl{\'a}zquez}, {La
  Barbera}, {Beasley}, \& {Pietrinferni}}]{vazdekis15}
{Vazdekis}, A., {Coelho}, P., {Cassisi}, S., {et~al.} 2015, \mnras, 449, 1177,
  \dodoi{10.1093/mnras/stv151}

\bibitem[{{Worthey} \& {Ottaviani}(1997)}]{worthey97}
{Worthey}, G., \& {Ottaviani}, D.~L. 1997, \apjs, 111, 377,
  \dodoi{10.1086/313021}

\bibitem[{{Worthey} {et~al.}(2014){Worthey}, {Tang}, \& {Serven}}]{worthey14}
{Worthey}, G., {Tang}, B., \& {Serven}, J. 2014, \apj, 783, 20,
  \dodoi{10.1088/0004-637X/783/1/20}

\bibitem[{{Xu} {et~al.}(2010){Xu}, {Narayanan}, \& {Walker}}]{xu10}
{Xu}, X., {Narayanan}, D., \& {Walker}, C. 2010, \apjl, 721, L112,
  \dodoi{10.1088/2041-8205/721/2/L112}

\bibitem[{{Zieleniewski} {et~al.}(2015){Zieleniewski}, {Houghton}, {Thatte}, \&
  {Davies}}]{zieleniewski15}
{Zieleniewski}, S., {Houghton}, R. C.~W., {Thatte}, N., \& {Davies}, R.~L.
  2015, \mnras, 452, 597, \dodoi{10.1093/mnras/stv1251}

\bibitem[{{Zieleniewski} {et~al.}(2017){Zieleniewski}, {Houghton}, {Thatte},
  {Davies}, \& {Vaughan}}]{zieleniewski17}
{Zieleniewski}, S., {Houghton}, R. C.~W., {Thatte}, N., {Davies}, R.~L., \&
  {Vaughan}, S.~P. 2017, \mnras, 465, 192, \dodoi{10.1093/mnras/stw2712}

\end{thebibliography}
\bibliographystyle{aasjournal}



\end{document}